\documentclass[11pt,a4paper]{article}
\pdfoutput=1
\usepackage{jheppub}

\allowdisplaybreaks
\usepackage{xcolor}
\usepackage{enumitem} 
\usepackage{amsmath}
\usepackage{subcaption}
\usepackage{graphicx}
\usepackage{esint}
\usepackage{slashed}
\usepackage{bigints}
\usepackage[english]{babel}
\usepackage{comment}
\usepackage{hyperref}

\usepackage{babel}
\begin{document}

\title{Revisiting supernova constraints on dark gauge bosons}

\author{Zun Wang$^{1}$, Yongchao Zhang$^{1,2}$}

\affiliation{$^1$School of Physics, Southeast University, Nanjing 211189, China}
\affiliation{$^2$Center for High Energy Physics, Peking University, Beijing 100871, China}

\emailAdd{wangzun@seu.edu.cn}
\emailAdd{zhangyongchao@seu.edu.cn}

\abstract{We revisit the supernova constraints on three well-motivated light gauge bosons: the dark photon, the $U(1)_{B-L}$ gauge boson, and the $U(1)_{L_\mu-L_\tau}$ gauge boson. For the dark photon and the $U(1)_{B-L}$ gauge boson, we re-evaluate the nucleon bremsstrahlung process within the one-pion exchange approximation, retaining the full momentum and mass dependence. 
The geometrical dependence of the decay and absorption of these gauge bosons inside the supernova core is also taken into account. 
Updated bounds on these models are derived based on the cooling of SN1987A and energy deposition constraints from low-energy supernovae (LESNe). 
Taking the SFHo-18.8 $18.8\,M_\odot$ profile, the SN1987A cooling constraints reach masses of about $230$ MeV, $280$ MeV, and $300$ MeV for the   dark photon, the $U(1)_{B-L}$ boson, and the $U(1)_{L_\mu-L_\tau}$ boson, while the corresponding LESN limits extend to about $450$ MeV, $440$ MeV, and $600$ MeV, respectively. 
The kinetic mixing parameter $\epsilon$ for dark photon,
the gauge coupling $g_{BL}$ in the $U(1)_{B-L}$ model,  and the gauge coupling $g_{\mu\tau}$ for the $U(1)_{L_\mu-L_\tau}$ model are constrained down to 
$5.8\times10^{-12}$, $2.9\times10^{-16}$, and $6.4\times10^{-12}$, respectively.
}

\maketitle

\section{Introduction}

In light of the absence of new heavy particles observed at the Large Hadron Collider, increasing attention has been devoted to the possibility that new physics beyond the Standard Model (BSM) may emerge at low energies, in particular below the GeV scale~\cite{Antel:2023hkf,Arza:2026rsl}. These light particles can be searched for in high-intensity laboratory experiments~\cite{Batell:2022dpx}. If sufficiently light, these particles can also be produced in compact astrophysical objects such as supernovae and neutron stars, where they may significantly affect the properties and evolution of these stars. This makes these astrophysical systems ideal natural laboratories for studying such particles~\cite{Raffelt:1996wa}. One of the main approaches to probing these new particles in supernovae is the stellar cooling argument: if light particles produced in the supernova core can freely escape, they may carry away excessive energy and thus alter the observed neutrino signals of SN1987A by Kamiokande II~\cite{Kamiokande-II:1987idp}, IMB~\cite{Bionta:1987qt} and the Baksan  collaboration~\cite{Alekseev:1987ej}. A complementary approach is to consider the energy deposition from particles that escape the core and subsequently decay or scatter in the stellar mantle~\cite{Fiorillo:2025yzf}, which can be constrained by observations of low-energy supernovae (LESNe)~\cite{Chugai:1999en,Pastorello:2003tc,Burrows:2020qrp,Kitaura:2005bt,Melson:2015tia,Radice:2017ykv,Muller:2018utr,Burrows:2019rtd,Stockinger:2020hse}.

In this paper, we revisit the supernova limits on the dark gauge bosons, which constitute one of the leading astrophysical constraints on these light BSM particles. In particular, we focus on three well-motivated $U(1)$ extensions of the Standard Model (SM). 
\begin{itemize}
    \item The dark photon $A'$~\cite{Holdom:1985ag} (see Refs.~\cite{Ilten:2018crw,Bauer:2018onh,Fabbrichesi:2020wbt,Caputo:2021eaa} for reviews). The dark photon arises from an extra $U(1)^\prime$ gauge group and interacts with the charged SM fermions exclusively through the kinetic mixing with the SM hypercharge, measured by the mixing parameter $\epsilon$.  
    As a generic gauge interaction, the dark photon has been extensively studied as a popular dark matter (DM)  candidate~\cite{Redondo:2008ec,Arias:2012az,McDermott:2019lch,Cyncynates:2024yxm,Zhang:2025pgb} or a dark force mediator, yielding a large variety of (low-energy) observational consequences in collider experiments~\cite{Bjorken:2009mm,Alekhin:2015byh,Gardner:2015wea,Chou:2016lxi,Feng:2017uoz,Ilten:2018crw,Berlin:2018pwi,MATHUSLA:2020uve,Belle-II:2022jyy,Bickendorf:2022buy,Brdar:2022vum,Araki:2023xgb,Zhou:2024aeu,Araki:2024uad,Davoudiasl:2025rpn,Araki:2026wfe,Angel:2026hfp,Lu:2026vsr,Deng:2026zna,Fabbrichesi:2026xnh}, DM experiments~\cite{An:2013yua,An:2014twa,Suzuki:2015sza,Bloch:2016sjj,Knirck:2018ojz,Baryakhtar:2018doz,Phipps:2019cqy,Nguyen:2019xuh,Hochberg:2019cyy,Brun:2019kak,SuperCDMS:2019jxx,XENON:2019gfn,DAMIC:2019dcn,SENSEI:2020dpa,Tomita:2020usq,FUNKExperiment:2020ofv,Dixit:2020ymh,Gelmini:2020kcu,XENON:2021qze,XENON:2022ltv,Manenti:2021whp,Chiles:2021gxk,Cervantes:2022yzp,DarkSide:2022knj,DOSUE-RR:2022ise,Fan:2022uwu,Ramanathan:2022egk,Bajjali:2023uis,BREAD:2023xhc,ALPHA:2022rxj,Levine:2024noa,Koppell:2025dmt,Fan:2024mhm,Demirci:2026oaf}, and neutrino scattering experiments~\cite{Lindner:2018kjo}. There are also extensive studies of the rich signatures or constraints from the astrophysical~\cite{Fedderke:2021rrm,Fedderke:2021aqo, Redondo:2008aa,An:2013yfc,Redondo:2013lna,Vinyoles:2015aba,Li:2023vpv,An:2020bxd,An:2020jmf,An:2023wij,An:2024wmc,     Bjorken:2009mm,Dent:2012mx,Kazanas:2014mca,Rrapaj:2015wgs,Chang:2016ntp,Sung:2021swd,Shin:2022ulh,Syvolap:2024hdh,Vogl:2024ack,Caputo:2025aac,Caputo:2025avc,Cheng:2026ugs,Fortin:2019npr,Hong:2020bxo,    Zechlin:2008tj,Bi:2020ths,Baryakhtar:2017ngi,Wadekar:2019mpc,Dubovsky:2015cca,An:2022hhb,Linden:2024fby,Cheng:2026zdc}  and cosmological~\cite{Jaeckel:2008fi,Fradette:2014sza,Berger:2016vxi,Pospelov:2018kdh,McDermott:2019lch,Ibe:2019gpv,Caputo:2020bdy,Caputo:2020rnx,Witte:2020rvb,Coffey:2020oir,Li:2020roy,Caputo:2022keo,Adshead:2022ovo,Pirvu:2023lch,Gan:2023wnp,Cyncynates:2023zwj,Cyncynates:2024yxm,Trost:2024ciu,Aramburo-Garcia:2024cbz,McCarthy:2024ozh,Hook:2025pbn,Caputo:2025avc,Xu:2025wlq} observations. 
    
    \item The $Z'$ gauge boson in the $U(1)_{B-L}$ model~\cite{Davidson:1978pm,Marshak:1979fm,Mohapatra:1980qe,Wetterich:1981bx} (see Refs.~\cite{Ilten:2018crw,Bauer:2018onh} for reviews).    
    The $U(1)_{B-L}$ gauge symmetry is a well-motivated extension of the SM that provides a natural framework to account for the tiny neutrino masses via the type-I seesaw mechanism~\cite{Minkowski:1977sc,Mohapatra:1979ia,Yanagida:1979as,Gell-Mann:1979vob,Glashow:1979nm}, while also accommodating DM  candidates~\cite{Marshak:1979fm,Mohapatra:1980qe}. In scenarios where the $U(1)_{B-L}$ gauge symmetry breaking occurs at a low scale, the associated gauge coupling $g_{BL}$ can be very small. This naturally yields a light gauge boson $Z'$ in the sub-GeV range without spoiling the viability of the seesaw mechanism, leading to rich phenomenology at the     collider experiments~\cite{Ilten:2018crw,NA64:2022yly,Asai:2022zxw,Asai:2023mzl,Felkl:2023nan,Davoudiasl:2025rpn},  neutrino  experiments~\cite{Lindner:2018kjo,Dev:2021qjj,Asai:2022zxw,Melas:2023olz,Felkl:2023nan}, and DM direct detection experiments~\cite{Harnik:2012ni,Demirci:2026oaf}. 
    There are also astrophysical limits from the Sun~\cite{Hardy:2016kme,Li:2023vpv}, supernovae~\cite{Hardy:2016kme,Asai:2022zxw,Shin:2022ulh,Akita:2023iwq,Vogl:2024ack,Knapen:2017xzo,Croon:2020lrf,Shin:2021bvz,Esseili:2025sbc}, and neutron stars~\cite{Hong:2020bxo,Shin:2021bvz}, as well as the cosmological constraints~\cite{Coffey:2020oir,Adshead:2022ovo,Barman:2026btd}.

    \item The $Z'$ gauge boson in the $U(1)_{L_\mu-L_\tau}$ model~\cite{Foot:1990mn,He:1990pn,He:1991qd} (see Ref.~\cite{Bauer:2018onh} for review).     
    The flavor-dependent gauge extension $U(1)_{L_\mu-L_\tau}$ of the SM is anomaly-free, with the corresponding gauge boson $Z'$ coupling exclusively to the second and third generations of leptons, i.e. the muon $\mu$, tau lepton $\tau$ and the accompanying neutrinos $\nu_{\nu,\,\tau}$. Such a setup naturally leads to the $\mu - \tau$ symmetric lepton mixing and introduces flavor specific non-standard neutrino interactions~\cite{Heeck:2011wj}.  
    There have been some constraints from the laboratory experiments~\cite{Andreas:2012mt,Abdullah:2018ykz,NA64:2024klw,CCFR:1991lpl, Kaneta:2016uyt,Bellini:2011rx,Kopp:2012dz,BaBar:2009lbr,BaBar:2014zli}, the Sun~\cite{Li:2023vpv}, supernova~\cite{Escudero:2019gzq,Croon:2020lrf,Akita:2023iwq,Manzari:2023gkt,Lai:2024mse,Vogl:2024ack,Blinov:2025aha,Das:2026uam}, neutron stars~\cite{Fiorillo:2026wso,Fiorillo:2026dqu}, and cosmological~\cite{Coffey:2020oir} observations.

\end{itemize}

Since these models feature distinct coupling properties, based on kinetic mixing ($\epsilon$), universal flavor ($g_{BL}$) and flavor-dependent ($g_{\mu\tau}$) interactions, respectively, the corresponding predominant production channels in the supernova core and potential signals and constraints differ significantly. 
\begin{itemize}
    \item 
    The production of $A'$ within the supernova core is dominated by the proton-neutron bremsstrahlung process $np \to np A'$.  
    The non-negligible abundance of $\pi^-$ inside the supernova cores~\cite{Fore:2019wib} is also important for the production of dark photon, although the contribution is subleading~\cite{Shin:2022ulh}. 
    In addition to the cooling bound from the observations of neutrinos events from SN1987A~\cite{Bjorken:2009mm,Dent:2012mx,Kazanas:2014mca,Rrapaj:2015wgs,Chang:2016ntp,Caputo:2025avc}, there are also constraints from the $\gamma$-rays and positron observations~\cite{Shin:2022ulh,Caputo:2025avc}, neutrino signals from the decays $A' \to \mu^+ \mu^-,\; \pi^+ \pi^-$~\cite{Syvolap:2024hdh}, the bounce shock and the subsequent supernova explosions~\cite{Caputo:2025aac}, fireball formation~\cite{Caputo:2025avc}, circumstellar medium surrounding the stars~\cite{Cheng:2026ugs,Cheng:2026zdc}, LESNe~\cite{Caputo:2025avc}, and neutron stars~\cite{Fortin:2019npr,Hong:2020bxo}. Some of the supernova limits on dark photon may also involve light DM particles~\cite{Sung:2021swd,Vogl:2024ack}.

    \item The $Z'$ boson in the $U(1)_{B-L}$ model couples both to protons and neutrons, which is different from the dark photon. 
    With the dominant nucleon bremsstrahlung production process of $NN \to NN Z'$ (with $N=p,\,n$) and the subdominant contributions of the $\pi^-$-induced processes~\cite{Shin:2022ulh}, the supernova limits on the $Z'$ boson are distinct from that on the dark photon~\cite{Hardy:2016kme,Asai:2022zxw,Shin:2022ulh,Akita:2023iwq,Vogl:2024ack,Knapen:2017xzo,Croon:2020lrf,Shin:2021bvz,Esseili:2025sbc} (see also Ref.~\cite{Hong:2020bxo,Shin:2021bvz} for the neutron star limits). The absence of high-energy neutrinos from the decay $Z' \to \nu \bar\nu$ from SN1987A is also used to constrain the coupling of $Z'$ boson~\cite{Akita:2023iwq}. 
    
    \item The production of the $Z'$ gauge boson in the $U(1)_{L_\mu - L_\tau}$ model is dominated by the neutrino coalescence $\nu \bar\nu \to Z'$ and the semi-Compton (sC) scattering 
    $\mu\gamma\to\mu Z'$ (the abundance of muon inside the supernova cores is significant~\cite{Bollig:2017lki})~\cite{Escudero:2019gzq,Croon:2020lrf,Akita:2023iwq,Manzari:2023gkt,Lai:2024mse,Vogl:2024ack,Blinov:2025aha,Das:2026uam}. 
    In addition, there are also constraints on the $Z'$ boson in this model from LESNe~\cite{Blinov:2025aha} and the SN1987A neutrinos~\cite{Akita:2023iwq,Blinov:2025aha}.
\end{itemize}

In this work, we revisit these supernova constraints on the dark photon $A'$, the $U(1)_{B-L}$ $Z'$ boson and the $Z'$ boson in the $U(1)_{L_\mu - L_\tau}$ model, by consistently treating the production and transport of light gauge bosons in a dense environment. We focus mainly on the constraints on these gauge bosons from the cooling of SN1987A and energy deposition in the LESNe. The geometric effects are fully taken into account, which are important for the absorption and decay of these gauge bosons inside the supernova core~\cite{Caputo:2022rca,Balaji:2022noj,Lai:2024mse,Ge:2024cto,Caputo:2021rux,Fiorillo:2025yzf,Li:2025beu,Das:2026uam}. 
For the nucleon bremsstrahlung production of the dark photon and the $U(1)_{B-L}$ $Z'$ gauge boson, we adopt the one-pion exchange (OPE) approximation, and retain the full momentum and gauge boson mass dependence up to the next-to-leading order~\cite{Dev:2020eam,Joseph:2026nut}. We also incorporate the in-medium plasma effect, which modifies the effective couplings, resulting in the leading contributions for the production of dark photon and $U(1)_{B-L}$ $Z'$ boson in large regions of parameter space~\cite{Hardy:2016kme}. 

The calculation of the $np$ bremsstrahlung production of dark photon is a direct generalization of that for the $Z'$ boson in the $U(1)_{B-L}$ model. As a result, we will first consider the case of the $U(1)_{B-L}$ $Z'$ boson, and then the dark photon.
In addition, there are some uncertainties associated with the supernova profiles, by roughly a factor of two, for the baryon density, temperature, and other quantities in the stellar core. Taking this into account, we adopt the supernova profiles Fischer 11.8$M_\odot$~\cite{Fischer:2016cyd}, SFHo-18.8 18.8$M_\odot$ and LS220 20$M_\odot$~\cite{Mirizzi:2015eza,Bollig:2017lki,Bollig:2020xdr} as estimates for the uncertainties of SN1987A bounds on these gauge bosons. If we take the baryon density and temperature to be constants inside the supernova core with size $R_{C} = 10$ km, i.e. $n_B=1.2 \times 10^{38}\; \mathrm{cm}^{-3}$ and $T = 30$ MeV, the calculations can be greatly simplified, which is a good reference for the corresponding SN1987A limits~\cite{Balaji:2022noj}. For the LESN limits, we take only the supernova profile of SFHo-18.8 18.8$M_\odot$. With these setups, the updated supernova constraints on these gauge bosons are as follows, taking the SFHo-18.8 $18.8M_\odot$ profile as a representative example.
\begin{itemize}
    \item The gauge coupling $g_{BL}$ is excluded by the SN1987A cooling within the range of $1.7 \times10^{-11}$ to $4.6\times10^{-8}$, with the gauge boson mass $m_{Z'}$ excluded up to 145 MeV for transverse modes (cf. Fig.~\ref{fig:BL}). For the longitudinal mode, the corresponding excluded coupling range is $2.9\times10^{-16}\lesssim g_{BL}\lesssim 5.8\times10^{-12}$, while the excluded mass range extends up to $m_{Z'}\simeq 280$ MeV (cf. Fig.~\ref{fig:BL}).  With the luminosity constraints over two orders of magnitude stronger, the LESN data have excluded larger regions of parameter space, with $1.9 \times10^{-12} \lesssim g_{BL} \lesssim 3.1\times10^{-7}$ and $m_{Z'} \lesssim 440$ MeV (cf. Fig.~\ref{fig:BL:LESNe}).
    
    \item For the dark photon $A'$, the kinetic mixing parameter $\epsilon$ is excluded by the SN1987A cooling within the range of $1.7\times 10^{-10}$ to $2\times 10^{-7}$, with the dark photon mass $m_{A'}$ constrained up to roughly $230$ MeV in the transverse modes (cf. Fig.~\ref{fig:DP}). No parameter space is excluded in the longitudinal mode. With the stronger luminosity constraints, the LESN data exclude a larger region of parameter space, with $5.8\times 10^{-12}\lesssim\epsilon\lesssim 6\times10^{-7}$ and $m_{A'}\lesssim450$ MeV (cf. Fig.~\ref{fig:DP:LESNe}).
    
    \item For the $U(1)_{L_\mu-L_\tau}$ $Z'$ gauge boson, the gauge coupling $g_{\mu\tau}$ is excluded by the SN1987A cooling within the range of $1.4\times 10^{-10}$ to $6.7\times 10^{-7}$, and the mass $m_{Z'}$ is excluded up to roughly $300$ MeV (cf. Fig.~\ref{fig:mutau:SN1987A}). The LESN constraints exclude more parameter space, within the ranges of $ 6.4\times 10^{-12}\lesssim g_{\mu\tau}\lesssim9\times 10^{-7}$ and $m_{Z'}\lesssim600$ MeV (cf. Fig.~\ref{fig:mutau:SN1987A}). 
    
    \item The updated supernova limits obtained in this paper are largely complementary to other constraints. Qualitatively speaking, the supernova observations are more sensitive to smaller couplings than the laboratory experiments, and can also extend the cosmological limits on the dark gauge bosons up to larger masses ranges (cf.~Figs.~\ref{fig:BL:all_limits}, \ref{fig:DP:all_limits} and \ref{fig:mutau:all_limits}). 
\end{itemize}

The paper is organized as follows. The geometric effects are discussed in  Section~\ref{sec.setups}. 
The constraints of SN1987A cooling and LESNe on the $Z'$ boson in the $U(1)_{B-L}$ model are derived in Section~\ref{sec:BL},  and the corresponding limits on the dark photon are obtained in Section~\ref{sec:DP}. 
In this section, we also collect the main differences of the supernova constraints on the $U(1)_{B-L}$ $Z'$ boson and dark photon. The constraints on the $U(1)_{L_\mu-L_\tau}$ gauge boson from the SN1987A cooling and the LESN observations are obtained in  Section~\ref{sec.Lmutau}, before we summarize the main results and conclude in Section~\ref{sec.conclusion}. 
The calculation details of the squared matrix elements for the $U(1)_{B-L}$ gauge boson are collected in Appendix~\ref{app:ampsq:BL}, while some additional details of the semi-Compton production of the $U(1)_{L_\mu-L_\tau}$ gauge boson are provided in Appendix~\ref{app:semi-Compton}.

\section{Setups: geometric effects}
\label{sec.setups}

The general energy emission rate per unit volume in the supernova core for a specific process of the particle $X$ can be calculated as~\cite{Chang:2016ntp} 
\begin{equation}
\label{eqn:emission_rate}
    {\cal Q}_X 
    \ = \ \int\frac{{\rm d}^3k}{(2\pi)^3} E_X \Gamma^\mathrm{prod}_{X}\mathrm{exp}\{-\tau_X\} \,,
\end{equation}
where $E_X$ is the energy of the particle $X$, $\Gamma^\mathrm{prod}_{X}$ is the production rate, 
and $\tau_X$ is the corresponding optical depth that measures the survival possibility of the particle $X$. 
For the gauge bosons such as the dark photon and the $Z'$ bosons in this paper, the energy emission rates in the transverse and longitudinal modes should be calculated separately. 
Both the production rate $\Gamma_X^{\rm prod}$ and the optical depth $\tau_X$ are proportional to the squared coupling $g_X$ involved, i.e. $\Gamma_X^{\rm prod} \propto g_X^2$ and $\tau_X \propto g_X^2$.\footnote{This scaling holds true when all the relevant production, absorption, scattering, and decay rates are dominated by interactions involving a single coupling \(g_X\). In the presence of several different couplings, the interference terms may lead to a more complicated dependence on the couplings.} 
As a result, when the coupling $g_X$ is small, the emission rate ${\cal Q}_X$ is dominated by the production rate $\Gamma^{\rm prod}_X$. In the regime of large couplings, the emission rate is predominantly modulated by the optical depth $\tau_X$ via the exponent factor $\exp\{-\tau_X\}$ in Eq.~(\ref{eqn:emission_rate}). 
The optical depth accumulated by a particle along its trajectory quantifies its survival probability in the stellar medium and is given by~\cite{Fiorillo:2025yzf,Caputo:2022rca,Balaji:2022noj},
\begin{equation}
\label{eqn:optiacal_depth}
    \tau_X(r_\ast,r,\phi) \ = \ \int_0^{{d}(r_\ast,r  ,\phi)} {\rm d}r^\prime \Big( \Gamma^\mathrm{decay}_{X}[L(r,\phi;r^\prime)]+\Gamma^\mathrm{abs}_{X}[L(r,\phi;r^\prime)] \Big) \,,
\end{equation}
where $r$ is the radial coordinate at which the particle is produced, $r_\ast$ denotes the radius at which the optical depth is evaluated, and $\Gamma_X^{\rm decay}$ and $\Gamma_X^{\rm abs}$ are the decay and absorption rates of $X$, respectively. 
The scattering of the particle $X$ with the medium particles, such as nucleons and electrons, could also contribute to the optical depth. However, the corresponding rates are at the higher order in the coupling $g_X$, proportional to $g_X^4$, and are thus subleading to the decay and absorption contributions in Eq.~(\ref{eqn:optiacal_depth}). For simplicity, we neglect the contributions from the scattering processes throughout this paper.

In Eq.~(\ref{eqn:optiacal_depth}), we have taken into account explicitly the geometric effects for both the decay and absorption of $X$ inside the supernova core, i.e. depending on where the particle $X$ is produced and its flight direction inside the star~\cite{Caputo:2022rca,Balaji:2022noj,Lai:2024mse,Ge:2024cto,Caputo:2021rux,Fiorillo:2025yzf,Li:2025beu,Das:2026uam}. For simplicity, we assume the supernova core is spherically symmetric. Following Ref.~\cite{Balaji:2022noj} (in particular Fig.~3 therein), the parameter $L$ in Eq.~(\ref{eqn:optiacal_depth}) quantifies the instantaneous distance between the particle $X$ and the geometric center of the stellar core, defined as
\begin{equation}
L\left(r, \phi ; r^{\prime}\right) \ \equiv \ \sqrt{r^2+r^{\prime 2}+2 r r^{\prime} \cos \phi} \,,
\end{equation}
where $\phi$ is the polar angle between the 3-momentum of $X$ and the $z$-direction (the $z$-direction is defined
as the direction from the stellar center to the production site of $X$), 
and $r^\prime$ is introduced as an auxiliary integration variable. The trajectory $X$ travels inside the star before decaying or being absorbed is
\begin{equation}
\label{eqn:geometry:d}
d(r_\ast,r, \phi) \ \equiv \ \sqrt{r_\ast^2-r^2 \sin ^2 \phi}-r \cos \phi \,,
\end{equation}
with $R_{\nu}$ being the neutrinosphere radius of the star.

Although the constraints from SN1987A on new physics scenarios are always very stringent, there are still some uncertainties on the progenitor neutron star (PNS) itself for the SN1987A explosion. In particular, there is roughly a factor of two for the mass of the PNS, and the corresponding uncertainties for the temperature and density profiles are also at the same order~\cite{Sukhbold:2015wba}. To estimate the profile dependence of the SN1987A limits, we consider the Fischer $11.8M_{\odot}$~\cite{Fischer:2016cyd}, SFHo-18.8 $18.8M_{\odot}$ and LS220 $20M_{\odot}$~\cite{Mirizzi:2015eza,Bollig:2017lki,Bollig:2020xdr}
profiles at $1$ sec after core bounce. For the SFHo-18.8 $18.8M_{\odot}$ and LS220 $20M_{\odot}$ profiles, we use the
radially dependent quantities provided by the corresponding simulations~\cite{Mirizzi:2015eza,Bollig:2017lki,Bollig:2020xdr}.\footnote{The full profile data for SFHo-18.8 $18.8M_{\odot}$ and LS220 $20M_{\odot}$ are available from the Garching Core-Collapse Supernova Archive: \url{https://wwwmpa.mpa-garching.mpg.de/ccsnarchive/archive.html}.}
The neutron and proton number densities are obtained from the baryon density and the corresponding fractions, while the chemical potentials of electrons, muons, and neutrinos are taken directly from the profiles. 
In the calculations below, we adopt the effective nucleon mass $m_N=m_{N,0}+\Sigma_{S}$ in dense nuclear matter, where $m_{N,0}$ is the nucleon mass in vacuum, and $\Sigma_{S}$ denotes the contribution of the scalar self-energy~\cite{Carenza:2019pxu,Hempel:2014ssa}.

\section{$U(1)_{B-L}$ gauge boson}
\label{sec:BL}

This section is devoted to the supernova limits on the $Z'$ boson in the $U(1)_{B-L}$ model. Section~\ref{subsec:plasma} is on the plasma effect which is crucial for the supernova limits. Then we provide details for the production of $Z'$ in the supernova core in Section~\ref{subsec:Zprime:prod}, the decay of $Z'$ in Section~\ref{subsec:Zprime:decay} and then the details for the emission rate in Section~\ref{subsec:BL:rate}. The SN1987A cooling limits are obtained in Section~\ref{subsec:BL:SN1987A}, and the LESN constraints are derived in Section~\ref{subsec:BL:LESN}.

\subsection{Plasma effect}
\label{subsec:plasma}

For the purpose of anomaly cancellation, three right-handed neutrinos (RHNs) are naturally introduced to the $U(1)_{B-L}$ gauge theory, which is essential to implement the type-I seesaw for tiny neutrino masses. However, for simplicity we assume all the RHNs are very heavy and decouple from the low-energy phenomena in this paper. The couplings of the $U(1)_{B-L}$ gauge boson to the SM quarks $q$, charged leptons $\ell$ and neutrinos $\nu$ can be written as 
\begin{equation}
{\cal L}_{BL} \ = \ - g_{BL} Z_\mu^\prime \left( \frac13 \bar{q} \gamma^\mu q - \bar\ell \gamma^\mu \ell - \bar\nu \gamma^\mu \nu \right) \,,
\end{equation}
where $g_{BL}$ is the universal gauge coupling in the Lagrangian, and we do not show explicitly the generation indices. The $Z'$ boson may also mix with the SM photon, e.g. via the SM or heavy fermions charged under both the SM gauge group and $U(1)_{B-L}$ or from the pure plasma effects in medium~\cite{Raffelt:1987im,Li:2023vpv,Shin:2022ulh}. For the sake of simplicity, we neglect such a kinetic mixing. Then the tree-level couplings of the $Z'$ boson to protons and neutrons are both $g_{BL}$, i.e. $g_{p}^{} = g_{n}^{} = g_{BL}$.

In the dense medium such as the supernova cores, the photon will acquire a nonzero effective mass from interacting with the dense charged particles in the medium, i.e.  the so-called plasma effect~\cite{Hardy:2016kme}. The effective mixing of $Z'$ with the SM photon via the medium effect will change dramatically the couplings of $Z'$ boson~\cite{Li:2023vpv}. In particular,  the effective coupling of $Z'$ to protons will receive significant contribution from the plasma effect, whereas the coupling of $Z'$ to neutrons is not affected, at least at the leading order~\cite{Hong:2020bxo}. 
In the supernova core, the plasma effect is dominated by electrons, and the effective coupling of a generic $Z'$ boson to protons can be written as~\cite{Braaten:1993jw,Hardy:2016kme}:  
\begin{equation}
\label{eqn:effive_coupling} 
\tilde{g}_{p}^{T,L} \ = \ 
\frac{{g_{p}^{}m_{Z^\prime}^2-\left(g_{p}^{}+g_{e}^{}\right)\Pi_{\gamma\gamma}^{T,L}}}{m_{Z^{\prime}}^2-\Pi_{\gamma\gamma}^{T,L}} \,,
\end{equation}
where $g_e$ is the coupling of $Z'$ in the $U(1)_{B-L}$ model to electrons, $T$ and $L$ stand for the transverse and longitudinal modes, respectively, $\Pi_{\gamma\gamma}^{T,L}$ are the photon self-energy terms, which are in general complex. The real parts in the $T$ and $L$ modes for the relativistic and degenerate plasma can be written as, respectively~\cite{Braaten:1993jw,Chang:2016ntp},
\begin{subequations}
\begin{align}
\label{eqn:real}
\operatorname{Re} \Pi_{\gamma\gamma}^{T} & \ = \ \frac{3 \omega_{\rm pl}^2}{2 v^2}\left[1-\frac{1-v^2}{2 v} \log \left(\frac{1+v}{1-v}\right)\right] \,, \\ \operatorname{Re} \Pi_{\gamma\gamma}^{L} & \ = \ \frac{3 \omega_{\rm pl}^2}{v^2}\left(1-v^2\right)\left[\frac{1}{2 v} \log \left(\frac{1+v}{1-v}\right)-1\right] \,,
\end{align}
\end{subequations}
where $v$ is the photon velocity, $\omega_{\rm pl}=\sqrt{4 \pi \alpha_{} n_e / E_F}$ is the plasma frequency, with $\alpha$ being the fine-structure constant, $n_e$ the electron number density, and the Fermi energy $E_F=[m_e^2+\left(3 \pi^2 n_e\right)^{2/3}]^{1/2}$ (with $m_e$ being the electron mass).
The imaginary part of the polarization tensor is relevant to the absorption of the SM photon in the plasma~\cite{Weldon:1983jn,An:2013yfc}. Assuming the photons are in local thermal equilibrium in the stars, we have~\cite{Chang:2016ntp} 
\begin{equation}
\operatorname{Im} \Pi_{\gamma\gamma}^{T,L}(\omega) \ = \ -\omega\left(1-e^{-\omega / T}\right) \Gamma_{\gamma,\,\mathrm{abs}}^{T,L}(\omega) \,,
\end{equation}
where $\Gamma_{\gamma,\,\mathrm{abs}}(\omega)$ is the re-absorption width of SM photons with energy $\omega$ in the medium. More details can be found e.g. in Ref.~\cite{Chang:2016ntp}.

\begin{figure}[t]
    \centering
    \includegraphics[width=0.49\textwidth]{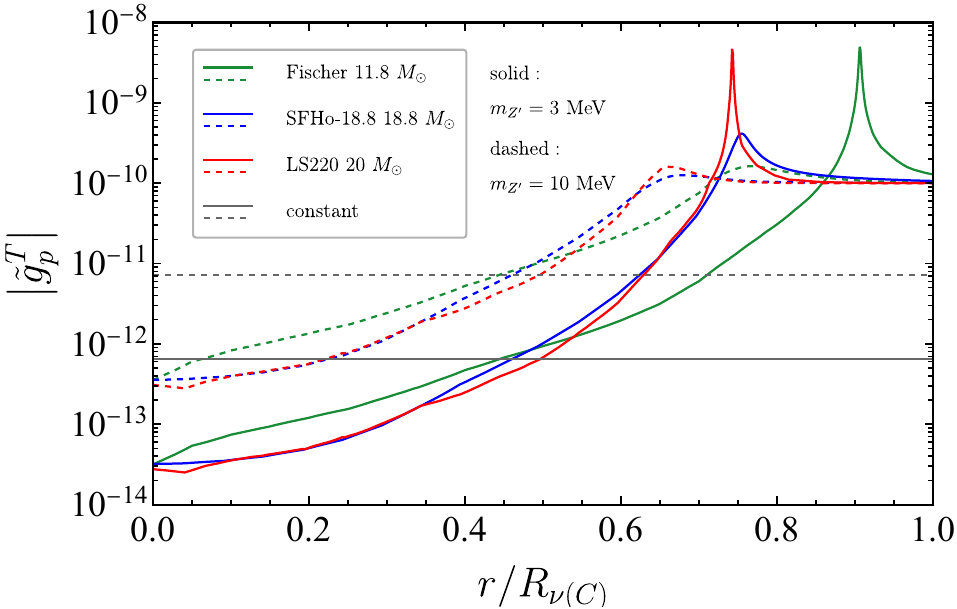} 
    \includegraphics[width=0.49\textwidth]{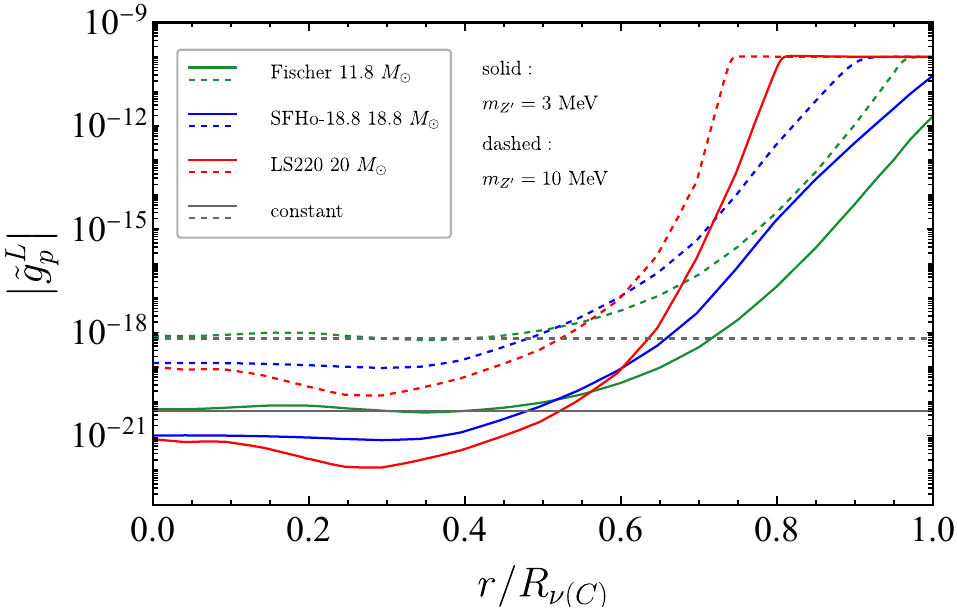}
    \caption{
    The absolute values of the effective couplings $|\tilde{g}_{p}^{T,L}|$ of the $U(1)_{B-L}$ $Z'$ boson  to protons in the transverse (left) and longitudinal (right) modes, as functions of the radius $r$ in unit of the neutrinosphere $R_\nu$ or the core size $R_C$ ($R_C$ is only for the constant profile case and $R_\nu$ for the other three profiles). 
    The values for the Fischer $11.8\,M_{\odot}$, SFHo-18.8 $18.8\,M_{\odot}$, LS220 $20\,M_{\odot}$ and constant supernova profiles are depicted in green, blue, red and gray, and the solid and dashed lines are for the masses $m_{Z'} = 3$ MeV and 10 MeV, respectively. We have set $g_p = 10^{-10}$ and $E_{Z'}=3T(r)$. See text for more details. 
    }
    \label{fig:coupling}
\end{figure}

In the $U(1)_{B-L}$ model, the couplings $g_{p}+ g_{e} = 0$, and the resulting effective coupling of $Z'$ to protons is proportional to the gauge boson mass $m_{Z^\prime}^2$ in Eq.~(\ref{eqn:effive_coupling}), i.e.
\begin{equation}
\label{eqn:effive_coupling2} 
\tilde{g}_{p}^{T,L}  = \frac{{g_{p}^{}m_{Z^\prime}^2}}{m_{Z^{\prime}}^2- {\rm Re} \Pi_{\gamma\gamma}^{T,L} - i {\rm Im} \Pi_{\gamma\gamma}^{T,L}} \,.
\end{equation}
One may expect a resonance at $\operatorname{Re}\Pi_{\gamma\gamma}^{T,L}\sim m^2_{Z^{\prime}}$ with the condition of $|\operatorname{Im}\Pi_{T,L}|\ll \operatorname{Re}\Pi_{\gamma\gamma}^{T,L}$.  
For illustration purposes, the absolute values of the effective couplings of $Z'$ to protons $|\tilde{g}_{p}^{T,L}|$ are shown in Fig.~\ref{fig:coupling} as function of the radius $r$ in unit of the neutrinosphere $R_\nu$.  The couplings in the transverse and longitudinal modes are shown in the left and right panels, and the lines for the Fischer $11.8\,M_{\odot}$, SFHo-18.8 $18.8\,M_{\odot}$, LS220 $20\,M_{\odot}$ profiles are depicted in green, blue and red, respectively. As a comparison, the effective couplings for the constant supernova profile case with
$n_B=1.2 \times 10^{38}\; \mathrm{cm}^{-3}$ and $T = 30$ MeV are presented as the gray lines, with the radius $r$ up to the core size of $R_C = 10$ km.  
We have taken the benchmark values of $m_{Z'} = 3$ MeV and 10 MeV, and the corresponding values are shown as the solid and dashed lines, respectively. 
In both panels we have set the coupling $g_{p}^{} = 10^{-10}$ and the $Z'$ boson energy $E_{Z'} = 3 T(r)$. 
It is very clear in Fig.~\ref{fig:coupling} that the resonance occurs only in the transverse mode, when the electron number density $n_e(r)$ is small enough such that $m_{Z^\prime}^2$ approaches ${\rm Re} \Pi_{\gamma\gamma}^{T} \propto n_e$. This occurs when $r \gtrsim 0.7 R_\nu$ for the profiles Fischer $11.8\,M_{\odot}$ and SFHo-18.8 $18.8\,M_{\odot}$, and $0.6 R_\nu \lesssim r \lesssim 0.9 R_\nu$ for LS220 $20\,M_{\odot}$, respectively, as seen in the left panels of Fig.~\ref{fig:coupling}. When it is away from the resonance, i.e. $|m_{Z^\prime}^2-{\rm Re} \Pi_{\gamma\gamma}^{T}| \gtrsim m_{Z^\prime}^2$, the effective coupling in the $T$ mode is suppressed by the gauge boson mass, i.e. $\tilde{g}_{p}^{T} \propto m_{Z^\prime}^2$.  In contrast, no resonance occurs in the constant profile case, since the electron number density $n_e$ is spatially uniform and therefore the resonance condition $m_{Z^\prime}^2 \simeq \text{Re}\,\Pi_{\gamma\gamma}^{T}$ cannot be satisfied at a particular radius.
For the longitudinal mode, there is no such a resonance throughout the neutrinosphere, as the imaginary part ${\rm Im} \Pi_{\gamma\gamma}^L$ can be large with respect to the corresponding real part ${\rm Re} \Pi_{\gamma\gamma}^L$, which differs from that in Refs.~\cite{Chang:2016ntp,Shin:2021bvz}.

\subsection{Production}
\label{subsec:Zprime:prod}

\begin{figure}[t]
\centering
\includegraphics[width=1\textwidth]{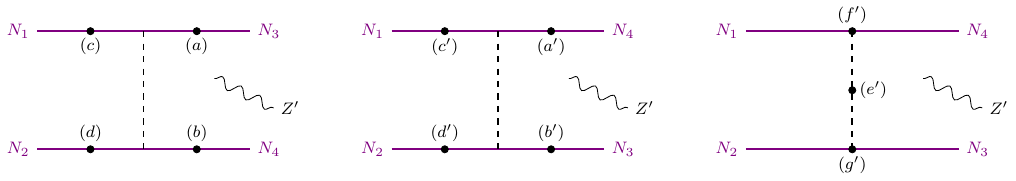}
\caption{Feynman diagrams for the production of the $Z'$ boson in the $U(1)_{B-L}$ model via the nucleon-nucleon bremsstrahlung process $N_1 +N_2\rightarrow N_3+N_4+Z'$ in the supernova core, with $N_i=p,n$. The left panel is for the $t$-channel, while the middle and right panels are for the $u$-channel. The $Z'$ boson can be attached to either of the nucleon lines $(a),\,(b),\,(c),\,(d)$ in the left panel and $(a'),\,(b'),\,(c'),\,(d')$ in the middle panel, or to the charged pion vertices $(e'),\, (f'),\ (g')$ in the right panel, as denoted by the blobs ($\bullet$). The couplings in the right panel exist only for the $np$ scattering process, which are induced by the interactions in Eq.~(\ref{eqn:coupling:Zprime:pion}).  
}
\label{fig:diagrams}
\end{figure}

The dominant production channel for the $U(1)_{B-L}$ gauge boson $Z'$ in the supernova core is the nucleon bremsstrahlung process:
\begin{equation}
\label{eqn:process:brem}
    N +N\rightarrow N+N+Z' \,,
\end{equation}
where $N = p,\,n$. In other words, $Z'$ can be produced in the scattering of $pp$, $nn$ and $np$ processes. In this work we take the OPE approximation, with the nucleons interaction mediated by a single pion~\cite{Dent:2012mx}. 
For the $np$ scattering, the $Z'$ boson couples not only to nucleons, but also to the charged pions in the $u$-channel. The relevant interactions up to the order of dimension-5 can be written as~\cite{Kroll:1953vq,Shin:2021bvz} 
\begin{equation}
\label{eqn:coupling:Zprime:pion}
{\cal L} \ = \ - \left( \tilde{g}_{p}^{} - g_{n}^{} \right) Z'_\mu \left[ \Big( (\partial^\mu \pi^+) \pi^- - \pi^+ (\partial^\mu \pi^-) \Big)
- \frac{f_{pn}}{m_\pi} \Big( \pi^+ \bar{p} \gamma^\mu \gamma_5 n - \pi^- \bar{n} \gamma^\mu \gamma_5 p \Big) \right] \,,
\end{equation}
where the pion-nucleon coupling $f_{pn} = \sqrt{2} f_{pp}$ with $f_{pp} \simeq 1$. In vacuum the couplings $g_{p}^{}$ and $g_{n}^{}$ are the same, and such couplings in Eq.~(\ref{eqn:coupling:Zprime:pion}) vanish. However, in the dense medium such as the supernova cores, the coupling of $Z'$ to protons receives extra contributions from the plasma effect (cf. Eq.~(\ref{eqn:effive_coupling2})), which makes the couplings in Eq.~(\ref{eqn:coupling:Zprime:pion}) non-vanishing. 
The corresponding Feynman diagrams for the nucleon bremsstrahlung production of the $Z'$ boson are presented in Fig.~\ref{fig:diagrams}. The left panel is for the diagrams in the $t$-channel, and the middle and right panels are for the $u$-channel diagrams. The $Z'$ boson can couple to either of the nucleon or pion lines or the vertices, labeled by the blobs in the diagrams $(a),\,(b),\,(c),\,(d)$,  $(a'),\,(b'),\,(c'),\,(d')$ and $(e'),\, (f'),\ (g')$. In total, there are 11 diagrams when the plasma effect is taken into account. 
Given the couplings in Eq.~(\ref{eqn:coupling:Zprime:pion}), the $Z'$ boson can also be produced in the supernova core via the $\pi^-$-induced process $\pi^- + p \to n + Z'$, which is similar to the production of axion in the supernova core~\cite{Carenza:2020cis,Carenza:2023lci}. However, the contributions from this process are found to be subleading~\cite{Shin:2022ulh}. For simplicity, we neglect the $\pi^-$-induced process in this paper.

For the squared matrix elements for the nucleon bremsstrahlung process in Eq.~(\ref{eqn:process:brem}), let us first consider the $np$ scattering. In light of different couplings of $Z'$ to protons and neutrons in the presence of the plasma effect, we define the following couplings 
\begin{equation}
\label{eqn:g0g1}
  g_0 \ \equiv \ \tilde{g}_{p}-g_{n}^{} \,, \qquad
  g_1 \ \equiv \ \tilde{g}_{p}+g_{n}^{} \,.
\end{equation}
As the nucleons are non-relativistic, in evaluating the nucleon propagators in the 8 diagrams in the left and middle panels of Fig.~\ref{fig:diagrams}, it is a good approximation that $p_i \cdot k_{Z^\prime} \simeq m_N E_{Z^\prime}$. Here $p_i$ (with $i=1,\,2,\,3,\,4$) are momenta for the incoming and outgoing nucleons, and $k_{Z'}$ is the momentum of $Z'$. However, if we take this approximation in the nucleon propagators at the beginning of calculations, the amplitudes squared for the four $t$-channel diagrams and the four $u$-channel diagrams turn out to be 
\begin{align}
\label{eq:ampsq:propto}
\left| {\cal M}_{a} + {\cal M}_{b} + {\cal M}_{c} + {\cal M}_{d} \right|^2 \,, \,\,\,
\left| {\cal M}_{a'} + {\cal M}_{b'} + {\cal M}_{c'} + {\cal M}_{d'} \right|^2 
 \,\,\, \propto  \,\,\, g_0^2 \,.
\end{align}
This implies that the amplitude squared $\sum_{} |{\cal M}|^2$ vanishes in the limit of $g_0 \to 0$, e.g. in the absence of the plasma effect when $m_{Z'} \gtrsim \omega_{\rm pl}$ (recall that the three diagrams in the right panel of Fig.~\ref{fig:diagrams} are all proportional to $g_0$; cf. Eq.~(\ref{eqn:coupling:Zprime:pion})). This is to some extent similar to the CP-even scalar case~\cite{Dev:2020eam,Joseph:2026nut}. 
Such a procedure can also be applied to the $pp$ and $nn$ bremsstrahlung processes (cf. Eq.~(\ref{eqn:ppnn:replacement})), 
and the corresponding amplitudes squared also vanish, with the corresponding diagrams $(a),\,(b),\,(c),\,(d)$ and $(a'),\,(b'),\,(c'),\,(d')$ in Fig.~\ref{fig:diagrams} canceling with each other.\footnote{The cancellation is also true for the production of dark photon via the subleading $pp$ bremsstrahlung process. 
In contrast, the leading $np$ bremsstrahlung process for the production of dark photon does not exhibit such a cancellation, since the couplings of dark photon to protons and neutrons are different.
}

It is very clear that the approximation above is not good enough.
In this approximation, some of the leading order contributions are dropped, which arise from the nucleons' relative momenta. 
By grouping the relevant propagator terms and utilizing momentum conservation, we can systematically extract these kinematic structures. One such representative relation is given by~\cite{Brinkmann:1988vi,Dev:2020eam}:
\begin{align}
\label{eqn:sim}
({k}\cdot {k}_{Z^\prime}) \left( \frac{1}{p_4 \cdot k_{Z'}} - \frac{1}{p_2 \cdot k_{Z'}} \right)
\ \simeq \ \frac{(\mathbf{k}\cdot {\mathbf{k}}_{Z^\prime})^2}{({p_2 \cdot k_{Z'}})({p_4 \cdot k_{Z'}})} \,,
\end{align}
where we have taken the approximation $k\cdot k_{Z^\prime} \simeq -\mathbf{k}\cdot \mathbf{k}_{Z^\prime}$ (with $k$ being the momentum transferred in the $t$-channel). After averaging the direction of $Z'$, it turns out that the terms in Eq.~(\ref{eqn:sim}) are at the order of ${\cal O} ({\bf k}^2/m_N^2)$. 
In the calculations, we have also kept the terms proportional to $m_{Z'}^2$ and $m_{Z'}^4$ in Taylor expansion of the nucleon and pion propagators in terms of the small parameters $m_{Z'}^2/2(p_i \cdot k_{Z'})$ and $m_{Z'}^2 / ({\bf l}^2 + m_\pi^2)$, that are important in particular when $Z'$ is heavy with respect to the temperature $T$. More details can be found in Appendix~\ref{appsub:preparation}.

Summing over all spins, the amplitudes squared for the $np$ process in the transverse and longitudinal modes can be expressed in a compact way as, respectively,
\begin{subequations}
\label{eqn:ampsq:np}
\begin{align}
\sum_{\rm spins}|\mathcal{M}_{np}|_{T}^2 \ \simeq \ &  g_{0}^2\left(\mathcal{T}^{(0,T)}+\mathcal{U}^{(0,T)}\right)+g_{1}^2\left(\mathcal{T}^{(1,T)}+\mathcal{U}^{(1,T)}\right)
\,, \\
\label{eqn:ampsq:np:L}
\sum_{\rm spins}|\mathcal{M}_{np}|_{L}^2 \ \simeq \ & g_{0}^2\left(\frac{\mathcal{T}^{(\alpha,L)}}{m_{Z^\prime}^2}+\mathcal{T}^{(0,L)}+\mathcal{U}^{(0,L)}\right)+g_{1}^2\left(\mathcal{T}^{(1,L)}+\mathcal{U}^{(1,L)}\right) 
\,.
\end{align}
\end{subequations} 
The ${\cal T}$ and ${\cal U}$ terms originate from the expansions of the nucleon and pion propagators, respectively. The lengthy coefficients for the $g_0^2$ and $g_1^2$ terms can be found in Appendices~\ref{appsub:g0} and \ref{appsub:g1}, respectively. 
Employing the approximations of  $\mathbf{k}^2 \sim \mathbf{l}^2 \sim (\mathbf{k\cdot l}) \sim 3 m_N T$ and $E_{Z'} \sim T$ (with $l$ the momentum transferred in the $u$-channel), 
the terms in Eq.~(\ref{eqn:ampsq:np}) 
can be compared in a straightforward way~\cite{Dent:2012mx}. For the $T$ mode terms, 
\begin{equation}
\label{eqn:term:comparison:1}
\frac{\mathcal{X}^{(1,T)}}{\mathcal{X}^{(0,T)}} 
\ \sim \ {\cal O} \left( \frac{T}{m_N} \right) \,.
\end{equation}
with ${\cal X} = {\cal T},\, {\cal U}$. 
For the $L$ mode, the terms proportional to $g_0^2$ and $g_1^2$ are of the same order as the corresponding terms in the $T$ modes, respectively, i.e. 
\begin{equation}
\label{eqn:term:comparison:2}
{\cal X}^{(0,L)} \ \sim \ {\cal X}^{(0,T)} \,, \qquad
{\cal X}^{(1,L)} \ \sim \ {\cal X}^{(1,T)} \,. 
\end{equation}
In addition, there is also the ${\cal T}^{(\alpha,L)}$ term in the amplitude squared for the $L$ mode that is proportional to $g_0^2/m_{Z^\prime}^2$, as seen in Eq.~(\ref{eqn:ampsq:np:L}). Ignoring the pion mass $m_\pi^2 \lesssim |{\bf l}|^2$, we can compare the terms proportional to $g_0^2/m_{Z^\prime}^2$ with the other terms, e.g. 
\begin{equation}
\label{mass denpent terms}
\frac{\mathcal{T}^{(\alpha,L)}/m^2_{Z^\prime}}{\mathcal{T}^{(0,L)}} \ \sim \ \mathcal{O}\left( \frac{T^2}{m_{Z^\prime}^2} \right) \,.
\end{equation}
It is obvious that this contribution depends highly on the $Z'$ boson mass. 
For a large $m_{Z^\prime}$, this contribution of ${\cal T}^{(\alpha,L)}$ is comparable to ${\cal T}^{(0,L)}$, while for a small mass $m_{Z^\prime} \ll T$, the ${\cal T}^{(\alpha,L)}$ term becomes dominant over the ${\cal T}^{(0,L)}$ term.

For the squared matrix elements in Eq.~(\ref{eqn:ampsq:np}), there are also the terms proportional to the couplings $g_0 g_1$. The corresponding coefficients are at the same order as that for the ${\cal X}^{(1,T)}$ and ${\cal X}^{(1,L)}$ terms. As a result, the $g_0g_1$ terms are always subleading to the $g_0^2$ or $g_1^2$ terms in Eq.~(\ref{eqn:ampsq:np}), and can thus be safely neglected. 
Let us now grab some of the qualitative features of the squared matrix elements in Eq.~(\ref{eqn:ampsq:np}). 
\begin{itemize}
    \item When the $Z'$ boson mass $m_{Z'} \lesssim \omega_{\rm pl}$, the amplitudes squared $\sum|{\cal M}_{np}|^2$ in the transverse and longitudinal modes are both dominated by terms proportional to $g_0^2$, which are vanishing in vacuum but nonzero due to the plasma corrections to the coupling of $Z'$ to protons. This is qualitatively in agreement with the results in Ref.~\cite{Shin:2021bvz}. 
    \item When the $Z'$ boson is heavy, i.e. $m_{Z'} \gtrsim \omega_{\rm pl}$, the plasma effect becomes inefficient and the coupling $g_0 \to 0$. Then the $g_1^2$ terms become the dominant contribution. 
\end{itemize}

The scattering of identical nucleons, i.e. the $pp$ and $nn$ bremsstrahlung processes, can also contribute to the production of $Z^\prime$. In this case, there are only the 8 Feynman diagrams, i.e. those in the left and middle panels of Fig.~\ref{fig:diagrams}. 
The amplitudes squared for the processes $pp\rightarrow ppZ^\prime$ and $nn\rightarrow nnZ^\prime$ can be obtained by making the following replacements in the equations of Appendix~\ref{app:ampsq:BL}: 
\begin{subequations}
\label{eqn:ppnn:replacement}
\begin{align}
    \sum_{\rm spins} |{\cal M}_{pp}|_{T,L}^2 \ = \ & \sum_{\rm spins} |{\cal M}_{np}|_{T,L}^2 (f_{nn} \to f_{pp},f_{pn} \to f_{pp},g_0\to0,g_1\to2\tilde{g}_p) \,, \\
    \sum_{\rm spins} |{\cal M}_{nn}|_{T,L}^2 \ = \ & \sum_{\rm spins} |{\cal M}_{np}|_{T,L}^2 (f_{pp} \to f_{nn},f_{pn} \to f_{nn},g_0\to0,g_1\to2g_n) \,.    
\end{align}
\end{subequations}
In particular, with identical nucleons in the initial and final states, the $g_0^2$ terms vanish, and there are only the $g_1$ terms left. Consequently, the $pp$ and $nn$ contributions are of the same order as the $np$ process for the mass range of $m_{Z'} \gtrsim \omega_{\rm pl}$.
In our calculations of the supernova limits below, the $pp$ and $nn$ processes are therefore taken into account.

At the amplitude squared level, our results are to some extent similar to those in Ref.~\cite{Shin:2021bvz}. However, there are still some differences in the calculation details.
\begin{itemize}
    \item We adopt the polarization sums for the transverse and longitudinal massive gauge bosons in the relativistic medium in Ref.~\cite{Chang:2016ntp}, which differs from that in Refs.~\cite{Shin:2021bvz,Shin:2022ulh}.  
    In particular, there are no terms proportional to $1/m_{Z'}^2$ in Ref.~\cite{Shin:2021bvz}, which are important when $m_{Z^\prime}$ is small. 
    \item There are no terms proportional to the mass $m_{Z^\prime}$ in Ref.~\cite{Shin:2021bvz}. Such terms are important when $Z'$ is heavy. 
\end{itemize}

\subsection{Decay}
\label{subsec:Zprime:decay}

In the supernova core, the $Z'$ boson can decay into the three neutrino eigenstates, and the charged leptons $e^+ e^-$ and $\mu^+ \mu^-$ if kinematically allowed. In the dense medium of the supernova core, the coupling of $Z'$ to neutrinos is not affected by the medium effects, i.e. $g^{}_{\nu} = -g_{BL}$. The effective couplings of the $Z^\prime$ boson to electrons and muons are exactly the same as that to protons in Eq.~(\ref{eqn:effive_coupling2}), up to the minus sign, i.e. $\tilde{g}_{e,\,\mu} = - \tilde{g}_{p}$. 
In the dense medium, the electron mass also receives significant contributions from the medium effects, which leads to the effective electron mass~\cite{Braaten:1991hg}: 
\begin{equation}
m_{e} \ = \ \frac12 m_{e,0}+\sqrt{\frac{1} {4} m_{e,0}^2 +\frac{\alpha}{2\pi}(\pi^2 T^2+\mu_e^2)} \,,
\end{equation}
where $m_{e,0} = 0.511$ MeV is the electron mass in vacuum, and $\mu_e$ the chemical potential of electron. As the muon mass $m_\mu$ is significantly higher than the typical temperature $T \sim 30$ MeV in the supernova core, the medium correction to $m_\mu$ is not important. 

Following Ref.~\cite{Chang:2016ntp}, the corresponding partial widths in the $T$ and $L$ modes for the neutrino, electron, and muon channels in the dense medium are, respectively,
\begin{subequations}
\label{eqn:width:Zprime}
\begin{align}
\label{neutrino coalescence}
\Gamma_{T,L} (Z' \rightarrow \nu_i \bar\nu_i) \ = \ &
\frac{g^{2}_{\nu} m_{Z^\prime}^{2}}{4\pi \sqrt{E_{Z'}^2-m_{Z^\prime}^{2}}} \nonumber \\
& \hspace{-30pt} \times \int_{x_{-}}^{x_{+}} {\rm d}x \frac{\frac{1}{2} {\sf m}_{T,L}^2(0,m_{Z^\prime},E_{Z'}, x E_{Z'}) }{\left[1+\exp \left(\frac{-x+\mu_{\nu_i} / E_{Z'}}{T / E_{Z'}}\right)\right] \left[1+ \exp \left(\frac{-{(1-x)
}+\mu_{\bar{\nu_i}} / E_{Z'}}{T/E_{Z'}}\right)\right]} \,, \\
\label{eqn:decay:Zprime:electron}
\Gamma_{T,L} (Z^\prime \rightarrow e^{+} e^{-}) \ = \ &  \frac{\tilde{g}_{e|T,L}^{2} m_{Z^\prime}^{2}}{4\pi \sqrt{E_{Z'}^2-m_{Z^\prime}^{2}}} 
\int_{x_{-}}^{x_{+}} {\rm d}x \frac{{\sf m}_{T,L}^2(m_{e,\,{\rm eff}},m_{Z^\prime},E_{Z'}, xE_{Z'})}{1+\exp \left(\frac{-x+\mu_e / E_{Z'}}{T / E_{Z'}}\right)} \, \Theta\left(m_{Z^\prime}^{}-2 m_{e}\right) \,, \\
\label{eqn:decay:Zprime:muon}
\Gamma_{T,L} (Z^\prime\rightarrow\mu^{+} \mu^{-}) \ = \ &  \frac{\tilde{g}_{\mu|T,L}^{2} m_{Z^\prime}^{2}}{4\pi \sqrt{E_{Z'}^2-m_{Z^\prime}^{2}}} \int_{x_{-}}^{x_{+}} {\rm d}x \, \frac{{\sf m}_{T,L}^2(m_{\mu},m_{Z^\prime},E_{Z'}, xE_{Z'})}{1+\exp \left(\frac{-x+\mu_\mu / E_{Z'}}{T / E_{Z'}}\right)} \, 
\Theta\left(m_{Z^\prime}^{}-2 m_\mu\right) \,, 
\end{align}
\end{subequations}
where $\mu_{\nu_i}$ and $\mu_{\bar{\nu}_i}$ are the chemical potentials of neutrinos and antineutrinos in the supernova core, respectively, obtained from the simulation profiles~\cite{Mirizzi:2015eza,Bollig:2017lki,Bollig:2020xdr}. In Eq.(\ref{eqn:decay:Zprime:electron}) and Eq.(\ref{eqn:decay:Zprime:muon}), $e^+$ and $\mu^+$ are not Pauli blocked and the distributions of $e^-$ and $\mu^-$ are included in the calculation. The dimensionless squared matrix elements in the $T$ and $L$ modes are, respectively,
\begin{subequations}
\begin{align}
{\sf m}_T^2 (m,M,E;E_1) & \ = \ \frac{m^2}{M^2} +z(M,E,E_1) \,, \\
{\sf m}_L^2 (M,E;E_1) & \ = \ 1-2z(M,E,E_1) \,,
\end{align}
\end{subequations}
with 
\begin{equation}
    z(M,E,E_1) \ \equiv \ \frac{E_1 E_2}{M^{2}}-\frac{(E E_1 -M^{2}/2 )(E E_2 -M^{2}/2)}{E^2-M^{2}} \,,
\end{equation}
and $E_2=E-E_1$. 
The limits of integration in Eq.~(\ref{eqn:width:Zprime}) are~\cite{Chang:2016ntp}:
\begin{eqnarray}
x_{ \pm} (m,M,E) 
\ \equiv \ \frac{1}{2} \pm \frac{1}{2} \sqrt{\left(1-\frac{4 m^2}{M^{2}}\right)\left(1-\frac{M^{2}}{E^2}\right)} \,.
\end{eqnarray}

\subsection{Emission rate}
\label{subsec:BL:rate}

The energy emission rates in the $T$ and $L$ modes per unit volume in the supernova core due to the nucleon bremsstrahlung process are, respectively~\cite{Dent:2012mx},
\begin{eqnarray}
\label{eqn:emission_rate:zprime}
{\cal Q}_{T,L} 
& \ = \ & \sum_{ij} \int\frac{d^3k_{Z^\prime}}{(2\pi)^3} E_{Z^\prime}\mathcal{S}
\Gamma^{\mathrm{prod}}_{ij|T,L}\mathrm{exp}\{-\tau_{T,L}\} \nonumber \\
& \ \simeq \ & \sum_{ij} \int \mathrm{d} \Pi_5 \mathcal{S}(2 \pi)^4 \delta^{(4)}\left(p_1+p_2-p_3-p_4-k_{Z^\prime}\right) E_{Z'} f_1 f_2 \sum_{\text{spins}}|\mathcal{M}_{ij}|_{T,L}^2 \mathrm{exp}\{-\tau_{T,L}\} \,, \nonumber \\ &&
\end{eqnarray}
where we have summed up the contributions from the $ij=np,\, pp,\, nn$ processes, $\mathrm{d} \Pi_5$ is the phase space factor, ${\cal S}$ is the symmetry factor for identical particles, being 1 for $np$ scattering process and 1/4 for the $pp$ and $nn$ processes, and $f_{1,2}$ is the non-relativistic Maxwell-Boltzmann distributions of the two incoming nucleons in the non-degenerate limit:
\begin{equation}
f_{1,2}(\mathbf{p} ; r) \ = \ \frac{n_{N}(r)}{2 }\left(\frac{2 \pi}{m_N T(r)}\right)^{3 / 2} \mathrm{e}^{-\mathbf{p}^2 / 2 m_N T(r)} \,,
\end{equation}
with ${\bf p}$ the 3-momentum of the nucleon. The nucleon number density $n_N(r)$ (with $N = p,\,n$) and temperature $T(r)$ are both functions of the radius $r$, depending on the supernova profile. For simplicity we have neglected the Pauli-blocking factors $(1-f_3)(1-f_4)$ for the nucleons in the final state. 
After the lengthy but straightforward calculations, the unit emission rate in Eq.~(\ref{eqn:emission_rate}) can be simplified to
\begin{eqnarray}
\label{eqn:emission_rate:dimensionless}
{\cal Q}_{T,L}(r) & \ = \ & \sum_{ij} \frac{T^{7/2}(r) n_i(r) n_j(r)}{2^{9} \pi^{7/2}
 m_N^{5/2}}  \nonumber \\ 
&& \times \int {\rm d} u {\rm d} v {\rm d} z {\rm d} x \sqrt{u v} \mathrm{e}^{-u}  x{\sqrt{x^2-q^2}} \delta(u-v-x){\cal S}\sum_{\text{spins}}|\mathcal{M}_{ij}|_{T,L}^2 \mathrm{exp}\{-\tau_{T,L}\} \,, \nonumber \\ && 
\end{eqnarray}
where $u,\,v,\,z,\,x,\,q$ are dimensionless variables defined in Eq.~(\ref{eqn:uvfunctions}), and $\sum_{\text{spins}}|\mathcal{M}|_{T,L}^2$ here represents the expression after substituting the variables according to Eqs.~(\ref{eqn:uvfunctions}), (\ref{eqn:dimensionlesskl}) and (\ref{eqn:variables:cdef}).

After being produced, the $Z'$ bosons can decay into neutrinos, electrons or muon pairs inside the star, if kinematically allowed. The details for the decay widths can be found in Section~\ref{subsec:Zprime:decay}. Alternatively, the $Z'$ bosons could also be absorbed by the stellar medium when escaping from the supernova core. It is expected that the absorption of $Z'$ is also dominated by the nucleon bremsstrahlung process $N + N + Z' \to N + N$, and the corresponding absorption rate is intimately correlated with the bremsstrahlung production. The absorption rates in the $T$ and $L$ modes for the nucleon inverse bremsstrahlung process can be written as, respectively, 
\begin{equation}
\label{eqn:zprime:abs}
\Gamma^\mathrm{abs}_{T,L} \ = \ \frac{1}{2 E_{Z'}} \sum_{ij} \int \mathrm{d} \Pi_4 \mathcal{S} \sum_{\rm spins}\left|\mathcal{M}_{ij}^{\prime}\right|^2_{T,L}(2 \pi)^4 \delta^{(4)}\left(p_1+p_2-p_3-p_4+k_S\right) f_1 f_2 \,,
\end{equation}
where ${\rm d}\Pi_4$ is the phase space for the nucleons in initial and final states, and $\sum |{\cal M}_{ij}^\prime|^2 = \sum |{\cal M}_{ij}|^2$ is the amplitudes squared for the inverse $ij=np,\,pp,\,nn$ bremsstrahlung process. Following the same procedure above as that for the emission rate, the absorption rate in Eq.~(\ref{eqn:zprime:abs}) can be simplified to
\begin{equation}
\Gamma^\mathrm{abs}_{T,L}(r) \ = \  \sum_{ij} \frac{n_i(r)n_j(r)}{64 \pi^{3 / 2} m_N^{5 / 2} T^{1 / 2}(r) x} \int_0^{\infty} \mathrm{d} u \int_q^{\infty} \mathrm{d} v \int_{-1}^1 \mathrm{~d} z \sqrt{u v} e^{-u} \delta(u-v+x) \mathcal{S} \sum_{\text {spins }}|\mathcal{\cal M}_{ij}|_{T,L}^2 \,.
\end{equation}
The absorption rate 
obeys the local thermal equilibrium by $\Gamma_{{T,L}}^{\mathrm{abs}}=e^{E_{Z^\prime}/T}\Gamma_{{T,L}}^{\mathrm{prod}}$.

By evaluating the decay and absorption of $Z'$ inside the star, it is found that for the parameter space of interest in this paper, the optical depths $\tau_{T,L}$ of $Z'$ in the $U(1)_{B-L}$ model are dominated by the absorption processes. This is also true for the case of dark photon in Section~\ref{sec:DP}.

\subsection{SN1987A cooling limits}
\label{subsec:BL:SN1987A}

\begin{figure}[t]
\centering
\includegraphics[width=0.49\textwidth]{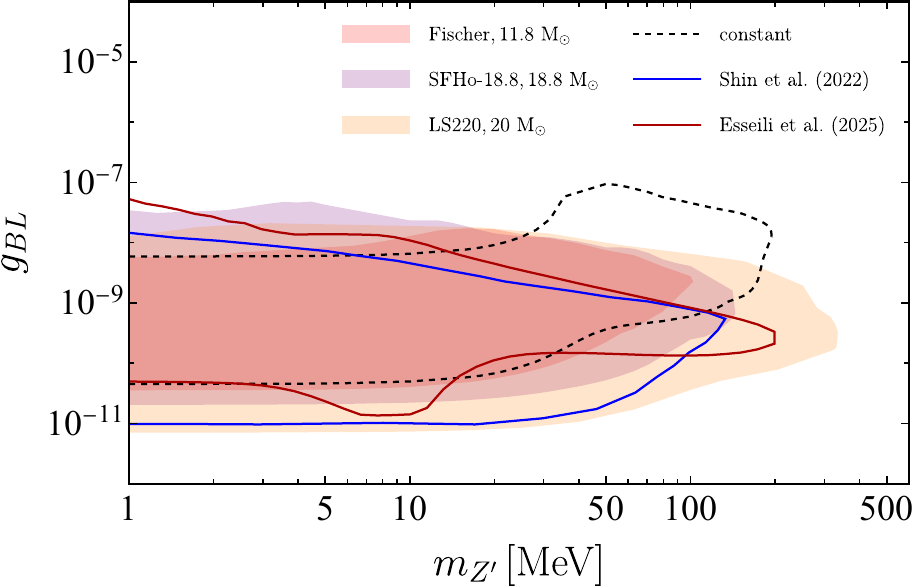}
\includegraphics[width=0.49\textwidth]{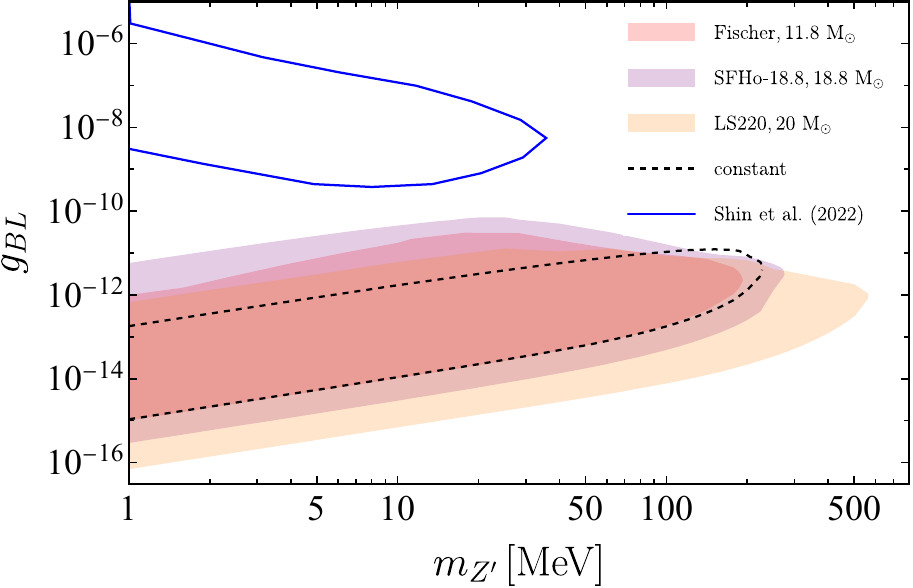}
\caption{SN1987A cooling limits on the transverse (left) and longitudinal (right) modes of the $U(1)_{B-L}$ gauge boson mass $m_{Z'}$ and coupling $g_{BL}$, assuming the Fischer $11.8 M_{\odot}$ (shaded red), SFHo-18.8 $18.8 M_{\odot}$ (shaded purple) and LS220 $20 M_{\odot}$ (shaded orange) supernova profiles. The dashed  black lines denote the limits based on the constant density $n_B=1.2 \times 10^{38}\; \mathrm{cm}^{-3}$ and temperature $T=30$ MeV within the core radius of $R_{C}=10$ km. The bounds from Shin et al. (blue)~\cite{Shin:2021bvz} and Esseili et al. (olive)~\cite{Esseili:2025sbc} are also shown for comparison.
See text for more details. 
}
\label{fig:BL}
\end{figure}

Integrating the unit emission rate ${\cal Q}_{T,L}$ in Eq.~(\ref{eqn:emission_rate:dimensionless}) over the supernova core, up to the radius $R_\nu$, 
we can obtain the total emission rate due to the emission of $Z^\prime$ from the supernova core, i.e. 
\begin{equation}
\label{eqn:rate:BL}
\int {\cal Q}_{T,L} \mathrm{d}V   \ = \ \int_{0}^{R_{\nu}} \mathrm{d}r \, 2\pi r^2 \int_{0}^{\pi}\mathrm{d}\phi\, \sin\phi \, {\cal Q}_{T,L}(r,\,\phi) \,.
\end{equation}
To set constraints on the $Z'$ boson $m_{Z'}$ and the gauge coupling $g_{BL}$, we adopt the ``Raffelt criterion'', which requires that the instantaneous luminosity due to the emission of light BSM particles should not exceed roughly ${\cal O}(10\%)$ of the measured neutrino luminosity $\mathcal{L_\nu}\simeq 3\times 10^{53}$ erg/s. To be specific, we take the constraint of~\cite{Raffelt:1996wa} 
\begin{equation}
\label{limit equation}
    \mathcal{L}_{Z^\prime} \ < \ 3\times 10^{52}\; \mathrm{erg/s} \,.
\end{equation}
The resultant supernova limits on transverse and longitudinal components of the $U(1)_{B-L}$ gauge boson $Z'$ are shown in the left and right panels of Fig.~\ref{fig:BL}, respectively. 
Here we have taken three different supernova profiles Fischer $11.8M_{\odot}$~\cite{Fischer:2016cyd}, SFHo-18.8 $18.8M_{\odot}$\cite{Mirizzi:2015eza} and LS220 $20M_{\odot}$~\cite{Bollig:2020xdr}, and the corresponding constraints are presented as the red, purple and orange shaded regions, respectively. As a comparison, the limit based on the constant density $n_B=1.2\times 10^{38}\, \mathrm{cm}^{-3}$, temperature $T=30$ MeV and the core size $R_{C}=10$ km is presented as the dashed black line. The geometric effect has been taken into account for both the three profiles and the constant case. 
\begin{itemize}
    \item As seen in the left panel of Fig.~\ref{fig:BL}, for the three profiles Fischer $11.8M_{\odot}$, SFHo-18.8 $18.8M_{\odot}$ and LS220 $20M_{\odot}$, the $U(1)_{B-L}$ gauge boson mass is excluded up to 100 MeV, 145 MeV and 330 MeV in the transverse mode, respectively, with respect to the mass exclusion range of 190 MeV for the constant profile case. 
    
    The lower boundaries are mainly determined by the production of $Z'$, with the coupling $g_{BL}$ excluded to $3.7\times 10^{-11}$, $1.7\times 10^{-11}$ and $7.1\times 10^{-12}$ for the three profiles, and $4.7\times 10^{-11}$ for the constant profile case. Comparing the results for the three realistic profiles, as the temperature and density is higher in the profile LS220 $20M_{\odot}$ than in Fischer $11.8M_{\odot}$ and SFHo-18.8 $18.8M_{\odot}$, the corresponding excluded region can extend to smaller values of $g_{BL}$.  
    
    The upper boundaries are dominated by the absorption of $Z'$ in the medium, with the subdominant effects from the decay of $Z'$, and the coupling $g_{BL}$ is excluded up to $1.7\times 10^{-8}$, $4.6\times 10^{-8}$ and $2.1\times 10^{-8}$ for the three profiles of Fischer $11.8M_{\odot}$, SFHo-18.8 $18.8M_{\odot}$ and LS220 $20M_{\odot}$, and up to $1.7\times 10^{-8}$ for the constant profile case. As clearly seen in the left panel of Fig.~\ref{fig:BL}, the limits based on the constant profile case is to some extent conservative when the $Z'$ boson is light, say $m_{Z'} \lesssim T \sim 30$ MeV; for a heavy $Z'$ with $m_{Z'} \gtrsim T$, the supernova profile starts to play an important role in determining the SN1987A constraints in the $T$ modes.

    \item 
    In the longitudinal mode, as seen in the right panel of Fig.~\ref{fig:BL},
    for the three supernova profiles Fischer $11.8M_{\odot}$, SFHo-18.8 $18.8M_{\odot}$ and LS220 $20M_{\odot}$ the gauge boson mass $m_{Z'}$ is excluded up to 190 MeV, 280 MeV and 560 MeV, respectively, and up to 230 MeV for the constant profile case. As stated in Section~\ref{subsec:Zprime:prod}, the production of the longitudinal mode of $Z'$ is dominated by the ${\cal T}^{(\alpha,L)}/m_{Z'}^2$ term; therefore the limits become stronger for small $m_{Z'}$. For the value of $m_{Z'} = 1$ MeV, the $g_{BL}$ ranges of $[1.1\times10^{-15},\, 10^{-12}]$, $[2.9\times10^{-16},\, 5.8\times10^{-12}]$ and $[7\times10^{-17},\, 6.7\times10^{-13}]$ are excluded for the three profiles, respectively. For the constant profile case, the excluded range is $g_{BL} \in[1.1\times 10^{-15},\, 1.8\times 10^{-13}]$. As seen in the  right panel of Fig.~\ref{fig:BL}, the SN1987A limit on the $L$ mode of $Z'$ is to some extent conservative if we adopt the constant profile case.  
\end{itemize}
If the supernova constraints on the transverse and longitudinal modes are combined together, there are almost no new excluded regions in the parameter space of $m_{Z'}$ and $g_{BL}$, as the limits in the transverse and longitudinal modes differ significantly.

The SN1987A limits on the $Z'$ boson in the $U(1)_{B-L}$ model have also been recently obtained in some other papers~\cite{Knapen:2017xzo,Croon:2020lrf,Shin:2021bvz,Esseili:2025sbc}. For comparison purposes, the bounds on the $T$ modes of $Z'$ from Shin et al.~\cite{Shin:2021bvz} and Esseili et al.~\cite{Esseili:2025sbc} are shown in the left panel of Fig.~\ref{fig:BL} as the blue and olive lines, respectively. 
The constraint for $m_{Z^\prime}\lesssim20$~MeV in Ref.~\cite{Esseili:2025sbc} 
relies on the soft radiation approximation (SRA) of nucleon-nucleon bremsstrahlung, while the results obtained in Ref.~\cite{Shin:2021bvz} are based on the OPE approximation. 
The limits from Ref.~\cite{Esseili:2025sbc} exhibit a dip around $m_{Z^\prime}\sim10$--$20$~MeV, which is, however, absent in Ref.~\cite{Shin:2021bvz} and our result, as the resonance width calculated under the SRA is underestimated~\cite{Shin:2021bvz}. 
The difference of our limits from Ref.~\cite{Shin:2021bvz} in the $T$ modes are mainly from the adopted supernova profiles and
quantitative differences in the squared matrix elements, e.g. the $m_{Z'}^2$ terms (cf.~Section~\ref{subsec:Zprime:prod}). 
In addition, the treatment of particle propagation also differs from that in Ref.~\cite{Shin:2021bvz}, as we adopt an exact geometrical treatment while Ref.~\cite{Shin:2021bvz} employs a geometrically averaged treatment.

The limit in Ref.~\cite{Knapen:2017xzo} was obtained by simply rescaling the supernova constraints on the dark photon and $U(1)_B$ gauge boson, and the bound in Ref.~\cite{Croon:2020lrf} was 
obtained by rescaling the $U(1)_B$ model limits in Ref.~\cite{Rrapaj:2015wgs} (the production of $Z'$ via the neutrino-pair coalescence $\nu \bar\nu \to Z'$ is also taken into account). Such simple rescaling is problematic when the plasma effect is taken into account for the effective coupling of a gauge boson to protons~\cite{Shin:2021bvz}. In particular, the leading order $g_{0}^2$ terms in the $U(1)_{B-L}$ gauge boson case are different from that of the dark photon case  
(cf. Section~\ref{sec:DP} below), and the bounds can not be simply rescaled from these models. Therefore, the limits from Refs.~\cite{Knapen:2017xzo,Croon:2020lrf} are not shown in the left panel of Fig.~\ref{fig:BL}.

The SN1987A limits on the longitudinal mode of $Z'$ were also obtained in Ref.~\cite{Shin:2021bvz}, which is presented as the blue line in the right panel of Fig.~\ref{fig:BL}. For the longitudinal mode, the leading terms proportional to ${\cal T}^{(\alpha,L)}/m_{Z'}^2$ that appear in our calculation are absent in Ref.~\cite{Shin:2021bvz}. This is the main reason for the significant difference between our longitudinal mode results and those of Ref.~\cite{Shin:2021bvz}.

\subsection{Low-energy supernova limits}
\label{subsec:BL:LESN}

The $U(1)_{B-L}$ gauge bosons escaping from the PNS can transport a large amount of energy into the stellar mantle, thereby increasing significantly the explosion energy of LESNe in some regions of the parameter space. Since the explosion energy of LESNe is only of order $E_{\mathrm{exp}} \sim 0.1\,\mathrm{bethe} = 10^{50}\,\mathrm{erg}$, even a modest amount of additional energy deposition can significantly modify the explosion dynamics, making LESNe a particularly sensitive probe of the weakly interacting exotic particles~\cite{Chugai:1999en,Pastorello:2003tc,Burrows:2020qrp,Kitaura:2005bt,Melson:2015tia,Radice:2017ykv,Muller:2018utr,Burrows:2019rtd,Stockinger:2020hse}. 
To derive the corresponding constraints, we employ the SFHo-18.8 $18.8M_\odot$ profile at a representative snapshot of $1\,\mathrm{s}$ post-bounce, and take the neutrinosphere radius $R_\nu$ as the reference boundary~\cite{Mirizzi:2015eza,Bollig:2017lki,Bollig:2020xdr}. The $U(1)_{B-L}$ gauge bosons produced within this radius are treated separately from those produced at larger radii, which are referred to as mantle production. The outer boundary of the mantle is taken to be the progenitor radius of $R_\ast \simeq 5.4 \times 10^7$ km. Consequently, the mantle region relevant for energy deposition is defined by $R_\nu < r < R_\ast$, and all the production, absorption, decay, and energy-transfer processes are integrated over this spatial region. The temperature in the mantle is around ${\cal O} ({\rm MeV})$, and as in the core the production of $U(1)_{B-L}$ $Z'$ bosons inside the mantle is dominated by the nucleon bremsstrahlung process.

Following Refs.~\cite{Fiorillo:2025yzf, Huang:2025xvo}, the energy deposited in the stellar mantle can be obtained from the balance between energy transported outward by the $Z'$ bosons produced inside the core and the energy taken away by the $Z'$ bosons generated in the mantle, i.e.
\begin{equation}
E_{\mathrm{dep}} \ = \ \Delta t \left( {\cal P}_{\mathrm{c}} - {\cal P}_{\mathrm{m}}\right) \,,
\end{equation}
where ${\cal P}_{\mathrm{c}}$ and ${\cal P}_{\rm m}$ denote the instantaneous luminosities associated with energy transported from the PNS core to the stellar mantle and the energy loss from the mantle, respectively, and $\Delta t=3~\mathrm{s}$ is the typical supernova explosion duration~\cite{Caputo:2022mah,Huang:2025rmy}. 
The instantaneous luminosity ${\cal P}_{\rm c}$ is  
\begin{align}
\label{eq:core}
{\cal P}_{\mathrm{c}} \ \simeq \ &
\int_{0}^{R_\nu} \mathrm{d}r \, 2\pi r^2
\int_{0}^{\pi} \mathrm{d}\phi\, \sin\phi \,
\Gamma^\mathrm{prod}_{Z'}
\left[
e^{-\tau_{Z^\prime}(R_\nu,r,\phi)}
-e^{-\tau_{Z^\prime}(R_{\ast},r,\phi)} 
-e^{-\tau_{Z^\prime}(R_\nu,r,\phi)}{p}_{c}^{(\nu)}(r,\phi)
\right] \,.
\end{align}
with
\begin{align}
\label{eq:pcnu}
{p}_{c}^{(\nu)}(r,\phi)
\ = \ &
\int_{{d}(R_\nu,r,\phi)}^{{d}(R_\ast,r,\phi)}
{\rm d}r^\prime\,
\Gamma_{Z^\prime}^{\rm decay}[L(r,\phi;r^\prime)]
{\rm BR}(Z^\prime\rightarrow\nu\bar{\nu})
\nonumber\\
&\times
\exp\left\{
-\int_{{d}(R_\nu,r,\phi)}^{r^\prime}
{\rm d}r^{\prime\prime}
\Big(
\Gamma^\mathrm{decay}_{Z^\prime}[L(r,\phi;r^{\prime\prime})]
+\Gamma^\mathrm{abs}_{Z^\prime}[L(r,\phi;r^{\prime\prime})]
\Big)
\right\} \,.
\end{align}
The first two exponential factors in Eq.~(\ref{eq:core}) represent the survival probabilities for a $Z'$ boson to reach the core boundary and the progenitor surface, respectively. Their difference accounts for the energy lost through the decay and absorption of $Z'$ within the stellar mantle. 
It should be noted that only the charged lepton decay products $Z' \to e^+ e^-,\, \mu^+ \mu^-$ contribute to the energy deposition. The neutrinos from the decay $Z' \to \nu_i \bar\nu_i$ deposit only a tiny fraction of energy into the mantle material. 
We therefore exclude the corresponding contribution by the third term in Eq.~(\ref{eq:core}), which is proportional to the factor of ${p}_{c}^{(\nu)}$ given in Eq.~(\ref{eq:pcnu}). 

The instantaneous luminosity loss from the mantle due to the $Z'$ bosons is~\cite{Fiorillo:2025yzf} 
\begin{equation}
{\cal P}_{\mathrm{m}} \ = \ {\cal P}_{\mathrm{m}}^{(1)} + P_{\mathrm{m}}^{(2)}+ {\cal P}_{\mathrm{m}}^{(\nu)} \, ,
\end{equation}
where ${\cal P}_{\mathrm{m}}^{(1)}$, ${\cal P}_{\mathrm{m}}^{(2)}$, and ${\cal P}_{\mathrm{m}}^{(\nu)}$ correspond to the energy carried by $Z'$ bosons produced in the mantle that are absorbed or decay within the core, escape the progenitor's outer boundary, and decay into neutrinos that instantaneously escape freely from the mantle, respectively. The contribution ${\cal P}_{\mathrm{m}}^{(1)}$ can be calculated as follows~\cite{Huang:2025xvo,Fiorillo:2025yzf}:
\begin{equation}
{\cal P}_{\mathrm{m}}^{(1)} \ = \ 
\int_{R_\nu}^{R_{\ast}} \mathrm{d}r\,
2\pi r^2 \int_{\phi_{\rm min}}^{\pi}
\mathrm{d}\phi\, \sin\phi\,
\Gamma^\mathrm{prod}_{Z'}\, e^{-\tau^\prime_{Z^\prime}(r,\phi)}
\left[ 1-e^{-\tau^{\prime\prime}_{Z^\prime}(r,\phi)} \right],
\end{equation}
where $\phi_{\rm min} = \pi-\arcsin(R_\nu/r)$, and 
\begin{subequations}
\begin{align}
\label{eqn:tau_prime}
\tau^\prime_{Z^\prime}(r,\phi) \ = \ & \int_0^{{s_1}(r,\phi)} {\rm d}r^\prime \Big( \Gamma^\mathrm{decay}_{Z^\prime}[L(r,\phi;r^\prime)])+\Gamma^\mathrm{abs}_{Z^\prime}[L(r,\phi;r^\prime)] \Big) \,, \\
\tau^{\prime\prime}_{Z^\prime}(r,\phi) \ = \ & \int_{{s_1}(r,\phi)}^{{s_2}(r,\phi)} {\rm d}r^\prime \Big( \Gamma^\mathrm{decay}_{Z^\prime}[L(r,\phi;r^\prime)]+\Gamma^\mathrm{abs}_{Z^\prime}[L(r,\phi;r^\prime)] \Big) \,,
\end{align}
\end{subequations}
are the optical depths accumulated from the $Z'$ boson production point in the mantle to the core and within the core, respectively. 
The two integration limits are given by 
\begin{equation}
s_{1,2} \ = \ -r\cos\phi\mp \sqrt{R_\nu^2-r^2(1-\cos^2\phi)} \,, 
\end{equation}
which represent the path lengths from the production point to the entry and exit points of the core, respectively. The term ${\cal P}_{\mathrm{m}}^{(2)}$ reads~\cite{Huang:2025xvo,Fiorillo:2025yzf}:
\begin{eqnarray}
{\cal P}_{\mathrm{m}}^{(2)}
\ = \ \int_{R_\nu}^{R_{\ast}} \mathrm{d}r\,
2\pi r^2
\int_{0}^{\pi}
\mathrm{d}\phi\, \sin\phi\,
\Gamma^\mathrm{prod}_{Z'}\,
e^{-\tau_{Z^\prime}(R_{\ast},r,\phi)} \,.
\end{eqnarray}
 The term ${\cal P}_{\mathrm{m}}^{(\nu)}$ is
\begin{align}
\label{eq:mantle_nu}
{\cal P}_{\mathrm m}^{(\nu)}
\ = \ &
\int_{R_\nu}^{R_*}\mathrm{d}r\,2\pi r^2
\int_0^\pi\mathrm{d}\phi\,\sin\phi\,
\Gamma_{Z'}^{\mathrm{prod}}
\int_0^{d(R_*,r,\phi)}
\mathrm{d}r'\,
\Gamma_{Z'}^{\mathrm{decay}}[L(r,\phi;r')]
{\rm BR}(Z'\to\nu\bar\nu)
\Theta(L>R_\nu)
\nonumber\\
&\times
\exp\left\{
-\int_0^{r'}\mathrm{d}r''
\Big(
\Gamma_{Z'}^{\mathrm{decay}}[L(r,\phi;r'')]
+
\Gamma_{Z'}^{\mathrm{abs}}[L(r,\phi;r'')]
\Big)
\right\} \,.
\end{align}
It is found that the contributions of  ${\cal P}_{\mathrm{m}}^{(1)}$ and ${\cal P}_{\mathrm{m}}^{(2)}$ to $E_{\rm dep}$ from the $Z'$ bosons produced in the stellar mantle is concentrated near the core surface and become relevant only in the large coupling regime. Numerically, the regions beyond approximately $1.2R_\nu$ make a negligible contribution to ${\cal P}_{\rm m}$. This reflects the suppression of the production of $Z'$ in the mantle, caused by the steep decrease in the temperature and baryon density outside the core. In addition, the solid angle subtended by the core decreases with increasing radius, further reducing the probability of mantle-produced $Z'$ bosons propagating into the core. Furthermore, the absorption of $Z'$ becomes also increasingly inefficient outside the core due to the rapidly decreasing matter density.

\begin{figure}[t]
\centering
\includegraphics[width=0.6\textwidth]{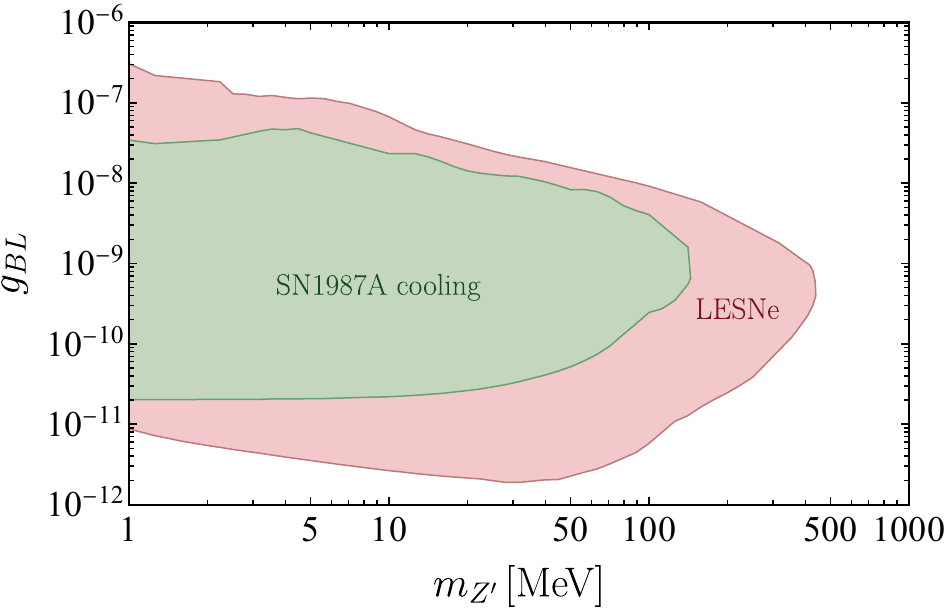}
\caption{LESN limits on the transverse modes of the $U(1)_{B-L}$ gauge boson mass $m_{Z'}$ and coupling $g_{BL}$, shown as the red shaded region. The green shaded region denotes the SN1987A cooling limit (cf. Fig.~\ref{fig:BL}). 
Both limits in this figure are based on the SFHo-18.8 $18.8M_\odot$ profile.
}
\label{fig:BL:LESNe}
\end{figure}

Adopting the supernova profile of SFHo-18.8 $18.8M_\odot$, the resulting LESN constraints on the transverse modes of $Z'$ in the $U(1)_{B-L}$ model are presented as the red shaded region in Fig.~\ref{fig:BL:LESNe}. As the LESN limit is more stringent than the SN1987A cooling bound by over two orders of magnitude, the excluded parameter space of $m_{Z'}$ and $g_{BL}$ is much larger than that by the cooling of SN1987A, which is shown as the green shaded region in Fig.~\ref{fig:BL:LESNe}, based on the same profile of SFHo-18.8 $18.8M_\odot$. 
In the weak coupling regime, the energy deposited in the mantle accounts for only a small fraction of the total energy carried away by the $Z'$ bosons escaping from the core. Furthermore, the lower boundary of the LESN limits for the $U(1)_{B-L}$ $Z'$ boson is reached when the mantle becomes optically thin regime, i.e. $\tau_{Z^\prime}(R_\ast,r,\phi)\lesssim 1$. 
After subtracting the energy carried away by neutrinos, the energy deposition probability within the square brackets of Eq.~(\ref{eq:core}) scales approximately as 
\begin{equation}
    e^{-\tau_{Z^\prime}(R_\nu,r,\phi)}-e^{-\tau_{Z^\prime}(R_{\ast},r,\phi)}\sim \tau_{Z^\prime}(R_{\ast},r,\phi) \ \propto \ g_{BL}^2 \,.
\end{equation}
As a result, the deposited energy for the $U(1)_{B-L}$ gauge boson scales as $E_{\rm dep}^{(Z')} \propto g_{BL}^4$, 
and the lower boundary of the LESN constraint exhibits a mass dependence due to the decay $Z' \to e^+ e^-$ for the mass range $m_{Z'} \lesssim T \sim 30$ MeV. 
It is also worth noting that this behavior differs qualitatively from the dark photon case (see Section~\ref{subsec:DP:LESNe} below for more details). 
Conversely, at sufficiently large couplings, the upper boundary emerges as the $Z'$ bosons become increasingly trapped inside the core due to efficient absorption, which severely suppresses outward energy transport. 
As seen in Fig.~\ref{fig:BL:LESNe}, the gauge coupling $g_{BL}$ is excluded within the range of $1.9\times 10^{-12} \lesssim g_{BL} \lesssim 3.1\times 10^{-7}$ for the gauge boson mass up to $m_{Z'} \simeq 440$ MeV, extending largely the cooling bounds. 
On the longitudinal mode, it turns out  that there is no bound from the LESNe. Throughout the parameter space of interest in  this paper, the stellar mantle remains essentially optically thin, and the deposited energy is not large enough to constrain the parameters $m_{Z'}$ and $g_{BL}$. 
Consequently, the LESN constraint is entirely dominated by the contribution from the transverse modes. 

\begin{figure}[t]
\centering
\includegraphics[width=0.6\textwidth]{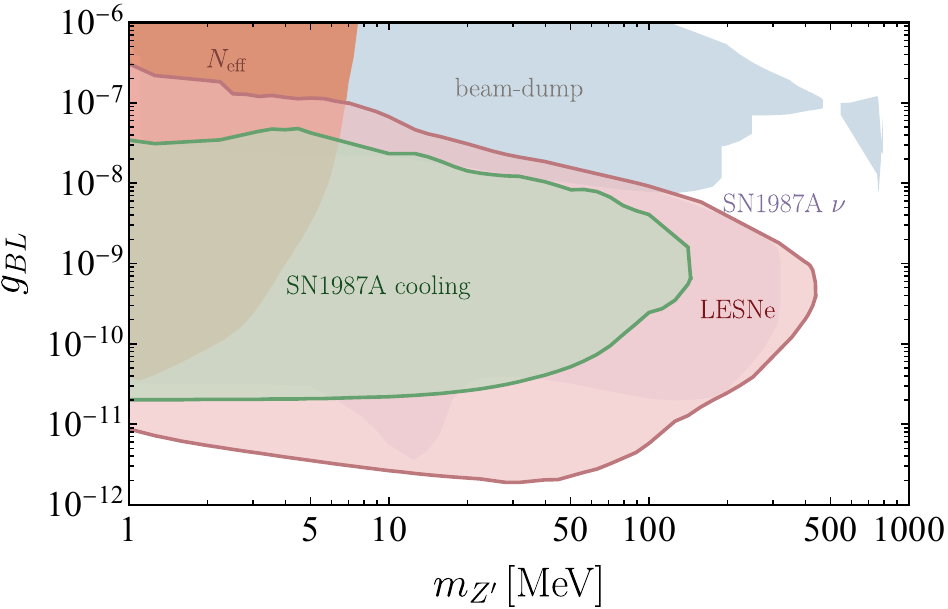}
\caption{Comparison of the supernova limits on the $U(1)_{B-L}$ gauge boson mass $m_{Z'}$ and coupling $g_{BL}$ obtained in this paper with other existing constraints. The shaded green and red regions are excluded by the SN1987A cooling and LESN constraints, respectively (cf.~Fig.~\ref{fig:BL:LESNe}). Other existing constraints include those from the beam-dump experiments (shaded gray, taken from Ref.~\cite{Esseili:2025sbc}), $N_{\rm eff}$ (shaded orange)~\cite{Esseili:2023ldf,Li:2023puz} and SN1987A neutrino observations (shaded purple)~\cite{Esseili:2025sbc}. See text for more details. 
}
\label{fig:BL:all_limits}
\end{figure}

The supernova limits obtained in this paper, based on the SFHo-18.8 $18.8M_{\odot}$ profile, are compared with other existing constraints on the $Z'$ boson in the $U(1)_{B-L}$ model in Fig.~\ref{fig:BL:all_limits}.
The gray shaded regions represent the constraints from the electron beam-dump experiments E137~\cite{Bjorken:1988as}, NA64~\cite{NA64:2018lsq,NA64:2019auh}, E141~\cite{Riordan:1987aw} and Orsay~\cite{Davier:1989wz}, and the proton beam-dump experiments FASER~\cite{FASER:2018eoc,FASER:2023tle}, LSND~\cite{Bauer:2018onh} and $\nu$-Cal~\cite{Tsai:2019buq}. Other weaker laboratory limits can be found in e.g. Refs.~\cite{Esseili:2025sbc,Ilten:2018crw,KLOE-2:2011hhj,NA482:2015wmo,COHERENT:2017ipa,BaBar:2017tiz,Ackermann:2025obx,COHERENT:2020iec,COHERENT:2021xmm}. 
The orange shaded region is excluded by the ${N_{\rm eff}}$ observations~\cite{Esseili:2023ldf,Li:2023puz}. The decay $Z' \to \nu_i \bar\nu_i$ could potentially contribute to the high-energy neutrinos from the SN1987A explosion. The existing neutrino events of SN1987A can then be used  to set limits on the coupling of $Z'$ in the $U(1)_{B-L}$ model. The corresponding constraint on $m_{Z'}$ and $g_{BL}$ is presented as the purple shaded region, which, however, lies totally beneath the LESN exclusion region~\cite{Esseili:2025sbc}. As clearly seen in Fig.~\ref{fig:BL:all_limits}, the supernova limits on the $Z'$ gauge boson in the $U(1)_{B-L}$ model are largely complementary to the laboratory and cosmological bounds, in the sense that the supernova limits are more sensitive to smaller values of $g_{BL}$ for the gauge boson mass $m_{Z'}$ ranging from 1 MeV to roughly $440$ MeV.\footnote{There are some extra supernova limits on dark photons, e.g. from the observations of 511 keV $\gamma$-ray lines and fireball (cf.~Fig.~\ref{fig:DP:all_limits})~\cite{Caputo:2025avc}. In light of the significant differences in the production of the $U(1)_{B-L}$ $Z'$ boson and dark photon in the supernova core, these limits on dark photon can {\it not} be directly applied to the $Z'$ boson by simple rescaling.}

\section{Dark photon}
\label{sec:DP}

The calculations of supernova limits on the $Z'$ boson in the $U(1)_{B-L}$ model in Section~\ref{sec:BL} can be generalized to the dark photon case in a straightforward way. After clarifying some calculation details in Section~\ref{subsec:DP:details}, the SN1987A limits are obtained in Section~\ref{subsec:DP:cooling} and the LESN constraints are derived in Section~\ref{subsec:DP:LESNe}. The main differences of the supernova limits on dark photon and the $U(1)_{B-L}$ $Z'$ boson are collected in Section~\ref{subsec:comparison}. 

\subsection{Calculation details}
\label{subsec:DP:details}

\begin{figure}[t]
\centering
\includegraphics[width=1\textwidth]{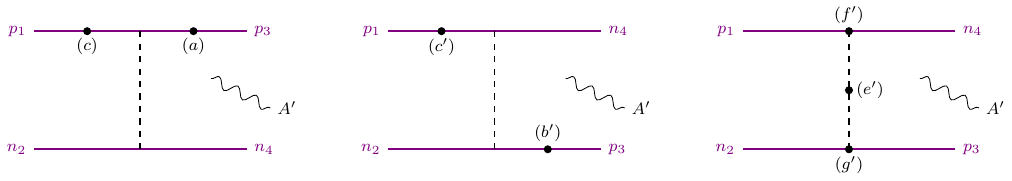}
\caption{Feynman diagrams for the production of the dark photon $A'$ via the $np$ bremsstrahlung process $p(p_1)+n(p_2) \rightarrow p(p_3)+n(p_4)+A'$ in the supernova core. The left panel is for the $t$-channel, while the middle and right panels are for the $u$-channel. The $A'$ boson can be attached to either of the proton lines $(a),\,(c)$ in the left panel and $(b'),\,(c')$ in the middle panel, or to the charged pion vertices $(e'),\, (f'),\ (g')$ in the right panel, as denoted by the blobs ($\bullet$). See text for more details. 
}
\label{fig:diagrams:DP}
\end{figure}

The couplings of the dark photon $A'$ with charged fermions in the SM can be induced by the low-energy effective kinetic mixing of $A'$ to the photon, which is parametrized as 
\begin{equation}
\mathcal{L}=-\frac{\epsilon}{2} F_{\mu \nu} F^{\prime \mu \nu} \,,
\end{equation}
Where $F_{\mu \nu}$ and $F_{\mu \nu}^{\prime}$ are the field strength of the SM photon and $A'$, respectively. All the couplings of $A'$ with the SM fermions are proportional to the kinetic mixing parameter $\epsilon$. 
The dark photon $A'$ can be produced in the supernova core via the $np$ and $pp$ bremsstrahlung processes. The corresponding Feynman diagrams for the $np$ process are presented in Fig.~\ref{fig:diagrams:DP}, with the dark photon $A'$ coupling to either the proton in the left and middle panels, or to the charged pions in the right panel. The couplings of $A'$ to charged pions at the dimension-4 and dimensions-5 orders can be obtained from that of the $Z'$ boson in Eq.~(\ref{eqn:coupling:Zprime:pion}) by taking $\tilde{g}_p \to e \tilde\epsilon$ and $g_{n} \to 0$, where $\tilde\epsilon$ is the plasma-dressed effective kinetic mixing parameter (cf. Eq.~(\ref{eqn:effive_coupling2}))~\cite{An:2013yfc,Chang:2016ntp}
\begin{equation}
\label{eqn:epsilon_effective}
\tilde{\epsilon}_{T,L} \ = \ \frac{\epsilon \, m_{A'}^2}{m_{A^{\prime}}^2- {\rm Re} \Pi_{\gamma\gamma}^{T,L} - i {\rm Im} \Pi_{\gamma\gamma}^{T,L}} \,.
\end{equation}
As implied by Eq.~(\ref{eqn:effive_coupling}), the effective kinetic mixing parameters $\tilde{\epsilon}_{T,L}$ are proportional to the dark photon mass $m_{A'}^2$, which has far-reaching implications for the supernova limits, as we will see below. Furthermore, as for the case of $\tilde{g}_{p}$ in Fig.~\ref{fig:coupling}, there is a significant resonant enhancement for the effective kinetic mixing parameter $\tilde\epsilon$ in the transverse modes, but not in the longitudinal mode.

It is then straightforward to derive the corresponding amplitudes squared for $np\rightarrow np A^\prime$ from that of the $Z'$ boson by taking the same replacement of $g_p \to e \epsilon$ and $g_{n} \to 0$, or equivalently $g_0,\,g_1 \to \tilde{g}_p \to e \tilde{\epsilon}$ in Eq.~(\ref{eqn:ampsq:np}), i.e.
\begin{subequations}
\label{eqn:ampsq:np:DP}
\begin{align}
\sum_{\rm spins}|\mathcal{M}_{np}^{(A')}|_{T}^2 \ \simeq & \ e^2 \tilde\epsilon^2 \left(\mathcal{T}^{(0,T)}+\mathcal{U}^{(0,T)}\right) \,, \\
\sum_{\rm spins}|\mathcal{M}_{np}^{(A')}|_{L}^2 \ \simeq & \ e^2 \tilde\epsilon^2 \left(\frac{\mathcal{T}^{(\alpha,L)}}{m_{A^\prime}^2}+\mathcal{T}^{(0,L)}+\mathcal{U}^{(0,L)}\right) \,.
\end{align}
\end{subequations}
The ${\cal X}^{(1,T)}$ and ${\cal X}^{(1,L)}$ terms are much smaller than the terms in Eq.~(\ref{eqn:ampsq:np:DP}) and have been neglected for simplicity (cf. Eqs.~(\ref{eqn:term:comparison:1}) and (\ref{eqn:term:comparison:2})). It is noticeable that the dark photon production in the $np$ bremsstrahlung process is always dominated by the ${\cal X}^{(0,T)}$, ${\cal X}^{(0,L)}$ and ${\cal T}^{(\alpha,L)}$ terms, regardless of the plasma effects, which is very different from the $U(1)_{B-L}$ $Z'$ boson case above. 
As seen in Eq.~(\ref{eqn:epsilon_effective}), when the plasma effect are taken into account, the effective kinetic mixing parameters $\tilde{\epsilon}_{T,L}$ are both suppressed by the dark photon mass via $m_{A'}^2$. As a result, when the dark photon is light, say e.g. $m_{A'} \lesssim T \sim 30$ MeV, the dependence of the amplitudes squared in the transverse and longitudinal modes on the dark photon mass is via, respectively~\cite{Li:2023vpv}, 
\begin{equation}
\label{eqn:ampsq:propto:mA}
|{\cal M}|_T^2 \ \propto \ \tilde{\epsilon}_T^2 \propto \epsilon^2 m_{A'}^4 \qquad \text{and} \qquad
|{\cal M}|_L^2 \ \propto \ \frac{\tilde{\epsilon}_L^2}{m_{A'}^2} \propto \epsilon^2 m_{A'}^2 \,. 
\end{equation}
This is dramatically different from the $U(1)_{B-L}$ gauge boson case, and will have far-reaching implications for the SN1987A limits on the dark photon, as we will see soon. 

For the $pp$ bremsstrahlung process $pp \to pp A'$, the amplitude squared can be obtained by taking $\tilde{g}_p \to e \tilde\epsilon$ and $g_n \to \tilde{g}_p \to e \tilde\epsilon$ in Eq.~(\ref{eqn:ampsq:np}), which leads to vanishing of the $g_0^2$ terms, and the amplitudes squared contain only the ${\cal X}^{(1,X)}$ terms. As a result, the $pp$ process is always subdominant to the $np$ process for the production of dark photon~\cite{Chang:2016ntp,Rrapaj:2015wgs}. 
For the sake of simplicity, we neglect the $pp$ bremsstrahlung process for dark photon in this paper.

If kinematically allowed, the dark photon $A'$ decays into charged leptons, i.e. $A' \to e^+ e^-,\, \mu^+ \mu^-$, and the corresponding widths can be obtained from the corresponding widths of $Z'$ in Eqs.~(\ref{eqn:decay:Zprime:electron}) and (\ref{eqn:decay:Zprime:muon}) by making the replacement of $\tilde{g}_e, \, \tilde{g}_\mu \to e \tilde\epsilon$, with the plasma affect and the Pauli-blocking factor taken into account. 
For the dark photon, there is an additional decay channel $A'\to\pi^+\pi^-$, and the corresponding partial decay width is given by~\cite{Syvolap:2024hdh} 
\begin{equation}
\Gamma_{T,L} (A' \to \pi^+ \pi^-) \ \simeq \ \frac{\delta_{T,L}}{12} \alpha \tilde{\epsilon}_{T,L}^2 m_{A'} 
\left( 1 - \frac{4m_{\pi}^2}{m_{A'}^2} \right)^{3/2}
|F_\pi (m_{A'}^2)|^2 \Theta (m_{A'} - 2m_\pi) \,,
\end{equation}
where $\delta_{T}=2$ and $\delta_L = 1$ denote the degree of freedom, and 
$F_\pi(s)$ is the electromagnetic pion form factor evaluated at $s=m_{A'}^2$.

\subsection{SN1987A cooling limits}
\label{subsec:DP:cooling}

\begin{figure}[t]
\centering
\includegraphics[width=0.49\textwidth]{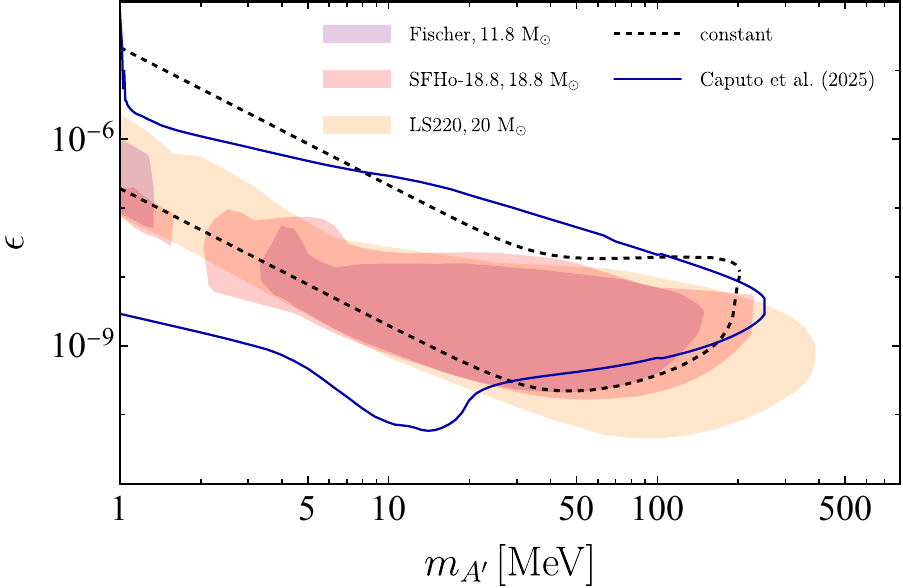}
\includegraphics[width=0.49\textwidth]{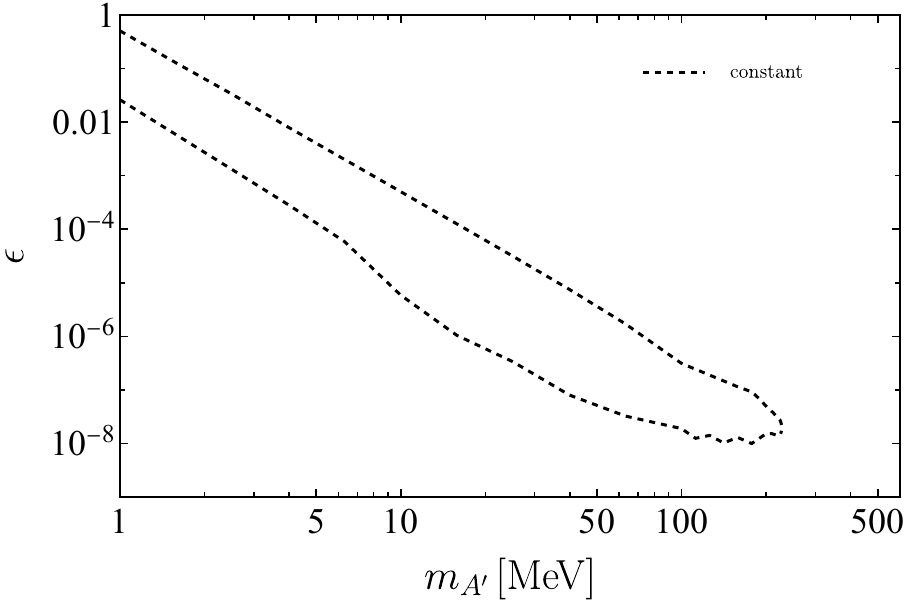}
\caption{SN1987A cooling limits on the transverse (left) and longitudinal (right) modes of the dark photon mass $m_{A'}$ and the kinetic mixing parameter $\epsilon$, assuming the Fischer $11.8 M_{\odot}$ (shaded purple), SFHo-18.8 $18.8 M_{\odot}$ (shaded red) and LS220 $20 M_{\odot}$ (shaded orange) supernova profiles.  The dashed black lines denote the limits based on the constant density $n_B=1.2 \times 10^{38}\; \mathrm{cm}^{-3}$ and temperature $T=30$ MeV within the core radius of $R_C=10$ km. 
In the right panel, no region is excluded if we adopt the three profiles above. 
The limit from Caputo et al. (blue)~\cite{Caputo:2025avc} is shown in the left panel for comparison. See text for more details. 
}
\label{fig:DP}
\end{figure}

The calculation procedure of the energy emission rate due to the dark photon $A'$ inside the supernova core is the same as that for the $U(1)_{B-L}$ gauge boson, based on Eq.~(\ref{eqn:rate:BL}) and the Raffelt criterion of ${\cal L}_{A'} < 3 \times 10^{52}$ erg/s. The resultant SN1987A cooling limits on the transverse and longitudinal components of dark photon $A'$ are shown in the left and right panels of Fig.~\ref{fig:DP}, respectively. As in the $U(1)_{B-L}$ $Z'$ boson case,  the three supernova profiles Fischer $11.8M_{\odot}$, SFHo-18.8  $18.8M_{\odot}$ and LS220 $20M_{\odot}$ are adopted, and the corresponding constraints are depicted as the red, purple and orange shaded regions, respectively. As a comparison, the limits based on the constant supernova profile are presented as the dashed black lines. The following features are observed for the SN1987A cooling limits on dark photon.
\begin{itemize}
    \item The excluded regions due to the emission of transverse dark photons from SN1987A are presented in the left panel of Fig.~\ref{fig:DP}. 
    When the dark photon is light, say $m_{A'} \lesssim T \sim 30$ MeV, the emission rate is suppressed by the dark photon mass $m_{A'}$, as seen in Eq.~(\ref{eqn:ampsq:propto:mA}). Furthermore, in the limit of small $m_{A'}$ with respect to all other kinematic scales, the dependence of emission rate on the dark photon mass is via the combination of $\epsilon\, m_{A'}^2$ in the effective mixing parameter in Eq.~(\ref{eqn:epsilon_effective}). For the constant profile case, the parameter space of $m_{A'}$ and $\epsilon$ within the range of 
    \begin{equation}
    \label{eqn:DP:scaling}
    2\times10^{-7} \lesssim \epsilon \, \left( \frac{m_{A'}}{\rm MeV} \right)^2 \lesssim 2\times10^{-5} 
    \end{equation}
    is excluded. 
    For the three profiles Fischer $11.8M_{\odot}$, SFHo-18.8 $18.8M_{\odot}$ and LS220 $20M_{\odot}$, although the constraints on $\epsilon$ also get weaker for light dark photon, the excluded regions are split into two or three noncontinuous areas, with  all the gaps being at $m_{A'} \sim $ few MeV.  As implied by the left panels of Fig.~\ref{fig:coupling}, this is due to the resonance in the effective mixing parameter $\tilde{\epsilon}_T$ outside the radius of $\sim 0.8 R_\nu$ for the profiles Fischer  $11.8M_{\odot}$ and SFHo-18.8 $18.8M_{\odot}$, and beyond $\sim 0.7 R_\nu$ for LS220 $20M_{\odot}$, respectively. As a result of the resonance effect, within certain regions of the parameter space there is a ``trapped sphere'' within $R_\nu$, which effectively prevents the emission of light dark photons from the star. Therefore, the constraints from the three supernova profiles are significantly weaker than the unrealistic constant case for a light dark photon.
    
    When the dark photon is relatively heavy, say $m_{A'} \gtrsim 30$ MeV, the three profiles are qualitatively similar to the constant profile case. For the profiles Fischer $11.8M_{\odot}$, SFHo-18.8 $18.8M_{\odot}$  and LS220 $20M_{\odot}$, the dark photon mass is excluded up to 160 MeV, 230 MeV and 490 MeV, with the kinetic mixing parameter $\epsilon$ constrained down to $2.2\times 10^{-10}$, $1.7\times 10^{-10}$ and $4.5\times 10^{-11}$, respectively. For the constant profile case, the mass $m_{A'}$ is constrained up to 200 MeV, and the parameter $\epsilon$ is excluded down to $2.2\times 10^{-10}$.
    
    \item Assuming the constant profile, the SN1987A constraint on the longitudinal dark photon is presented in the right panel of Fig.~\ref{fig:DP}. The dark photon mass is excluded up to 230 MeV, and the kinetic mixing parameter $\epsilon$ to {$1.3\times10^{-8}$}. Again, as implied by Eq.~(\ref{eqn:ampsq:propto:mA}), the constraint on the parameter $\epsilon$ gets weaker when the dark photon is light. It turns out that the narrow range of 
    \begin{equation}
    0.025 \lesssim  \epsilon \left( \frac{m_{A'}}{\rm MeV} \right)^{\eta} \lesssim 0.50 
    \end{equation}
    is excluded, with the index $\eta \sim 3$. The index $\eta$ deviates from a simple linear scaling ($\eta = 1$) as implied by Eq.~(\ref{eqn:ampsq:propto:mA}), because for sufficiently large mixing the lower bound in the right panel of Fig.~\ref{fig:DP} is already affected by the absorption and decay processes. Compared to the constant profiles case, the luminosity of dark photon can not reach the luminosity limit in all three profiles we adopted in this work, therefore no parameter space is excluded. 
    The weaker limits on the longitudinal mode can be qualitatively understood as follows: 
    Unlike in the inner core region, the effective kinetic mixing parameter $\tilde\epsilon_L$ is not suppressed at the outer regions of the realistic supernova profiles, as seen in the right panel of Fig.~\ref{fig:coupling}.
    This leads to efficient trapping of the produced dark photons, thereby efficiently suppressing the cooling constraints. 
    In contrast, such strong trapping effect is absent in the constant profile case, which leads to the excluded region in the right panel of Fig.~\ref{fig:DP}. 
\end{itemize}

The SN1987A cooling limits on the dark photon $A'$ has been extensively studied in the literature~\cite{Bjorken:2009mm,Dent:2012mx,Kazanas:2014mca,Rrapaj:2015wgs,Chang:2016ntp,Caputo:2025avc}. 
The limits in Refs.~\cite{Dent:2012mx,Kazanas:2014mca} are obtained under the OPE approximation. In these calculations, the partial cancellation of different Feynman diagrams is not fully taken into account, and the plasma effect and the geometric effect are not taken into consideration. The SRA is adopted for the limits obtained in Refs.~\cite{Chang:2016ntp,Caputo:2025avc}, taking into account the plasma effect. The multiple coherent scattering effects are also included in Ref.~\cite{Caputo:2025avc}, which are significant in the high-density medium~\cite{Landau:1953um,Migdal:1956tc}.
For comparison purposes, the latest results based on the SRA method in Ref.~\cite{Caputo:2025avc} in the transverse mode are presented as the solid blue line in the left panel of Fig.~\ref{fig:DP}. 
As stated in Section~\ref{subsec:Zprime:prod}, the squared amplitude for the production of dark photon calculated in the SRA is underestimated when the dark photon is heavy. This is the reason why our constraints in Fig.~\ref{fig:DP} tend to be more stringent than that in Ref.~\cite{Caputo:2025avc} for $m_{A'} \gtrsim T \sim 30$ MeV. As in the case of the $U(1)_{B-L}$ gauge boson, the dip at around $m_{A'} \sim 10$ MeV for the limits from Ref.~\cite{Caputo:2025avc} is due to the underestimated resonance width in the SRA method~\cite{Shin:2021bvz}.

\subsection{Low-energy supernova limits}
\label{subsec:DP:LESNe}
The LESNe constraints on dark photons can be obtained using the same calculation procedure, as well as the same core and mantle setups, as that for the $U(1)_{B-L}$ $Z'$ boson in Section~\ref{subsec:BL:LESN}, adopting again the  SFHo-18.8 $18.8M_\odot$ profile at $1\mathrm{s}$ post-bounce. As in the $U(1)_{B-L}$ case, the contribution from dark photons produced in the stellar mantle is concentrated near the core surface, with essentially all of the mantle contribution originating within approximately $\sim1.2R_\nu$. This contribution becomes relevant only in the large coupling regime. One key difference from the $U(1)_{B-L}$ $Z'$ case is that dark photons do not decay into neutrinos. Therefore, the neutrino contributions in Eqs.~(\ref{eq:pcnu}) and (\ref{eq:mantle_nu}) vanish for the dark photon case. For $m_{A'}>2m_e$, dark photons decay to $e^+e^-$, $\mu^+\mu^-$, and $\pi^+\pi^-$, when kinematically allowed, are assumed to deposit their energy in the stellar mantle.

\begin{figure}[t]
\centering
\includegraphics[width=0.6\textwidth]{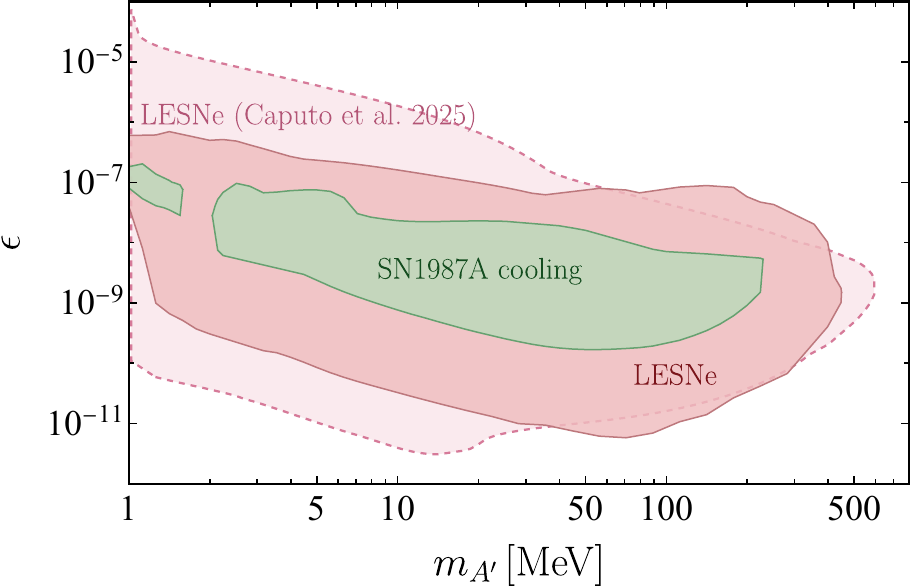}
\caption{
LESN limits on the dark photon mass $m_{A'}$ and the kinetic mixing parameter $\epsilon$ in the transverse modes, shown as the red shaded region. The green shaded region denotes the corresponding SN1987A cooling limit (cf.~Fig.~\ref{fig:DP}). Both limits in this figure are based on the SFHo-18.8 $18.8M_\odot$ profile. The pink shaded region is the LESN limit from Caputo et al.~\cite{Caputo:2025avc}. 
}
\label{fig:DP:LESNe}
\end{figure}

The resulting LESN constraint on the transverse modes of dark photon is presented as the red shaded region in Fig.~\ref{fig:DP:LESNe}, in comparison with the SN1987A cooling bound in green based on the same supernova profile of SFHo-18.8 $18.8M_\odot$. As in the $Z'$ boson case, the lower boundary is determined mainly by the decay of $A'$ in the mantle, while the upper boundary is by the absorption of dark photons by the mantle materials. 
In the weak coupling regime near the lower boundary, the dark photon decay length, while much larger than the core radius, is significantly smaller than the progenitor radius $R_\ast$. Consequently, although the dark photons can escape freely from the core without being trapped ($\tau_{A^\prime}(R_\nu,r,\phi)\ll 1$), they are still optically thick with respect to the stellar mantle ($\tau_{A^\prime}(R_\ast,r,\phi)\gg 1$). This ensures that the core-produced dark photons decay almost entirely  before reaching the stellar surface. The deposited energy is therefore determined predominantly by the total dark photon luminosity escaping the core, rather than being suppressed by a partial decay probability. Then, for the dark photon case, the factor within the square bracket of Eq.~(\ref{eq:core}) can be simplifies to 
\begin{equation}
    e^{-\tau_{A^\prime}(R_\nu,r,\phi)}-e^{-\tau_{A^\prime}(R_{\ast},r,\phi)} \ \sim \ 1 \,.
\end{equation}
The corresponding energy deposition scales as $E_{\rm dep}^{(A')} \propto \epsilon^2$, which is different from the $Z'$ boson case in Section~\ref{subsec:BL:LESN}.  
It is found that, for even smaller $\epsilon$, it is also possible that the decay factor $\tau_{Z_{A^\prime}}(R_\ast, r, \phi) \lesssim 1$, which is the same as the $Z'$ case above. However, the corresponding parameter space of $\epsilon$ is below the LESN constraints for dark photon. 
As seen in Fig.~\ref{fig:DP:LESNe}, the lower boundary of the LESN constraint extends down to $\epsilon \sim 5.8\times10^{-12}$, improving the SN1987A cooling bound by about two orders of magnitude, and the upper boundary go up to $\epsilon \sim 6\times 10^{-7}$, with the dark photon mass $m_{A'}$ excluded up to 450 MeV. As a result of the more stringent luminosity constraint, LESN observations exclude a much larger parameter space of $m_{A'}$ and $\epsilon$ than the conventional SN1987A cooling bound. 
No LESN constraint is obtained on the longitudinal mode of the dark photon, as the production rate of longitudinal dark photons in the core is highly suppressed by the effective kinetic mixing parameter $\tilde\epsilon$ (cf. the effective coupling in the panel of Fig.~\ref{fig:coupling}), thus too small to make a significant contribution to the total energy deposition in the stellar mantle. 

For comparison purposes, the LESN constraint obtained in Ref.~\cite{Caputo:2025avc} based on the SRA method is also shown in Fig.~\ref{fig:DP:LESNe}, as the shaded region surrounded by the dashed pink line. 
As in the case of SN1987A cooling (cf. Fig.~\ref{fig:DP}), for a relatively light dark photon with mass $m_{A'} \lesssim T\sim$ 30 MeV, the LESN limits in the OPE method are to some extent conservative than in SRA, while for a heavy dark photon with mass $m_{A'} \gtrsim T$ a wider range of $\epsilon$ tend to be excluded in the OPE approximation.  

\begin{figure}[t]
\centering
\includegraphics[width=0.6\textwidth]{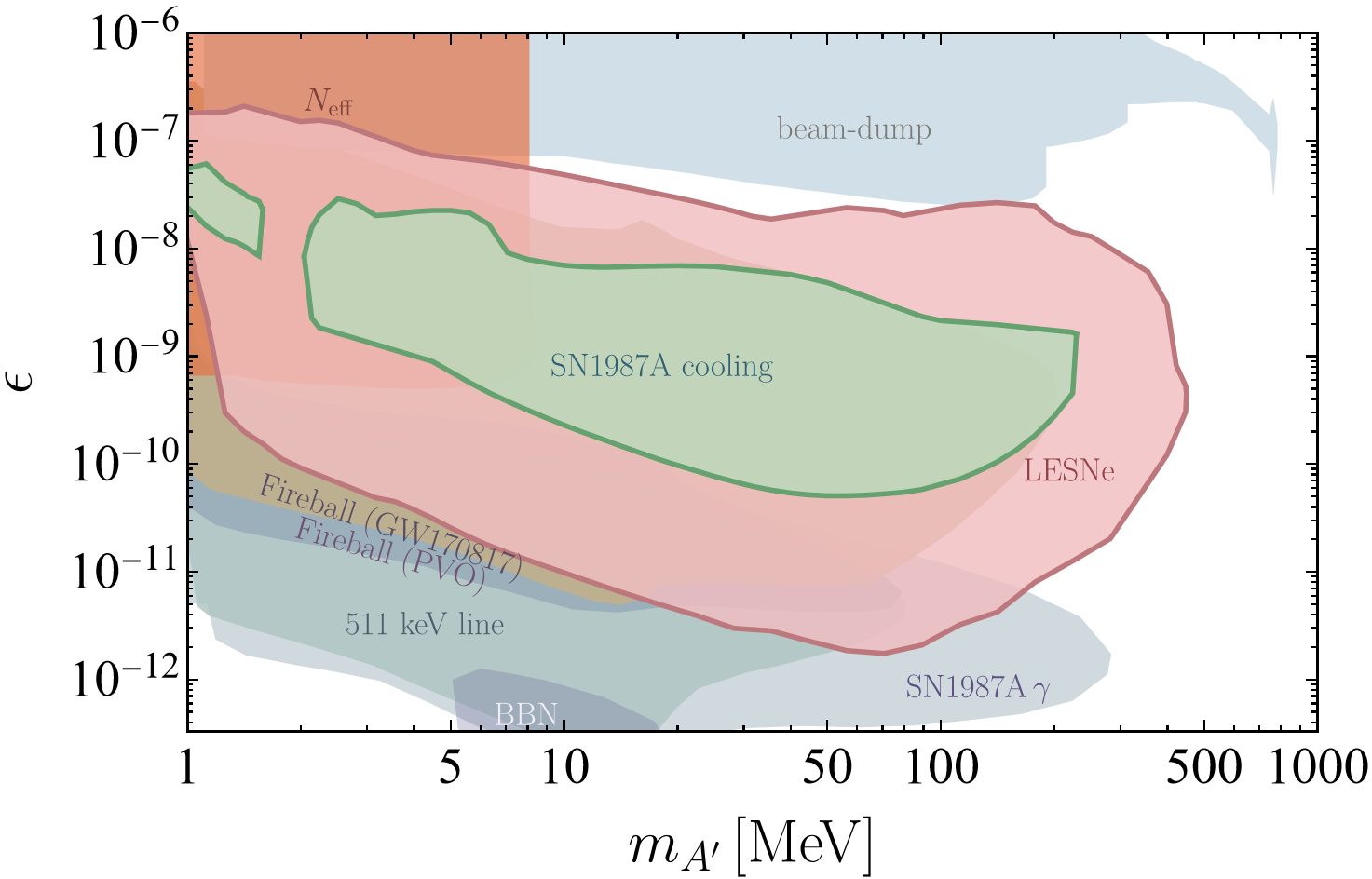}
\caption{Comparison of the supernova limits on the dark photon mass $m_{A'}$ and the kinetic mixing parameter $\epsilon$ obtained in this paper with other existing constraints. 
The shaded green and red regions are excluded by the SN1987A cooling and LESN constraints, respectively (cf.~Fig.~\ref{fig:DP:LESNe}). Other existing constraints include those from the beam-dump experiments (shaded gray, taken from Ref.~\cite{Esseili:2025sbc}), $N_{\mathrm{eff}}$ (shaded orange)~\cite{Ibe:2019gpv}, SN1987A $\gamma$-ray observations (shaded bluish-gray), the 511~keV line (shaded teal), fireball observations associated with GW170817 (shaded brown) and SN1987A via PVO (shaded blue), and BBN (shaded dark gray)~\cite{Caputo:2025avc}. See text for more details. 
}
\label{fig:DP:all_limits}
\end{figure}

For the dark photon with mass $m_{A'} > 1$ MeV, the recent thorough compilations of existing limits can be found e.g. in Refs.~\cite{Caputo:2025avc,Esseili:2025sbc}. The complementarity of the SN1987A limits on dark photon obtained in this paper, using the SFHo-18.8 $18.8M_{\odot}$ profile, with other existing constraints on dark photon is presented in Fig.~\ref{fig:DP:all_limits}. The constraints from the leading beam-dump experiments are shown as the gray shaded region; see Refs.~\cite{Caputo:2021eaa,Esseili:2025sbc} for more details and other weaker laboratory constraints. The bound from the $N_{\rm eff}$ observations is presented as the orange shaded region~\cite{Ibe:2019gpv}. There are additional constraints on dark photon from the observations of the SN1987A  $\gamma$-ray, 511~keV lines, the fireball observations associated with GW170817 and SN1987A via the Pioneer Venus Orbiter (PVO), which are depicted as the bluish-gray, teal, brown and blue shaded regions~\cite{Caputo:2025avc}. The limits on dark photon from the BBN data are presented as the dark gray shaded region~\cite{Caputo:2025avc}. More limits on dark photon, for instance for the sub-MeV scenarios, can be found, e.g. in Ref.~\cite{Caputo:2021eaa}.

\subsection{Comparing the $U(1)_{B-L}$ $Z'$ boson and dark photon limits}
\label{subsec:comparison}

In this section, we collect the main differences of the supernova limits on the  $U(1)_{B-L}$ $Z'$ boson and dark photon. 
As detailed above, the production of $Z'$ boson in the $U(1)_{B-L}$ model and dark photon $A'$ are both dominated by the nucleon bremsstrahlung process, and the plasma effect is crucial for the supernova limits. However, the $Z'$ boson couples directly to neutrons and neutrinos while the dark photon does not, which leads to the following differences for the production of these two gauge bosons in  the supernova core and the resulting supernova limits. 
\begin{itemize}
    \item In light of the plasma effect, when the $Z'$ boson is light, the production of $Z'$ is dominated by the $g_0^2$ terms. In other words, the main contributions are from the different couplings of $Z'$ with protons and neutrons. The matrix element squared in Eq.~(\ref{eqn:ampsq:np:DP}) for the production of dark photon in the leading $np$ bremsstrahlung process is essentially also from the different couplings of $A'$ to protons and neutrons, which turn out to be the effective kinetic mixing parameter $\tilde\epsilon$.
    \item Although the effective coupling of $Z'$ to protons $\tilde{g}_p$ in Eq.~(\ref{eqn:effive_coupling2}) is suppressed by its mass $m_{Z'}$ when $Z'$ is light, the combination $g_0 = \tilde{g}_p - g_n$ does not. In the limit of $m_{Z'} \to 0$ we have instead $g_0 \to g_n = g_{BL}$. As a result, the lower boundaries for the SN1987A cooling limits on the $Z'$ boson are flat (cf. Fig.~\ref{fig:BL}). On the contrary, the effective kinetic mixing parameter for dark photon $\tilde\epsilon \propto m_{A'}^2$ (cf. Eq.~(\ref{eqn:epsilon_effective})), and the SN1987A cooling limits on $\epsilon$ get weaker when the dark photon is light, roughly scales as $m_{A'}^{-2}$ in the transverse mode (cf. Eq.~(\ref{eqn:DP:scaling}) and Fig.~\ref{fig:DP}). 

    \item 
    After being produced in the supernova core, the \(Z'\) bosons can decay into both neutrinos and charged leptons. The decay channels into neutrinos provide an invisible component, which does not contribute to the energy deposition in the stellar mantle. As a result, the energy deposition of $Z'$ inside the mantle is to some extent suppressed by the branching fraction ${\rm BR}(Z' \to \nu\bar\nu)$. As a comparison, the energy carried by the decay products of dark photon is all deposited in the mantle materials. In other words, the dark photon is more efficient in transport energy from the core to mantle, with respect to the $Z'$ boson.

     \item At the lower boundaries for the LESN limits on the $Z'$ boson and dark photon, the dependence of $E_{\rm dep}$ has a different dependence on the coupling $g_{BL}$ or the kinetic mixing parameter $\epsilon$, i.e. $E_{\rm dep} \propto g_{BL}^4$ and $E_{\rm dep} \propto \epsilon^2$, which is a direct consequence of the optical depth $\tau_{Z',\,A'}$ around the lower boundaries. 
\end{itemize}

\section{$U(1)_{L_\mu - L_\tau}$ gauge boson}
\label{sec.Lmutau}
This section is mainly on the supernova limits on the flavor-specific $U(1)_{L_\mu - L_\tau}$ model, which is clearly distinct from the dark photon and $U(1)_{B-L}$ $Z'$ boson above. After sketching some details of this model in Section~\ref{subsec:model:mutau}, the production, decay and absorption in the supernova core is discussed in Section~\ref{subsec:prod:mutau}. Then the supernova limits on the $Z'$ boson in this model are obtained in Section~\ref{subsec:limit:mutau}.

\subsection{Model}
\label{subsec:model:mutau}

In the $U(1)_{L_\mu - L_\tau}$ gauge extension of SM, the corresponding gauge boson $Z'$ couples exclusively to muons, tau leptons and the corresponding neutrinos $\nu_{\mu,\,\tau}$ at the tree level. The relevant interactions are
\begin{equation}
\mathcal{L}_{\mu\tau} \ = \ 
- g_{\mu\tau} Z'_{\mu} \bigl( \bar{\mu} \gamma^\mu \mu + \bar{\nu}_\mu \gamma^\mu P_L \nu_\mu - \bar{\tau} \gamma^\mu \tau - \bar{\nu}_\tau \gamma^\mu P_L \nu_\tau \bigr) \,,
\end{equation}
where $g_{\mu\tau}$ is the gauge coupling constant, and $P_L$ is the left-handed chirality operator. 
The $Z'$ boson may also mix with the SM photon (and the $Z$ boson), e.g. from the muon and tauon loops or via the kinetic mixing terms, with the effective term written as 
\begin{equation}
{\cal L}_{\mu\tau} \ \supset \ - \frac{\epsilon_{\mu\tau}}{2} F_{\mu\nu} Z^{\prime\mu\nu} \,,
\end{equation}
with $Z^{\prime \mu\nu}$ being the standard field strength tensor for $Z'$. 
The irreducible contribution to the mixing of $Z'$ with photon is from the muon and tau lepton loops, and the corresponding contribution to the kinetic mixing parameter $\epsilon_{\mu\tau}$ depends on the momentum transferred $q$~\cite{Kamada:2015era}: 
\begin{equation}
\label{eqn:epsilon_qsq}
\epsilon_{\mu\tau} (q^2) \ = \ - \frac{e g_{\mu\tau}}{2\pi^2} \int_{0}^1 {\rm d}x \, x(1-x) \log \left( \frac{m_\tau^2 - x(1-x)q^2}{m_\mu^2 - x(1-x)q^2} \right) \,.
\end{equation}
At the low-energy scale, it is a good approximation to take $m_\mu \gg |q|$, which leads to the momentum-independent kinetic mixing parameter~\cite{Escudero:2019gzq}
\begin{equation}
\label{eqn:epsilon:70}
\epsilon_{\mu\tau} \ = \ - \frac{e g_{\mu\tau}}{12\pi^2}  \log \left( \frac{m_\tau^2}{m_\mu^2} \right) 
\ \simeq \ - \frac{g_{\mu\tau}}{70} \,, 
\end{equation}
which is called the ``natural kinetic mixing'' case. Then the kinetic mixing term induce the couplings of the $Z'$ boson in the $U(1)_{L_\mu - L_\tau}$ model to all the charged particles in the SM. For our purpose in this paper, the relevant couplings are to electrons and protons, i.e. 
\begin{equation}
{\cal L}_{\mu\tau} \ \supset \ - e \epsilon_{\mu\tau} \ Z_\mu^\prime \left( \bar{p} \gamma^\mu p - \bar{e} \gamma^\mu e \right) \,.
\end{equation} 
In UV-complete theories, there might also be other contributions to the kinetic mixing parameter $\epsilon_{\mu\tau}$, for instance the heavy fermions charged under both the electromagnetic $U(1)$ and $U(1)_{L_\mu - L_\tau}$. Since the $U(1)_{L_\mu-L_\tau}$ gauge boson has no direct coupling to
electrons, the contribution from the plasma effect to the effective
mixing of the $Z'$ boson with the SM photon through the photon
self-energy $\Pi_{\gamma\gamma}^{}$ is absent at the leading order. Therefore, the effective couplings of the $Z'$ boson in the $U(1)_{L_\mu - L_\tau}$ model to charged particles do not receive corrections from the plasma effect, and are identical for the transverse and longitudinal modes.

\subsection{Production, decay and absorption}
\label{subsec:prod:mutau}

\begin{figure}[t]
\centering
\includegraphics[width=0.65\textwidth]{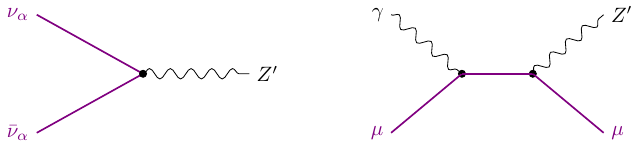}
\caption{Feynman diagrams for the production of $Z'$ in the $U(1)_{L_\mu - L_\tau}$ model in the supernova core, via the neutrino coalescence $\nu_\alpha + \bar{\nu}_\alpha \rightarrow Z^\prime$ ($\alpha = \mu,\,\tau$, left) and the semi-Compton process $\gamma + \mu \rightarrow \mu + Z^\prime$ (right).}
\label{fig:diagramsofLmutau}
\end{figure}

In the supernova core, the production of $Z'$ boson in the $U(1)_{L_\mu - L_\tau}$ model is dominated by the neutrino coalescence $\nu_\alpha + \bar{\nu}_\alpha \rightarrow Z^\prime$ (with $\alpha = \mu,\,\tau$) and the semi-Compton process $\gamma + \mu \rightarrow \mu + Z^\prime$, for both the cases of $\epsilon_{\mu\tau}=0$ and the natural kinetic mixing with a non-vanishing $\epsilon_{\mu\tau}$. The corresponding Feynman diagrams are shown in the left and right panels of Fig.~\ref{fig:diagramsofLmutau}, respectively. 
In presence of $\epsilon_{\mu\tau} \neq0$, the semi-Compton process involving an electron is also possible, i.e. $\gamma + e \to e + Z'$. However, this channel is highly suppressed by the loop-level coupling of $Z'$ with electrons.  
In the case of $\epsilon_{\mu\tau}=0$, there is also the bremsstrahlung process $\mu + p \to \mu + p + Z'$ with $Z'$ emitted from the muon line. However, this process is highly suppressed by the fine structure constant and the three-body phase space with respect to the semi-Compton process~\cite{Croon:2020lrf}. The muon coalescence process $\mu^+ + \mu^- \to Z'$ is also in principle possible, which is, however, highly suppressed by the negligibly small abundance of $\mu^+$ inside the supernova cores. 

For the neutral kinetic  mixing case with $\epsilon_{\mu\tau}\neq0$, the $np$ bremsstrahlung process $n+p \to n+p+Z'$ is also possible. 
For the production of $Z'$ in the supernova core, the momentum $q^2 = m_{Z'}^2$ in Eq.~(\ref{eqn:epsilon_qsq}). In the limit of small $Z'$ mass, the kinetic mixing parameter $\epsilon_{\mu\tau}$ does not depend on $m_{Z'}$, as seen in Eq.~(\ref{eqn:epsilon:70}), then the couplings of $Z'$ to protons are much like the dark photon, up to the overall rescaling factor of $\simeq -1/70$. However, one should note that when the $Z'$ mass $m_{Z'}$ becomes large with respect to $m_\mu$, the induced kinetic mixing $\epsilon_{\mu\tau}(m_{Z'}^2)$ increases significantly, and becomes complex when $m_{Z'} > 2m_\mu$ as $m_\mu^2 - x (1-x)m_{Z'}^2<0$ for $0<x<1$ in Eq.~(\ref{eqn:epsilon_qsq}). The absolute value $|\epsilon_{\mu\tau}|$ reaches maximal when $m_{Z'} = 2m_\mu$, which is roughly $1.47$ times larger than that in the small limit of $m_{Z'}$, i.e.
\begin{equation}
\frac{\epsilon_{\mu\tau} (m_{Z'} = 2m_\mu)}{\epsilon_{\mu\tau}(m_{Z'}=0)} \ \simeq \ 1.47 \,.
\end{equation}
In the calculations below, we will take into account the dependence of $\epsilon_{\mu\tau}$ on the $Z'$ mass, which is significant for a heavy $Z'$ with mass $m_{Z'}\gtrsim100$ MeV. 
In presence of the coupling to electrons and protons, the $Z'$ bosons can also be produced from the $ep$ bremsstrahlung process $e^- +p \to e^-+p+Z'$, which is, however, subdominant to the $np$ bremsstrahlung process above, as in the case of dark photon.

Let us now move on to the production rates for the neutrino coalescence, semi-Compton and $np$ bremsstrahlung processes. 
The production rate for the neutrino coalescence process can be calculated via the detailed balance with the corresponding decay process $Z' \to \nu_i \bar\nu_i$. 
The total production rate in this channel can be decomposed into the transverse and longitudinal modes
\begin{equation}
\label{eqn:T_nunuZ:total}
    \Gamma({\nu_\alpha\bar{\nu_\alpha}\rightarrow Z'}) \ = \  \Gamma_T({\nu_\alpha\bar{\nu_\alpha}\rightarrow Z'})+ \Gamma_L({\nu_\alpha\bar{\nu_\alpha}\rightarrow Z'}) \,,
\end{equation}
with the flavor index $\alpha = \mu,\,\tau$, and
\begin{equation}
\label{eqn:T_nunuZ}
    \Gamma_{T,L}({\nu_\alpha\bar{\nu_\alpha}\rightarrow Z'}) \ = \ e^{-E_{Z^\prime}/T} \Gamma_{T,L}({Z'\rightarrow\nu_\alpha\bar{\nu_\alpha}}) \,. 
\end{equation}
The decay rates $\Gamma_{T,L} (Z' \to \nu_\alpha \bar\nu_\alpha)$ in the transverse and longitudinal modes can be obtained from Eq.~(\ref{neutrino coalescence}) by the simple replacement of $g' \to g_{\mu\tau}$. The rates in Eq.~(\ref{eqn:T_nunuZ}) are derived for a single neutrino flavor; both the neutrinos of muon and tauon flavors are taken into account to set limits on the $Z'$ boson.  
For our purpose, the chemical potential of muon neutrinos is extracted from the SFHo-18.8 $18.8M_\odot$ profile to calculate the corresponding number densities of $\nu_\mu,\,\bar\nu_\mu$~\cite{Sukhbold:2017cnt}, whereas the tau neutrinos are treated as non-degenerate with a negligible chemical potential.

In the low energy limit of $E_{Z'} \ll m_\mu$, the semi-Compton process is quite similar to the standard Compton process, up to the different couplings involved. In light of the temperature $T < m_\mu$ inside the supernova core, we can take the limit of $E_\gamma \ll m_\mu$, with $E_\gamma$ being the incoming photon energy. In this limit, the muon absorbs only a small fraction of the photon energy in the scattering, and the photon energy is almost transferred to the light $Z'$ in the final state, i.e. $E_{Z'} \simeq E_\gamma$. In this limit, the cross sections for the semi-Compton process in the transverse and longitudinal modes are, respectively, 
\begin{subequations}
\label{eqn:xsec:sC}
\begin{align}
\sigma_{\mathrm{sC}}^T \ = \ & \frac{2\pi \alpha \alpha_{\mu\tau} m_{Z^\prime}^2 }{m_\mu^2 E_\gamma^2} \sqrt{1-\frac{m_{Z^\prime}^2}{E_{\gamma}^2}} \,, \label{eqn:xsec:sC:T} \\
\sigma_{\mathrm{sC}}^L \ = \ & \frac{8\pi\alpha \alpha_{\mu\tau}}{3m_\mu^2} \sqrt{1-\frac{m_{Z^\prime}^2}{E_{\gamma}^2}} \,,
\label{eqn:xsec:sC:L}
\end{align}
\end{subequations}
where $\alpha_{\mu\tau} \equiv g_{\mu\tau}^2/4\pi$.
Then the corresponding production rates in the transverse and longitudinal modes are, respectively, 
\begin{equation}
\label{eqn:Gamma:sC:prod}
\Gamma_{\mathrm{sC}| \, T, L}^{\mathrm{prod}} \ = \ \sigma_{\mathrm{sC}}^{T, L} \times \frac{n_\mu F_{\mathrm{deg}}}{e^{E_{Z'}/T} - 1} \,,
\end{equation}
where $n_\mu$ is the muon number density in the supernova core as a function of radius $r$~\cite{Sukhbold:2017cnt}. The factor $F_{\mathrm{deg}}$ in Eq.~(\ref{eqn:Gamma:sC:prod}) is the degeneracy factor for the Pauli-blocking effect for the muon in the final state, as a function of radius $r$ within the supernova core, and can be found in Refs.~\cite{Bollig:2020xdr,Lai:2024mse}. It should be noted that the plasma effect is not important for the production of $Z'$ in the semi-Compton process, as the $Z'$ boson does not couple directly to electrons, and the muon mass is significantly larger than the typical temperature $T$ in the supernova core. 
The total production rate in the semi-Compton process is obtained by summing over all polarization modes, i.e. 
\begin{equation}
\Gamma_{\mathrm{sC}}^{\mathrm{prod}} \ = \ \Gamma_{\mathrm{sC}| \, T}^{\mathrm{prod}} + \Gamma_{\mathrm{sC}| \, L}^{\mathrm{prod}} \,.
\end{equation}  
As seen in Eq.~(\ref{eqn:xsec:sC:T}), in the light $Z'$ regime, the transverse cross section suffers from an explicit $m_{Z'}^2/E_\gamma^2 \ll 1$ suppression, and the production of $Z'$ in the semi-Compton process is dominated by the longitudinal mode. More details can be found in Appendix~\ref{app:semi-Compton}. 

For the $np$ bremsstrahlung process, the \(U(1)_{L_\mu-L_\tau}\) gauge boson couples to protons through the loop-induced kinetic mixing with the SM photon, with an effective coupling \(e\epsilon_{\mu\tau}\), while its coupling to the neutron vanishes at leading order. The bremsstrahlung matrix elements can then be obtained directly from that for dark photon in Eq.~(\ref{eqn:ampsq:np:DP}) by the replacement of \(\tilde{\epsilon}\to \epsilon_{\mu\tau}\). The plasma effect is not important in this channel, as the coupling of $Z'$ to electrons arises at the 1-loop order in the $U(1)_{L_\mu - L_\tau}$ model.
The calculation procedure of the resulting production rate $\Gamma_{\rm brem}^{\rm prod}$ in this channel is the same as that for the $U(1)_{B-L}$ $Z'$ boson and dark photon above.

Regarding the decay of the $Z'$ boson, in addition to $Z' \to \nu_\mu \bar{\nu}_\mu,\, \nu_\tau \bar\nu_\tau$,  
the decays $Z' \to e^+ e^-,\, \mu^+ \mu^-$ 
are also possible if kinematically allowed, i.e. $m_{Z'} > 2m_\ell$ ($\ell = e, \mu$). The corresponding partial widths are given by Eqs.~(\ref{eqn:decay:Zprime:electron}) and~(\ref{eqn:decay:Zprime:muon}), with $\tilde{g}_{e}$ replaced by $e\epsilon_{\mu\tau}$ for the electron channel and $\tilde{g}_\mu$ by $g_{\mu\tau}$ for the muon channel, respectively. As mentioned above for the semi-Compton production, here the plasma effect is not important for the decay of $Z'$.
The absorption rate through the inverse semi-Compton process $Z' + \mu^- \to \mu^- + \gamma$ can be obtained by the principle of detailed balance:
\begin{equation}
\Gamma_{\mathrm{sC}}^{\mathrm{abs}} \ = \ \frac{2}{3}e^{E_{\gamma}/T} \Gamma_{\mathrm{sC}}^{\mathrm{prod}} \,,
\end{equation}
where the factor of $2/3$ accounts for the polarization averaging when replacing the initial photon by a massive $Z'$ state.

\subsection{Supernova limits}
\label{subsec:limit:mutau}

\begin{figure}[t]
\centering
\includegraphics[width=0.6\textwidth]{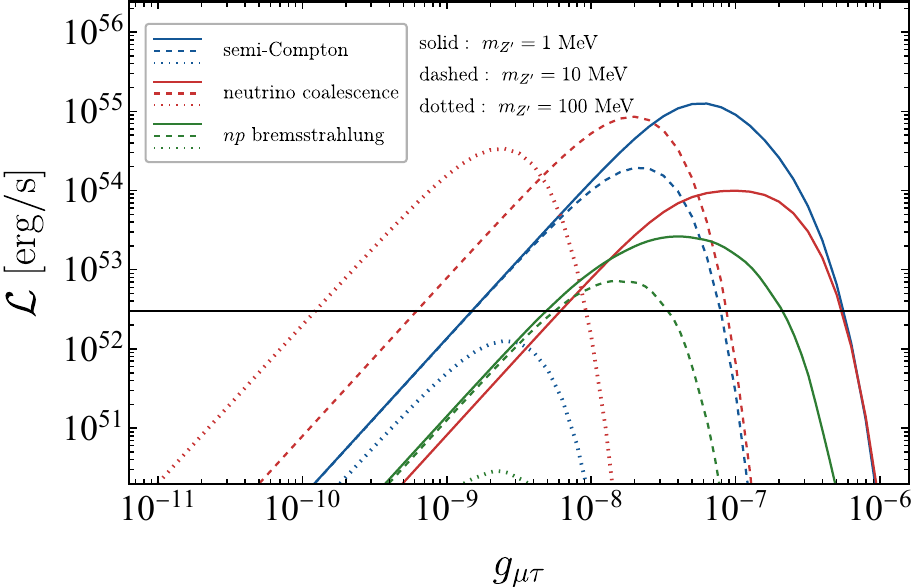}
\caption[2]{Luminosities of the $U(1)_{L_\mu-L_\tau}$ gauge boson in the channels of semi-Compton, neutrino coalescence, and $np$
bremsstrahlung processes, as functions of the coupling $g_{\mu\tau}$. The solid, dashed and dotted lines are for the masses of $m_{Z'} = 1$, 10 and 100 MeV, respectively. 
}
\label{L_mu-L_tau luminosity}
\end{figure}

Following the procedure outlined in Sec.~\ref{sec.setups}, the luminosity due to the $U(1)_{L_\mu-L_\tau}$ gauge boson can be written as 
\begin{equation}
    {\cal L} 
    \ = \ \int\frac{{\rm d}^3k}{(2\pi)^3} E_{Z'} \left(\Gamma_{\mathrm{sC}}^{\mathrm{prod}}+\Gamma({\nu_\alpha\bar{\nu_\alpha}\rightarrow Z'}) + \Gamma_{\rm brem}^{\rm prod}\right)\mathrm{exp}\{-\tau_{Z'}\} \,.
\end{equation}
where the optical depth $\tau_{Z'}$ accounts for the geometric effects and attenuation due to all relevant absorption and decay processes along the $Z'$ propagation trajectory.
Let us first compare the production channels of $Z'$ in the $U(1)_{L_\mu - L_\tau}$ model. Taking again the supernova profile of SFHo-18.8 $18.8 M_{\odot}$, the resulting luminosities due to $Z'$ in the neutrino coalescence, semi-Compton, and $np$ bremsstrahlung processes are depicted as the red, blue and green lines in Fig.~\ref{L_mu-L_tau luminosity}, respectively, as functions of the gauge coupling $g_{\mu\tau}$. For illustration purposes, we have chosen three benchmark values of the gauge boson mass, i.e. $m_{Z'} = 1\text{ MeV}$, $10\text{ MeV}$, and $100\text{ MeV}$, which are presented as the solid, dashed and dotted lines, respectively. It is rather clear in this figure that the semi-Compton scattering dominates in the lighter mass regime with $m_{Z^\prime} \lesssim 10\text{ MeV}$, whereas neutrino coalescence becomes the dominant channel when $m_{Z^\prime} \gtrsim 10\text{ MeV}$. Here for simplicity we adopt the natural kinetic mixing scenario
with $\epsilon_{\mu\tau}=-g_{\mu\tau}/70$, and it is found that the $np$ bremsstrahlung process is always subdominant throughout the parameter space of interest.  
Even if one uses the full momentum dependent expression for $\epsilon_{\mu\tau}(q^2)$ in Eq.~(\ref{eqn:epsilon_qsq}), the resulting luminosity in  the $np$ bremsstrahlung process does not change significantly. 
Therefore, we will neglect the $np$ bremsstrahlung process, and consider only the neutrino coalescence and semi-Compton processes for the production of $Z'$ in the $U(1)_{L_\mu - L_\tau}$ model in the supernova core.

\begin{figure}[t]
\centering
\includegraphics[width=0.6\textwidth]{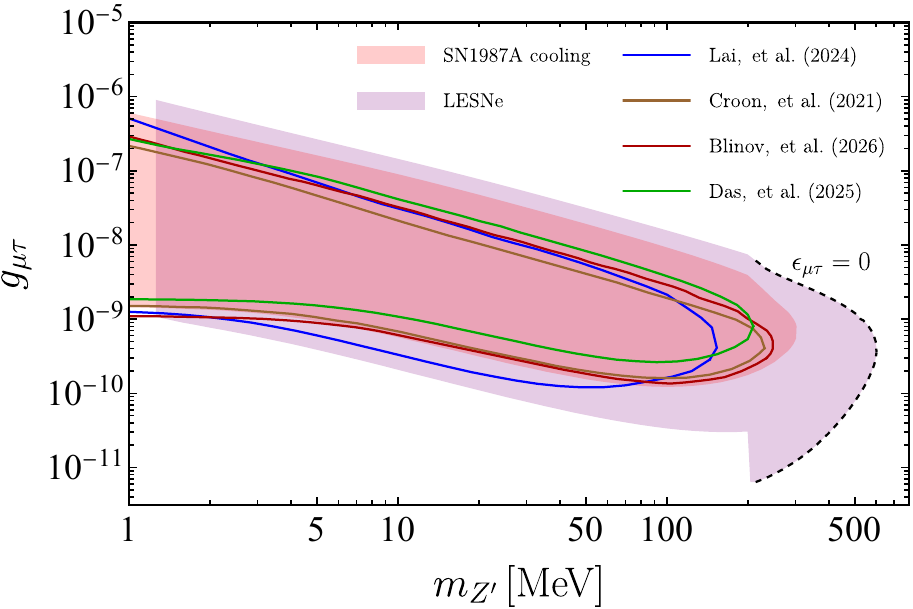}
\caption[2]{Supernova limits on the $U(1)_{L_\mu-L_\tau}$ gauge boson mass $m_{Z'}$ and coupling $g_{\mu\tau}$, based on the SFHo-18.8 $18.8 M_{\odot}$ profile. The pink and purple shaded regions represent the SN1987A cooling and LESN constraints obtained in this work, respectively. The dashed black line corresponds to the LESN limit in the special case of $\epsilon_{\mu\tau}=0$.  
The bounds from Lai et al. (blue)~\cite{Lai:2024mse}, Croon et al. (brown)~\cite{Croon:2020lrf}, Blinov et al. (red)~\cite{Blinov:2025aha} and Das et al. (green)~\cite{Das:2026uam} are also shown for comparison. See text for more details. 
}
\label{fig:mutau:SN1987A}
\end{figure}

Combining the contributions from the $T$ and $L$ modes, the resulting SN1987A cooling limit on the $Z'$ boson in the $U(1)_{L_\mu -L_\tau}$ model is presented as the pink shaded region in Fig.~\ref{fig:mutau:SN1987A}. For the sake of concreteness, we have taken the natural kinetic mixing assumption of $\epsilon_{\mu\tau} = -g_{\mu\tau}/70$. However, it is found that the SN1987A cooling limits remain essentially unchanged regardless of whether the kinetic mixing $\epsilon_{\mu\tau}$ is taken into account. The $Z'$ boson mass is excluded up to 300 MeV, with the coupling $g_{\mu\tau}$ constrained up to $1.2\times 10^{-10}$.  For comparison, the limits from Lai et al.~\cite{Lai:2024mse}, Croon et al.~\cite{Croon:2020lrf}, Das et al.~\cite{Das:2026uam} and Blinov et al.~\cite{Blinov:2025aha} are shown as the blue, brown, green and red lines, respectively. Our lower boundary is quite close to those obtained in these papers, in particular Refs.~\cite{Blinov:2025aha,Das:2026uam}.
At the upper boundary, larger values of $g_{\mu\tau}$ are excluded in our calculations.  
This originates from the different treatments of the trapping region and the adopted stellar profiles, which affect the optical depth calculations. 

The LESN limits on the $Z'$ boson in the $U(1)_{L_\mu-L_\tau}$ model depend largely on the decay $Z'\to\ell^+\ell^-$ (with $\ell=e,\mu$) inside the stellar mantle. For concreteness, let us first consider the natural kinetic mixing scenario with $\epsilon_{\mu\tau}=-g_{\mu\tau}/70$. The resulting LESN limits are shown as the purple shaded region in Fig.~\ref{fig:mutau:SN1987A}. 
The $Z'$ mass is excluded up to 600 MeV, with $g_{\mu\tau}$ constrained down to $6.4\times 10^{-12}$. 
For the mass range of $2m_e < m_{Z'} < 2m_\mu$, the energy deposition of $Z'$ inside the mantle is driven by the decay $Z' \to e^+e^-$, whereas for $m_{Z'} > 2m_\mu$, the energy deposition is dominated by the channel $Z' \to \mu^+\mu^-$. As the coupling of $Z'$ to electrons is entirely from the loop-induced kinetic mixing $\epsilon_{\mu\tau}$, the corresponding branching fraction ${\rm BR}(Z' \to e^+ e^-)$ is highly suppressed, and there is a prominent kink structure around $m_{Z'} \simeq 2m_\mu$ for the LESN limit. 
If we adopt instead the momentum-dependent kinetic mixing parameter $\epsilon_{\mu\tau}(q^2)$, the resulting modification to the LESN limits is negligible.
It should be noted that the LESN constraint only applies for $m_{Z'} > 2m_e$, as below this threshold the $Z'$ boson decays exclusively into neutrinos $\nu_{\mu,\,\tau}$, and there are no effective decay or absorption channels for the $Z'$ bosons to efficiently deposit energy into the stellar mantle. 
For comparison, we also consider the case of $\epsilon_{\mu\tau}=0$, in which the $Z'$ boson does not couple to electrons. Consequently, the energy deposition of $Z'$ through the charged lepton decay channels becomes possible only for $m_{Z'}>2m_\mu$, when the $Z'\to\mu^+\mu^-$ channel opens. The corresponding LESN limits are indicated by the dashed black line. 

\begin{figure}[t]
\centering
\includegraphics[width=0.7\textwidth]{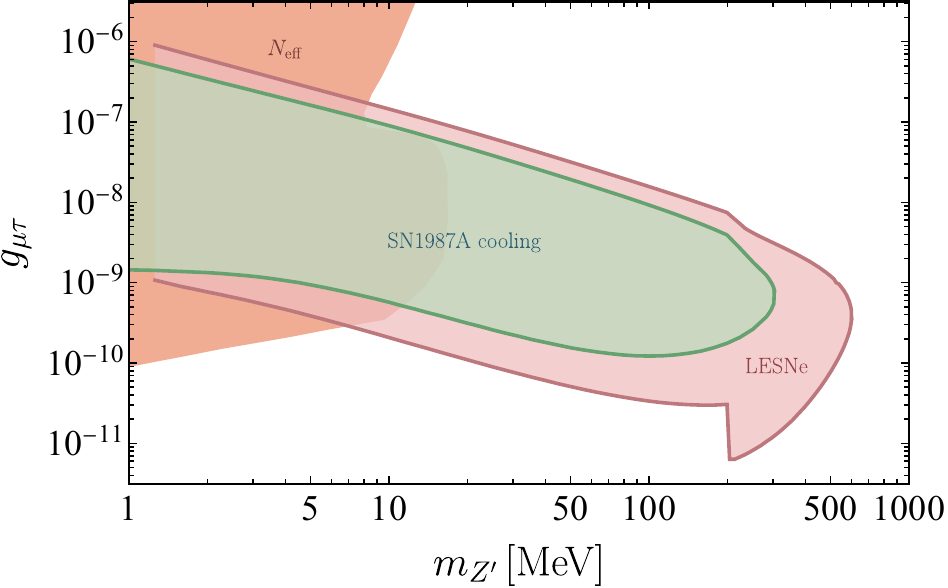}
\caption{Comparison of the supernova limits on the $U(1)_{L_\mu-L_\tau}$ gauge boson mass $m_{Z'}$ and coupling $g_{\mu\tau}$ from SN1987A cooling (shaded green) and LESNe (shaded red) obtained in this work (cf.~Fig.~\ref{fig:mutau:SN1987A}) with that from $N_{\rm eff}$ (shaded orange)~\cite{Escudero:2019gzq}. 
See text for more details. 
}
\label{fig:mutau:all_limits}
\end{figure}

The comparison of the supernova limits on the $Z'$ boson in the $U(1)_{L_\mu-L_\tau}$ model obtained in this paper with other constraints is presented in Fig.~\ref{fig:mutau:all_limits}. 
For concreteness, we adopt the natural kinetic-mixing scenario with $\epsilon_{\mu\tau}=-g_{\mu\tau}/70$. 
Within the parameter space of interest in this paper, the relevant bound arises from the limit of $N_{\rm eff}$~\cite{Escudero:2019gzq}. 
In light of the current limits on $\Delta N_{\rm eff}$ from the Planck data~\cite{Planck:2018vyg}, we take the limit of $\Delta N_{\rm eff} < 0.2$ from the bottom panel of Fig.~2 in Ref.~\cite{Escudero:2019gzq}. 
Other constraints, including those from the laboratory experiments~\cite{Andreas:2012mt,Abdullah:2018ykz,NA64:2024klw,CCFR:1991lpl,Kaneta:2016uyt,Bellini:2011rx,Kopp:2012dz,BaBar:2009lbr,BaBar:2014zli}, white dwarf cooling~\cite{Dreiner:2013tja} and the BBN data~\cite{Kamada:2015era}, are more sensitive to large values of $g_{\mu\tau}$, and are thus not shown in this figure. It is clear in Fig.~\ref{fig:mutau:all_limits} that the supernova limits on the $U(1)_{L_\mu - L_\tau}$ $Z'$ boson exclude large regions of parameter space beyond  the $N_{\rm eff}$ bounds of $m_{Z'} \gtrsim 15$ MeV, up to the $Z'$ mass of roughly 600 MeV.

\section{Conclusion}
\label{sec.conclusion}
In this work, we have revisited the supernova constraints on light BSM gauge bosons. This paper is mainly on the limits from the cooling of SN1987A and the energy deposition in the LESNe, and we focus primarily on three representative scenarios, i.e. the dark photon, the $U(1)_{B-L}$ gauge boson, and the $U(1)_{L_\mu-L_\tau}$ gauge boson. 
In obtaining the supernova limits, the in-medium plasma effects on the effective couplings, the supernova profiles and the geometric effects associated with energy transport inside the stars are systematically taken into account. 

For the $U(1)_{B-L}$ gauge boson scenarios, the leading production channel in the supernova core is the $np$, $pp$ and $nn$ nucleon bremsstrahlung process. Within the framework of OPE approximation, we retain the full momentum dependence of the emitted gauge boson on both nucleon and pion propagators, 
and consistently keep the gauge boson mass terms, which are important for the large mass range of $m_{Z^\prime} \gtrsim 3T$.
The resulting SN1987A cooling constraints depend largely on the supernova profiles adopted. For the profiles of Fischer $11.8 M_\odot$, SFHo-18.8 $18.8 M_\odot$, and LS220 $20 M_\odot$, the $Z'$ boson mass is excluded up to $100$, $145$, and $330~\mathrm{MeV}$, and the coupling $g_{BL}$ is constrained down to $3.7\times 10^{-11}$, $1.7\times10^{-11}$ and $7.1\times 10^{-12}$ in the transverse modes, respectively. As seen in Fig.~\ref{fig:BL}, smaller values of $g_{BL}$ are excluded in the longitudinal mode, i.e. down to $1.1\times10^{-15}$, $2.9\times 10^{-16}$ and $7\times 10^{-17}$ for $m_{Z'} >1$ MeV, with the gauge boson mass constraints up to $190$, $280$, and $560~\mathrm{MeV}$, respectively, for these three profiles. The excluded ranges of $g_{BL}$ are to some extent narrower in the longitudinal mode.
As a result of the more stringent constraints, the LESN observations exclude a significantly larger parameter space of $m_{Z'}$ and $g_{BL}$: based on the SFHo-18.8 $18.8 M_\odot$ profile, the LESN limit is dominated by the transverse modes, with the exclusion range of $1.9\times10^{-12}\lesssim g_{BL}\lesssim3.1\times10^{-7}$
for masses up to
$m_{Z'}\simeq440~\mathrm{MeV}$ (cf. Fig.~\ref{fig:BL:LESNe}).

Unlike the $U(1)_{B-L}$ $Z'$ boson case, the production of the dark photon $A'$ in the supernova core is always dominated by the $np$ bremsstrahlung process.
As the plasma-dressed kinetic mixing parameter $\tilde\epsilon$ is suppressed by the dark photon mass when $A'$ is light, the corresponding supernova limits get weaker accordingly. Regarding the SN1987A cooling limits in the transverse modes, for the Fischer $11.8 M_\odot$, SFHo-18.8 $18.8 M_\odot$, and LS220 $20 M_\odot$ profiles, the excluded dark photon mass ranges extend up to $160$, $230$, and $490~\mathrm{MeV}$, and the kinetic mixing parameter $\epsilon$ is excluded down to $2.2\times 10^{-10}$, $1.7\times 10^{-10}$ and $4.5\times 10^{-11}$, respectively (cf. Fig.~\ref{fig:DP}). On the LESN limits, based again on the SFHo-18.8 $18.8 M_\odot$ profile, the parameter $\epsilon$ is extended further down to $5.8\times10^{-12}$, with the dark photon masses excluded up to $m_{A'}\simeq450~\mathrm{MeV}$ in the transverse modes (cf. Fig.~\ref{fig:DP:LESNe}). No parameter space of $m_{A'}$ and $\epsilon$ is excluded in the longitudinal mode as a result of the suppression of effective kinetic mixing parameter by $m_{A'}$ when the plasma effect is taken into account.

On the $Z'$ gauge boson in the $U(1)_{L_\mu-L_\tau}$ model, its production in the supernova core is dominated by the neutrino coalescence and semi-Compton processes. Adopting the SFHo-18.8 $18.8 M_\odot$ profile, the SN1987A cooling constraint excludes the $Z'$ boson masses up to approximately $300~\mathrm{MeV}$, with the gauge coupling constrained down to $1.2\times10^{-10}$. The LESN constraint extends the mass exclusion to approximately $600~\mathrm{MeV}$ and the coupling to roughly $6.4\times10^{-12}$ (cf.~Fig.~\ref{fig:mutau:SN1987A}). 

The supernova limits on these dark gauge bosons obtained in this paper are largely complementary to other laboratory, astrophysical, and cosmological constraints (cf. Figs.~\ref{fig:BL:all_limits}, \ref{fig:DP:all_limits} and \ref{fig:mutau:all_limits}). 
We would like to emphasize that the OPE approximation for the supernova limits on the $U(1)_{B-L}$ gauge boson and dark photon can be further improved by incorporating systematically various nuclear effects~\cite{Carenza:2019pxu}. Furthermore, taking into account the time-dependent supernova evolution may also help to strengthen all the supernova limits in this paper. In addition, with the continued development of realistic supernova simulations, the interplay between particle transport and stellar dynamics could potentially provide new opportunities for probing light weakly coupled BSM particles. 

\section*{Acknowledgments}

The authors would like to thank Rabindra Mohapatra, Bhaskar Dutta, P. S. Bhupal Dev and Xun-Jie Xu for the enlightening discussions. This work  is supported by the National Natural Science Foundation of China (NSFC) under Grant No. 12175039, the State Key Laboratory of Dark Matter Physics and the ``Fundamental Research Funds for the Central Universities''. The numerical calculations are partially supported by the Big Data Computing Center of Southeast University.

\appendix

\section{The matrix elements squared for the $U(1)_{B-L}$ gauge boson}
\label{app:ampsq:BL}

\subsection{Preparation}
\label{appsub:preparation}

For the production of $Z'$ in the $U(1)_{B-L}$ model via the $np$ bremsstrahlung process $p(p_1) + n(p_2) \to p(p_3) + n(p_4) + Z' (k_{Z'})$, there are in total 11 diagrams in the OPE approximation, as shown in Fig.~\ref{fig:diagrams}. Following the convention in Refs.~\cite{Dent:2012mx,Dev:2020eam}, the amplitudes for the four diagrams in the left panel of Fig.~\ref{fig:diagrams} in the $t$-channel can be written as, respectively, 
\begin{subequations}
\label{eqn:amp:Mt}
\begin{align}
{\cal M}_a \ = \ & \frac{4 m_N^2}{m_\pi^2} f_{pp} f_{nn} \tilde{g}_{p}
	\frac{1}{k^2 - m_\pi^2} \frac{1}{(p_3 + k_{Z^\prime})^2 - m_N^2} \nonumber \\
	& \times \Big[ \bar{u} (p_3) \gamma_\mu \left( \slashed{p}_3 + \slashed{k}_{Z'} + m_N \right) \gamma_5 u(p_1) \Big] \Big[ \bar{u} (p_4) \gamma_5 u(p_2) \Big] \varepsilon^\mu (k_{Z^\prime}) \,, \\
\label{eqn:amp:Mb}
    {\cal M}_b \ = \ & \frac{4 m_N^2}{m_\pi^2} f_{pp} f_{nn} g_{n}
	\frac{1}{(k-k_{Z^\prime})^2 - m_\pi^2} \frac{1}{(p_4 + k_{Z^\prime})^2 - m_N^2} \nonumber \\
& \times \Big[ \bar{u} (p_3) \gamma_5 u(p_1) \Big] \Big[ \bar{u} (p_4)  \gamma_\mu \left( \slashed{p}_4 + \slashed{k}_{Z'} + m_N \right) \gamma_5 u(p_2) \Big] \varepsilon^\mu (k_{Z^\prime}) \,, \\
 {\cal M}_c \ = \ & \frac{4 m_N^2}{m_\pi^2} f_{pp} f_{nn} \tilde{g}_{p}
	\frac{1}{k^2 - m_\pi^2} \frac{1}{(p_1 - k_{Z^\prime})^2 - m_N^2} \nonumber \\
& \times \Big[ \bar{u} (p_3) \gamma_5 \left( \slashed{p}_1 - \slashed{k}_{Z'} + m_N \right) \gamma_\mu u(p_1) \Big] \Big[ \bar{u} (p_4) \gamma_5 u(p_2) \Big] \varepsilon^\mu (k_{Z^\prime}) \,, \\
\label{eqn:amp:Md}
    {\cal M}_d \ = \ & \frac{4 m_N^2}{m_\pi^2} f_{pp} f_{nn} g_{n}
	\frac{1}{(k-k_{Z^\prime})^2 - m_\pi^2} \frac{1}{(p_2 - k_{Z^\prime})^2 - m_N^2} \nonumber \\
& \times \Big[ \bar{u} (p_3) \gamma_5 u(p_1) \Big] \Big[ \bar{u} (p_4) \gamma_5 \left( \slashed{p}_2 - \slashed{k}_{Z'} + m_N \right) \gamma_\mu u(p_2) \Big] \varepsilon^\mu (k_{Z^\prime}) \,,
\end{align}
\end{subequations}
where $\tilde{g}_{p}$ is the effective coupling of $Z'$ to protons with the plasma effect, $k \equiv p_2 - p_4$ is the momentum transferred in the $t$-channel, and $\varepsilon^\mu (k_{Z'})$ is the polarization vector for the $Z'$ boson with momentum $k_{Z'}$. The four diagrams in the middle panel of Fig.~\ref{fig:diagrams} are in the $u$-channel, and the corresponding amplitudes are, respectively, 
\begin{subequations}
\label{eqn:amp:Mu}
\begin{align}
{\cal M}_{a^\prime} \ = \ & \frac{4 m_N^2}{m_\pi^2} f_{pn}^2 {g}_{n}
	\frac{1}{l^2 - m_\pi^2} \frac{1}{(p_4 + k_{Z^\prime})^2 - m_N^2} \nonumber \\
& \times \Big[ \bar{u} (p_4) \gamma_\mu \left( \slashed{p}_4 + \slashed{k}_{Z'} + m_N \right) \gamma_5 u(p_1) \Big] \Big[ \bar{u} (p_3) \gamma_5 u(p_2) \Big] \varepsilon^\mu (k_{Z^\prime}) \,, \\
\label{eqn:amp:Mbprime}
{\cal M}_{b^\prime} \ = \ & \frac{4 m_N^2}{m_\pi^2} f_{pn}^2 \tilde{g}_{p}
	\frac{1}{(l-k_{Z^\prime})^2 - m_\pi^2} \frac{1}{(p_3 + k_{Z^\prime})^2 - m_N^2} \nonumber \\
& \times \Big[ \bar{u} (p_4) \gamma_5 u(p_1) \Big] \Big[ \bar{u} (p_3)  \gamma_\mu \left( \slashed{p}_3 + \slashed{k}_{Z'} + m_N \right) \gamma_5 u(p_2) \Big] \varepsilon^\mu (k_{Z^\prime}) \,, \\
	{\cal M}_{c^\prime} \ = \ & \frac{4 m_N^2}{m_\pi^2} f_{pn}^2 \tilde{g}_{p}
	\frac{1}{l^2 - m_\pi^2} \frac{1}{(p_1 - k_{Z^\prime})^2 - m_N^2} \nonumber \\
& \times \Big[ \bar{u} (p_4) \gamma_5 \left( \slashed{p}_1 - \slashed{k}_{Z'} + m_N \right) \gamma_\mu u(p_1) \Big] \Big[ \bar{u} (p_3) \gamma_5 u(p_2) \Big] \varepsilon^\mu (k_{Z^\prime}) \,, \\
\label{eqn:amp:Mdprime}
{\cal M}_{d^\prime} \ = \ & \frac{4 m_N^2}{m_\pi^2} f_{pn}^2 g_{n}
	\frac{1}{(l-k_{Z^\prime})^2 - m_\pi^2} \frac{1}{(p_2 - k_{Z^\prime})^2 - m_N^2} \nonumber \\
& \times \Big[ \bar{u} (p_4) \gamma_5 u(p_1) \Big] \Big[ \bar{u} (p_3) \gamma_5 \left( \slashed{p}_2 - \slashed{k}_{Z'} + m_N \right) \gamma_\mu u(p_2) \Big] \varepsilon^\mu (k_{Z^\prime}) \,,
\end{align}
\end{subequations}
where the momentum transferred in the $u$-channel is $l \equiv p_2 - p_3$. The three diagrams in the right panel of Fig.~\ref{fig:diagrams} are purely from the plasma effect $g_0=\tilde{g}_p - g_n \neq 0$, and the corresponding amplitudes are, respectively,
\begin{subequations}
\begin{align}
\label{eqn:amp:Meprime}
    {\cal M}_{e^\prime} \ = \ & \frac{4 m_N^2}{m_\pi^2} f_{pn}^2 g_0 \frac{1}{l^2 - m_\pi^2}
	\frac{1}{(l-k_{Z^\prime})^2 - m_\pi^2} \nonumber \\
& \times \Big[ \bar{u} (p_4) \gamma_5 u(p_1) \Big] \Big[ \bar{u} (p_3) \gamma_5  u(p_2) \Big](k_{Z^\prime}-2l)_{\mu}\varepsilon^\mu(k_{Z^\prime}) \,, \\
\label{eqn:amp:Mfprime}
    {\cal M}_{f^\prime} \ = \ & \frac{2 m_N}{m_\pi^2} f_{pn}^2 g_0
	\frac{1}{l^2 - m_\pi^2} 
	\Big[ \bar{u} (p_4) \gamma_5 u(p_1) \Big] \Big[ \bar{u} (p_3) \gamma_5  u(p_2) \Big] \varepsilon^\mu (k_{Z^\prime}) \,, \\
\label{eqn:amp:Mgprime}
    {\cal M}_{g^\prime} \ = \ & -\frac{2 m_N}{m_\pi^2} f_{pn}^2 g_0
	\frac{1}{(l-k_{Z^\prime})^2 - m_\pi^2}\Big[ \bar{u} (p_4) \gamma_5 u(p_1) \Big] \Big[ \bar{u} (p_3) \gamma_5   u(p_2) \Big] \varepsilon^\mu (k_{Z^\prime}) \,.  
\end{align}
\end{subequations}

In the non-relativistic limit, it is a good approximation that 
$k^2 \simeq -\mathbf{k}^2$, $l^2 \simeq -\mathbf{l}^2$, and $k \cdot l \simeq -\mathbf{k} \cdot \mathbf{l}$, with ${\bf k}$ and ${\bf l}$ referring to the corresponding 3-momenta. In light of the temperature $T \sim 30$ MeV in the supernova core, we have the hierarchical structure for the mass and energy scales~\cite{Dent:2012mx}:
\begin{equation}
m_{Z^\prime}^2,\, \mathbf{k}_{Z^\prime}^2,\, E_{Z^\prime}^2 \ \ll \ |\mathbf{k}|^2,\, |\mathbf{l}|^2,\, \mathbf{k} \cdot \mathbf{l} \ \sim \ 3m_N T \ \ll \ m_N^2 \,.
\end{equation}
In the nucleon propagators of the 8 diagrams in the left and middle panels of Fig.~\ref{fig:diagrams}, there is the factor of $(p_i \pm k_{Z^\prime})^2 - m_N^2 = \pm 2(p_i \cdot k_{Z^\prime})+m_{Z^\prime}^2$, with the minus sign for $i=1,\,2$ and plus sign for $i=3,\,4$. As the nucleons are non-relativistic, it is a good approximation that $p_i \cdot k_{Z^\prime} \simeq m_N E_{Z^\prime}$. However, if we take this approximation in the nucleon propagators at the beginning of calculations, the amplitudes squared for the four $t$-channel diagrams and the four $u$-channel diagrams are proportional to $g_0^2$ (cf. Eq.~(\ref{eq:ampsq:propto})). 
In the limit of $g_0 =\tilde{g}_p - g_{n} \to 0$, i.e. in the absence of the plasma effect when $m_{Z'} \gtrsim \omega_{\rm pl}$, the amplitude squared vanishes. 

To retain the leading-order non-vanishing terms, we need to keep the terms as that in Eq.~(\ref{eqn:sim}). Let us now go into more details. The numerator in Eq.~(\ref{eqn:sim}) can be re-written as 
\begin{equation}
(\mathbf{k}\cdot {\mathbf{k}}_{Z^\prime})^2 \ = \ |{\bf k}|^2 |{\bf k}_{Z'}|^2 
( \hat{\bf k} \cdot \hat{\bf k}_{Z'} )^2 \,,
\end{equation}
with $\hat{\bf p} \equiv {\bf p}/|{\bf p}|$ being the unit vector representing the direction of the corresponding 3-momentum ${\bf p}$. Following Ref.~\cite{Brinkmann:1988vi}, we can evaluate the non-vanishing interference terms between different diagrams by performing the averaging over the direction ${\bf k}_{Z'}$ of the $Z^\prime$ bosons. Specifically, the most general angular averaging involving both ${\bf k}$ and ${\bf l}$ can be written in the form of
\begin{equation}
\langle (\hat{\mathbf{k}} \cdot \hat{{\mathbf{k}}}_{Z^\prime})^n (\hat{\mathbf{l}} \cdot \hat{{\mathbf{k}}}_{Z^\prime})^m \rangle \,.
\end{equation}
For any integers $n, m$, the averaging vanishes if $n+m$ is odd. For the even cases $n+m = 2,\, 4$ and 6, the averaging rules turn out to be
\begin{subequations}
\begin{align}
\label{eqn:rule:2-1}
\langle (\hat{\mathbf{k}} \cdot {\hat{\mathbf{k}}}_{Z^\prime})^2 \rangle = \langle (\hat{\mathbf{l}} \cdot{\hat{\mathbf{k}}}_{Z^\prime})^2 \rangle & \ = \  \frac{1}{3} \,, \\
\langle (\hat{\mathbf{k}} \cdot {\hat{\mathbf{k}}}_{Z^\prime})(\hat{\mathbf{l}} \cdot {\hat{\mathbf{k}}}_{Z^\prime}) \rangle & \ = \  \frac{1}{3} \cos\theta_{kl} \,, \\
\langle (\hat{\mathbf{k}} \cdot {\hat{\mathbf{k}}}_{Z^\prime})^4 \rangle = \langle (\hat{\mathbf{l}} \cdot {\hat{\mathbf{k}}}_{Z^\prime})^4 \rangle & \ = \  \frac{1}{5} \,, \\
\langle (\hat{\mathbf{k}} \cdot {\hat{\mathbf{k}}}_{Z^\prime})^3 (\hat{\mathbf{l}} \cdot {\hat{\mathbf{k}}}_{Z^\prime}) \rangle = \langle (\hat{\mathbf{k}} \cdot {\hat{\mathbf{k}}}_{Z^\prime}) (\hat{\mathbf{l}} \cdot {\hat{\mathbf{k}}}_{Z^\prime})^3 \rangle & \ = \  \frac{1}{5} \cos\theta_{kl} \,, \\
\langle (\hat{\mathbf{k}} \cdot {\hat{\mathbf{k}}}_{Z^\prime})^2 (\hat{\mathbf{l}} \cdot {\hat{\mathbf{k}}}_{Z^\prime})^2 \rangle & \ = \  \frac{1}{15} (2 + \cos^2\theta_{kl}) \,, \\
\langle (\hat{\mathbf{k}} \cdot {\hat{\mathbf{k}}}_{Z^\prime})^6 \rangle = \langle (\hat{\mathbf{l}} \cdot {\hat{\mathbf{k}}}_{Z^\prime})^6 \rangle & \ = \  \frac{1}{7} \,, \\
\langle (\hat{\mathbf{k}} \cdot {\hat{\mathbf{k}}}_{Z^\prime})^5 (\hat{\mathbf{l}} \cdot {\hat{\mathbf{k}}}_{Z^\prime}) \rangle = 
\langle (\hat{\mathbf{k}} \cdot {\hat{\mathbf{k}}}_{Z^\prime}) (\hat{\mathbf{l}} \cdot {\hat{\mathbf{k}}}_{Z^\prime})^5 \rangle & \ = \  \frac{1}{7} \cos\theta_{kl} \,, \\
\langle (\hat{\mathbf{k}} \cdot {\hat{\mathbf{k}}}_{Z^\prime})^4 (\hat{\mathbf{l}} \cdot {\hat{\mathbf{k}}}_{Z^\prime})^2 \rangle = \langle (\hat{\mathbf{k}} \cdot {\hat{\mathbf{k}}}_{Z^\prime})^2 (\hat{\mathbf{l}} \cdot {\hat{\mathbf{k}}}_{Z^\prime})^4 \rangle & \ = \  \frac{1}{35} (2 + 3 \cos^2\theta_{kl}) \,, \\
\langle (\hat{\mathbf{k}} \cdot {\hat{\mathbf{k}}}_{Z^\prime})^3 (\hat{\mathbf{l}} \cdot {\hat{\mathbf{k}}}_{Z^\prime})^3 \rangle & \ = \  \frac{1}{35} (6 \cos\theta_{kl} - \cos^3\theta_{kl}) \,.
\end{align}
\end{subequations}
Here $\theta_{kl}$ denotes the relative angle between the 3-momenta ${\bf k}$ and ${\bf l}$, and the averaging rules related by the interchanging of $\mathbf{k}$ and $\mathbf{l}$ are implicitly implied. 
Applying the averaging rules in Eq.~(\ref{eqn:rule:2-1}), and then substituting the non-relativistic approximation $p_i \cdot k_{Z^\prime} \simeq m_N E_{Z^\prime}$, the terms in Eq.~\eqref{eqn:sim} can be simplified to the non-vanishing term, at the order of 
\begin{equation}
\frac{\mathbf{k}^2 \mathbf{k}^2_{Z^\prime}}{3m_N^2 E_{Z^\prime}^2} \sim {\cal O} ({\bf k}_{}^2/m_N^2) \,,
\end{equation}
rather than vanishing.

As a result of the careful treatment procedure above, in the amplitudes in Eqs.~(\ref{eqn:amp:Mt}) and (\ref{eqn:amp:Mu}), the leading order terms $2(p_i \cdot k_{Z^\prime})$ in the nucleon propagators largely cancel with each other. Then the $m_{Z'}^2$ term in the factor of $\pm 2(p_i \cdot k_{Z^\prime})+m_{Z^\prime}^2$ becomes important, in particular when $Z'$ is heavy (or in other words $m_{Z'} \sim E_{Z'}$). 
To retrieve the corresponding leading-order terms, we need to perform a systematic Taylor expansion of the nucleon propagators in terms of the ratio $m_{Z^\prime}^2 /2 (p_i \cdot k_{Z^\prime})$, in the following way
\begin{align}
\label{eqn:expansion:nucleon}
    \frac{1}{(p_i \pm k_{Z^\prime})^2 - m_N^2} \ = \ \frac{1}{\pm 2(p_i \cdot k_{Z^\prime})+m_{Z^\prime}^2} 
    \simeq\pm \frac{1}{2(p_i\cdot k_{Z^\prime})} \left[ 1\mp\frac{{m_{Z^\prime}^2}}{2(p_i\cdot k_{Z^\prime})}+\frac{{m_{Z^\prime}^4}}{4(p_i\cdot k_{Z^\prime})^2} \right] \,.
\end{align}
For the purpose of this work, it suffices to expand up to the order of $m_{Z'}^4$, and we neglect the higher order terms beyond $m_{Z'}^4$. 
The pion propagators can be categorized into two types based on whether the exchanged momentum carries the momentum $k_{Z'}$. The amplitudes ${\cal M}_{b,\,d,\,b',\,d',\,e',\,g'}$ contain the $Z^\prime$ four-momentum, as seen in Eqs.~(\ref{eqn:amp:Mb}), (\ref{eqn:amp:Md}), (\ref{eqn:amp:Md}), (\ref{eqn:amp:Mdprime}), (\ref{eqn:amp:Meprime}) and (\ref{eqn:amp:Mgprime}). 
For the $u$-channel diagrams $(b')$, $(d')$, $(e')$ and $(g')$, we can expand the pion propagator in the following way, as in Eq~(\ref{eqn:expansion:nucleon}) above:
\begin{eqnarray}
\label{eqn:expansion:pion}
\frac{1}{(l-k_{Z^\prime})^2-m_\pi^2} & \ = \ & -\frac{1}{{\bf l}^2 + m_\pi^2 -2({\bf l}\cdot {\bf k}_{Z'}) -m_{Z^\prime}^2 } \nonumber\\ 
& \ \simeq \ & -\frac{1}{{\bf l}^2 + m_\pi^2} \left( 1 + \frac{2({\bf l}\cdot {\bf k}_{Z'})}{{\bf l}^2+m_\pi^2} + \frac{m_{Z^\prime}^2}{{\bf l}^2+m_\pi^2} \right) \,.
\end{eqnarray}
For the $t$-channel diagrams $(b)$ and $(d)$, one only needs to replace ${\bf l}$ by ${\bf k}$. 
This expansion is superfluous for most terms as the higher-order corrections in Eq.~(\ref{eqn:expansion:pion}) yield negligible contributions. However, the contributions are at the leading order for the interference terms between the amplitudes $(a')$, $(b')$, $(c')$, $(d')$ and $(e')$, $(f')$ and $(g')$ in the longitudinal mode. By contrast, the pure $t$-channel terms, together with the $t$-$u$ interference terms summed over the longitudinal polarization, vanish identically because of the Lorentz structure of the corresponding amplitudes.

For the purpose of phase space integration, it is convenient to define the following dimensionless parameters~\cite{Giannotti:2005tn,Dent:2012mx,Dev:2020eam}: 
\begin{align}
\label{eqn:uvfunctions}
& u  \ \equiv \  \frac{\mathbf{p}_i^2}{m_N T} \,, \quad 
v  \ \equiv \  \frac{\mathbf{p}_f^2}{m_N T} \,, \quad 
y  \ \equiv \  \frac{m_\pi^2}{m_N T} \,, \nonumber \\
& x  \ \equiv \  \frac{E_{Z'}}{T} \,, \quad
q  \ \equiv \  \frac{m_{Z^\prime}}{T} \,, \quad
z  \ \equiv \  \cos \theta_{i f} \,,
\end{align}
where
\begin{equation}
\mathbf{p}_1 \ \equiv \ \mathbf{P}+\mathbf{p}_i \,, \quad 
\mathbf{p}_2 \ \equiv \ \mathbf{P}-\mathbf{p}_i \,, \quad
\mathbf{p}_3 \ \equiv \ \mathbf{P}+\mathbf{p}_f \,, \quad
\mathbf{p}_4 \ \equiv \ \mathbf{P}-\mathbf{p}_f \,.
\end{equation}
In the limit of $|{\bf p}_i| \gg {E}_{Z'}$, ${\bf P}$ is the total 3-momentum of nucleons involved in the initial and final states. In (Eq.~\ref{eqn:uvfunctions}) $\theta_{if}$ is the angle between ${\bf p}_i$ and ${\bf p}_f$. It is then straightforward to obtain the following relations:
\begin{subequations}
\label{eqn:dimensionlesskl}
\begin{align}
\mathbf{k}^2 & \ = \ (u+v-2 z \sqrt{u v}) m_N T  \,, \\
\mathbf{k}^2+m_\pi^2 & \ = \ (u+v+y-2 z \sqrt{u v}) m_N T  \,, \\
\mathbf{l}^2 & \ = \ (u+v+2 z \sqrt{u v}) m_N T  \,, \\
\mathbf{l}^2+m_\pi^2 & \ = \ (u+v+y+2 z \sqrt{u v}) m_N T  \,, \\
\mathbf{k}\cdot\mathbf{l} & \ = \ (u-v)m_N T \,.
\end{align}
\end{subequations}
For the sake of cleanness in the calculations below, we define further the following dimensionless ratios:
\begin{subequations}
\label{eqn:variables:cdef}
\begin{align}
c_k \ \equiv \ & \frac{\mathbf{k}^2}{m_N E_{Z^\prime}} \ = \ \frac{u+v-2 z \sqrt{u v}}{x} \,, \\
c_l \ \equiv \ & \frac{\mathbf{l}^2}{m_N E_{Z^\prime}} \ = \ \frac{u+v+2 z \sqrt{u v}}{x}  \,, \\
c_{kl} \ \equiv \ & \frac{\mathbf{k}\cdot \mathbf{l}}{m_N E_{Z^\prime}} \ = \ \frac{u-v}{x} \,, \\
d_k \ \equiv \ & \frac{m_N E_{Z^\prime}}{\mathbf{k}^2+m_{\pi}^2} \ = \ \frac{x}{u+v+y-2z\sqrt{uv}}  \,, \\
d_l \ \equiv \ & \frac{m_N E_{Z^\prime}}{\mathbf{l}^2+m_{\pi}^2} \ = \ \frac{x}{u+v+y+2z\sqrt{uv}}  \,, \\
e_k \ \equiv \ & \frac{{\bf k}^2}{{\bf k}^2+m_{\pi}^2} \ = \ \frac{u+v-2z\sqrt{uv}}{u+v+y-2z\sqrt{uv}} \,, \\
e_l \ \equiv \ & \frac{{\bf l}^2}{{\bf l}^2+m_{\pi}^2} \ = \ \frac{u+v-2z\sqrt{uv}}{u+v+y+2z\sqrt{uv}} \,, \\
f_{k} \ \equiv \ & \frac{{\bf k \cdot l}}{{\bf k}^2+m_{\pi}^2} \ = \ \frac{u-v}{u+v+y-2z\sqrt{uv}} \,, \\
f_{l} \ \equiv \ & \frac{{\bf k \cdot l}}{{\bf l}^2+m_{\pi}^2} \ = \ \frac{u-v}{u+v+y+2z\sqrt{uv}} \,. 
\end{align}
\end{subequations}

\subsection{$g_0^2$ terms}
\label{appsub:g0}

Using the polarization sums in a relativistic medium~\cite{Chang:2016ntp}, 
\begin{equation}
\label{eq.polarsum}
P_{\mu \nu}^{T,L} \ \equiv \ \sum_{T,L} \varepsilon_\mu (k_{Z'}) \varepsilon_\nu (k_{Z'}) \ = \
\begin{cases}
\left(1-\delta_{\mu 0}\right)\left(1-\delta_{\nu 0}\right)\left(\delta_{ij} - \dfrac{{({\bf k}_{Z^\prime})}_i {({\bf k}_{Z^\prime})}_j}{\mathbf{k}_{Z^\prime}^2}\right) \,, & (T) \,, \\
-g_{\mu\nu} + \dfrac{{(k_{Z^\prime})}_\mu {(k_{Z^\prime})}_\nu}{m_{Z^\prime}^2} - P_{\mu\nu}^T \,, & (L) \,,
\end{cases}
\end{equation}
the amplitudes squared in the $T$ and $L$ can be expressed in the form of Eq.~(\ref{eqn:ampsq:np}). In practice, the easiest way to calculate the longitudinal mode terms is the following:
\begin{align}
\label{eqn:squaredamplitudeforLmode}
\sum_{\rm spins}|\mathcal{M}|_{L}^2\ = \ &\sum_{\rm spins}|\mathcal{M}|_{\mathrm{tot}}^2-\sum_{\rm spins}|\mathcal{M}|_{T}^2 \,,
\end{align}
with the total amplitude squared written in terms of the couplings $g_{0}$ and $g_{1}$: 
\begin{align}
\sum_{\rm spins}|\mathcal{M}|_{\mathrm{tot}}^2 \ \simeq \ g_{0}^2\left(\frac{\mathcal{T}^{(\alpha,L)}}{m_{Z^\prime}^2}+\mathcal{T}^{(0,\mathrm{tot})}+\mathcal{U}^{(0,\mathrm{tot})}\right)+g_{1}^2\left(\mathcal{T}^{(1,\mathrm{tot})}+\mathcal{U}^{(1,\mathrm{tot})}\right) \,.
\end{align}

The ${\cal T}^{(\alpha,L)}$ term can be decomposed into the following form:
\begin{align}
\label{eqn:T:alphaL}
{\cal T}^{(\alpha,L)} \ = \ & \left(\frac{2m_N}{m_\pi}\right)^{4} \biggl\{
    \frac{16f_{pn}^4}{{\bf l}^2 + m_\pi^2}\biggl[e_l^2\mathcal{T}^{(\alpha,L)}_u
+e_l^3\mathcal{T}^{(\alpha,L)}_{e^\prime}
+2e_l\mathcal{T}^{(\alpha,L)}_{f^\prime g^\prime}\nonumber \\&
+2e_l^2\mathcal{T}^{(\alpha,L)}_{e^\prime (f^\prime g^\prime)}
+e_l^2\mathcal{T}^{(\alpha,L)}_{u e^\prime}
+e_l\mathcal{T}^{(\alpha,L)}_{u (f^\prime g^\prime)}\biggl]
\biggl\}\left({E_{Z^\prime}^2-m_{Z^\prime}^2}\right) \,,
\end{align}
where the subscript $u$ is for the four $u$-channel diagrams in the middle panel of Fig.~\ref{fig:diagrams}, and $e',\,f',\,g'$ are for the diagrams $(e'),\,(f'),\,(g')$ in the right panel of Fig.~\ref{fig:diagrams}, respectively. The terms with two subscripts are for the corresponding interference contributions. The dimensionless coefficients in Eq.~(\ref{eqn:T:alphaL}) are, respectively,
\begin{subequations}
\begin{align}
\mathcal{T}^{(\alpha,L)}_u & \ = \ 
\mathcal{T}^{(\alpha,L)}_{e^\prime} \ = \ \mathcal{T}^{(\alpha,L)}_{f^\prime g^\prime} \ = \ -\mathcal{T}^{(\alpha,L)}_{e^\prime (f^\prime g^\prime)} \ = \ \frac{1}{3}\, , \\
\mathcal{T}^{(\alpha,L)}_{u e^\prime} & \ = \ 
- \mathcal{T}^{(\alpha,L)}_{u (f^\prime g^\prime)} \ = \  -\frac{2}{3} \,.
\end{align}
\end{subequations}
The decomposition of the term ${\cal T}^{(0,{\rm tot})}$ is:
\begin{align}
\mathcal{T}^{(0,\mathrm{tot})} 
\ = \ & \left(\frac{2m_N}{m_\pi}\right)^{4} \biggl\{\left.
     \frac{f_{nn}^2 f_{pp}^2 {\bf k}^2}{({\bf k}^2 + m_\pi^2)^2} \mathcal{T}^{(0,\mathrm{tot})}_t 
    + \frac{f_{pn}^4 {\bf l}^2}{({\bf l}^2 + m_\pi^2)^2} \mathcal{T}^{(0,\mathrm{tot})}_u
    + \frac{f_{pp} f_{nn} f_{pn}^2 {\bf k}^2}{({\bf k}^2 + m_\pi^2)({\bf l}^2 + m_\pi^2)} \mathcal{T}^{(0,\mathrm{tot})}_{tu} \right.
     \nonumber \\
    & + \frac{16 f_{pn}^4}{{\bf l}^2 + m_\pi^2}\left[ e_l^3 \mathcal{T}^{(0,\mathrm{tot})}_{e^\prime}+e_l\mathcal{T}^{(0,\mathrm{tot})}_{f^\prime g^\prime}+e_l^2\mathcal{T}^{(0,\mathrm{tot})}_{e^\prime (f^\prime g^\prime)}+e_l^2\mathcal{T}^{(0,\mathrm{tot})}_{u e^\prime}+e_l\mathcal{T}^{(0,\mathrm{tot})}_{u (f^\prime g^\prime)}\right]
    \nonumber \\
&+{8 f_{nn} f_{pn}^2 f_{pp}}\left[\frac{\mathbf{k \cdot l}}{\left(\mathbf{k}^2+m_{\pi}^2\right) \left(\mathbf{l}^2+m_{\pi}^2\right)}\left(\mathcal{T}^{(0,\mathrm{tot})}_{t e^\prime}
+\mathcal{T}^{(0,\mathrm{tot})}_{t (f^\prime g^\prime)}\right)\right]\biggl\} \,,
\end{align}
with the subscript $t$ referring to the four $t$-channel diagrams in the left panel of Fig.~\ref{fig:diagrams}. The corresponding dimensionless coefficients are, respectively,
\begin{subequations}
\begin{align}
\mathcal{T}^{(0,\mathrm{tot})}_{t} \ = \ & \frac{8}{3} c_k^2 \left(1+\frac{q^2}{2x^2}\right) \,, \\ 
\mathcal{T}^{(0,\mathrm{tot})}_{u} \ = \ & \frac83
\left( 8 - 4c_{kl} +  c_{kl}^2 \right) \left(1+\frac{q^2}{2x^2}\right) \,, \\
\mathcal{T}^{(0,\mathrm{tot})}_{tu} \ = \ & \frac83 
\left( 2c_{kl}+ c_{k}c_{l}-2c_{kl}^2  \right)
\left(1+\frac{q^2}{2x^2}\right) \,, \\
\mathcal{T}^{(0,\mathrm{tot})}_{e^\prime} \ = \ &\frac{1}{2}\mathcal{T}^{(0,\mathrm{tot})}_{f^\prime g^\prime} \ = \ -\frac{1}{2}\mathcal{T}^{(0,\mathrm{tot})}_{e^\prime (f^\prime g^\prime)}\ = \ 1 \,,\\
\mathcal{T}^{(0,\mathrm{tot})}_{t e^\prime} \ = \ &\left(c_{k}c_{l}-2c_{kl}^2\right)d_l\,,\\
\mathcal{T}^{(0,\mathrm{tot})}_{t (f^\prime g^\prime)} \ = \ & c_k\,,\\
\label{eqn:T0tot:ue}
 \mathcal{T}^{(0,\mathrm{tot})}_{u e^\prime} \ = \ & -2 +c_{kl}
 \,, \\
\label{eqn:T0tot:ufg}
\mathcal{T}^{(0,\mathrm{tot})}_{u (f^\prime g^\prime)} \ = \ & 4 - c_{kl}
\,.
\end{align}
\end{subequations}
The $g_0^2$ terms ${\cal U}^{(0,{\rm tot})}$ is shorter than ${\cal T}^{(0,{\rm tot})}$: 
\begin{align}
    {\cal U}^{(0,{\rm tot})} \ = \ & \frac{8}{3}\left(\frac{2m_N}{m\pi}\right)^4\biggl\{\frac{f_{pn}^4}{\mathbf{l}^2+m_\pi^2}\left[e_l^2\mathcal{U}^{(0,\mathrm{tot})}_{u1}+e_l^3\mathcal{U}^{(0,\mathrm{tot})}_{u2}\right]\nonumber \\ &
    +\frac{f_{nn} f_{pp} f_{pn}^2 \left(\mathbf{k}\cdot \mathbf{l}\right)}{\left(\mathbf{k}^2+m_\pi^2\right)\left(\mathbf{l}^2+m_\pi^2\right)}\mathcal{U}^{(0,\mathrm{tot})}_{tu}\biggl\} \left( 1 - \frac{m_{Z^\prime}^2}{E_{Z^\prime}^2} \right) \,,
\end{align}
where the corresponding dimensionless coefficients are, respectively,
\begin{subequations}
\begin{align}
\mathcal{U}^{(0,\mathrm{tot})}_{u1} \ = \ & 2\left(2-c_{kl}\right) \, ,\\
\mathcal{U}^{(0,\mathrm{tot})}_{u2} \ = \ &-2 \, ,\\
\mathcal{U}^{(0,\mathrm{tot})}_{tu} \ = \ & d_l\left(c_{kl}^2-c_k c_l\right) \, .
\end{align}
\end{subequations}
Let us now move on to the transverse terms. The term ${\cal T}^{(0,T)}$ can be decomposed as:
\begin{align}
    \mathcal{T}^{(0,T)} \ = \ & \frac43 \left(\frac{2m_N}{m_\pi}\right)^{4} \left\{ \frac{{f^2_{nn}}{f^2_{pp}}\mathbf{k}^2}{\left(\mathbf{k}^2+m_\pi^2\right)^2}\mathcal{T}^{(0,T)}_t+ \frac{f^4_{pn} \mathbf{l}^2}{\left(\mathbf{l}^2+m_\pi^2\right)^2}\mathcal{T}^{(0,T)}_u+\frac{{f_{nn}f_{pp}f^2_{pn} \mathbf{k}^2}}{{\left(\mathbf{k}^2+m_\pi^2\right)\left(\mathbf{l}^2+m_\pi^2\right)}}\mathcal{T}^{(0,T)}_{tu} \nonumber \right.
    \\&+\frac{8 f_{pn}^4}{\mathbf{l}^2+m_\pi^2}\left[e_l^3\mathcal{T}^{(0,T)}_{e^\prime}+e_l\mathcal{T}^{(0,T)}_{f^\prime g^\prime}  +e_l^2\mathcal{T}^{(0,T)}_{e^\prime (f^\prime g^\prime)}+e_l^2\mathcal{T}^{(0,T)}_{u e^\prime}+e_l\mathcal{T}^{(0,T)}_{u (f^\prime g^\prime)}\right]\nonumber \\
    &\left.+4{ f_{nn} f_{pn}^2 f_{pp}}\left[\frac{\mathbf{k \cdot l}}{ \left(\mathbf{k}^2+m_{\pi}^2\right)\left(\mathbf{l}^2+m_{\pi}^2\right)}\left(\mathcal{T}^{(0,T)}_{t e^\prime}+\mathcal{T}^{(0,T)}_{t (f^\prime g^\prime)}\right)\right]\right\} \,.
\end{align}
The corresponding dimensionless coefficients are, respectively,
\begin{subequations}
\begin{align}
\mathcal{T}^{(0,T)}_{t} \ = \ & c_k^2 \,, \\
    \mathcal{T}^{(0,T)}_{u} \ = \ &14-4c_{kl}+c_{k}c_{l} + \frac{q^2}{x^2} 
    \left(2-4c_{kl}+c_{k}c_{l} \right)  \,, \\
    \mathcal{T}^{(0,T)}_{tu} \ = \ &2c_{kl}+c_{k}c_{l}-2c_{kl}^2 \,, \\
    \mathcal{T}^{(0,T)}_{e^\prime} \ = \ &\frac{1}{2}\mathcal{T}^{(0,T)}_{f^\prime g^\prime} \ = \ -\frac{1}{2}\mathcal{T}^{(0,T)}_{e^\prime (f^\prime g^\prime)} \ = \  1 \,,\\
    \mathcal{T}^{(0,T)}_{t e^\prime} \ = \ & \left(c_k c_l- 2 c_{kl}^2\right)d_l \,, \\
     \mathcal{T}^{(0,T)}_{t (f^\prime g^\prime)} \ = \ & c_k \,, \\
    \mathcal{T}^{(0,T)}_{u e^\prime} \ = \ & -2+c_{kl} \,, \\
    \mathcal{T}^{(0,T)}_{u (f^\prime g^\prime)} \ = \ &-4+c_{kl} \,.
\end{align}
\end{subequations} 
The ${\cal U}^{(0,T)}$ term is comparatively much simpler:
\begin{align}
    {\cal U}^{(0,T)} \ = \ -\frac{8}{3}\left(\frac{2m_N}{m_\pi}\right)^4 \frac{f_{pn}^4 \mathbf{l}^2}{\left(\mathbf{l}^2+m_\pi^2\right)^2}
    \left( 1 - \frac{m_{Z^\prime}^2}{E_{Z^\prime}^2} \right) d_l\left(c_{l} + c_{kl} \right) \,.
\end{align}

\subsection{$g_1^2$ terms}
\label{appsub:g1}

For the $g_1^2$ terms, only the $t$-channel amplitudes ${\cal M}_{a,b,c,d}$, the $u$-channel amplitudes ${\cal M}_{a^\prime,b^\prime,c^\prime,d^\prime}$ and the cross terms between them contribute, which can be written as
\begin{align}
\mathcal{T}^{(1,\mathrm{tot})}\ = \ & \left( \frac{2 m_N}{m_\pi} \right)^{4} 
\left[ \frac{f_{nn}^2f_{pp}^2}{ \left(\mathbf{k}^2+m_\pi^2\right)^2}\mathcal{T}^{(1,\mathrm{tot})}_t+\frac{f_{pn}^4}{ \left(\mathbf{l}^2+m_\pi^2\right)^2}\mathcal{T}^{(1,\mathrm{tot})}_u \right.\nonumber \\
& \left.+\frac{f_{nn}f_{pn}^2f_{pp}}{ \left(\mathbf{k}^2+m_\pi^2\right) \left(\mathbf{l}^2+m_\pi^2\right)}\mathcal{T}^{(1,\mathrm{tot})}_{tu}\right] \,.
\end{align}
The corresponding dimensionless coefficients are decomposed in terms of the gauge boson energy $E_{Z'}$ and mass $m_{Z'}$ via
\begin{align}
    \mathcal{T}_{i}^{(1,\mathrm{tot})} \ = \ & E_{Z^\prime}^2\mathcal{C}_{i1}^{(1,\mathrm{tot})}+m_{Z^\prime}^2 \left[ \mathcal{C}_{i2}^{(1,\mathrm{tot})}+\frac{m_{Z^\prime}^2}{E_{Z^\prime}^2}\mathcal{C}_{i3}^{(1,\mathrm{tot})} \right] \,, 
\end{align}
where the index $i = t,\,u,\,tu$, 
and the dimensionless coefficients are, respectively, 
\begin{subequations}
\begin{align}
    \mathcal{C}_{t1}^{(1,\mathrm{tot})} \ = \ & \mathcal{C}_{u1}^{(1,\mathrm{tot})} \ = \  \frac{1}{15} c_k^2 \Big[ 16 + 8\left(12c_{kl} + c_{k}c_{l}\right)- 7c_{kl}^2 \Big] \,, \\
    \mathcal{C}_{t2}^{(1,\mathrm{tot})} \ = \ &\frac{4}{15} c_k \Big[ 67c_k - 10\left(c_l-3c_{kl}\right) - 98c_kc_{kl} + 6c_k^2\left( c_kc_l + 3c_l^2\right) \big] \,, \\
    \mathcal{C}_{t3}^{(1,\mathrm{tot})} \ = \ &\frac{8}{15} c_k \left[ -13c_k + 5\left(c_l - 3c_{kl}\right) + 22c_k c_{kl} - 2c_k\left(c_kc_l + 2 c_{kl}^2\right) \right] \,, \\
    \mathcal{C}_{u2}^{(1,\mathrm{tot})} \ = \ & \frac{4}{15} c_l \Big[ -5\left(c_k+2c_{kl}\right) + 67c_l - 58c_{kl}c_l + 6c_l\left(c_k c_l + 3c_{kl}^2 \right)\Big] \,, \\
    \mathcal{C}_{u3}^{(1,\mathrm{tot})} \ = \ &\frac{4}{15} c_l \Big[ 5\left(c_k+ 2c_{kl}\right) - 26c_l  + 64c_{kl}c_l - 4 c_l\left(2c_k c_l + c_{kl}^2 \right)\Big] \,, \\
    \mathcal{C}_{tu1}^{(1,\mathrm{tot})} \ = \ &\frac{8}{15}\Big[ -10c_{k} c_l + 4\left(4c_{k}c_{l}c_{kl} - 7c_{kl}^3 \right) 
    + c_k^2 c_l^2 - 9c_{kl}^2c_{k} c_{l}  + 14 c_{kl}^4 \Big] \,, \\
    \mathcal{C}_{tu2}^{(1,\mathrm{tot})} \ = \ &\frac{4}{15}\Big[ 60c_k - 10c_k\left( c_{kl} - 6 c_l\right) - 105c_{kl}^2 + 4c_{kl}\left(58c_{kl}^2 - 31c_k c_l \right) \Big. \nonumber \\ 
    &\Big. + 6\left(c_k^2 c_l^2 + c_{kl}^2c_k c_l - 6c_{kl}^4 \right)\Big] \,, \\
    \mathcal{C}_{tu3}^{(1,\mathrm{tot})} \ = \ &\frac{4}{15}\Big[ -10c_k + 5\left(2c_k c_{kl} + c_k c_l + 3c_{kl}^2\right) +8c_{kl}\left( 4c_k c_l - 7c_{kl}^2\right)\Big. \nonumber \\ 
    &\Big. - 4\left(2c_k^2 c_l^2 - 3c_k c_lc_{kl}^2 - 2c_{kl}^4 \right)\Big] \,.
\end{align}
\end{subequations}
The decomposition of ${\cal U}^{(1,{\rm tot})}$ is based on the dimensionless parameters $e_{k,\,l}$ and $f_{k,\,l}$, via
\begin{align}
    {\cal U}^{(1,{\rm tot})} \ = \ & \left(\frac{2m_N}{m_\pi}\right)^4 \biggl\{  \frac{f_{pp}^2f_{nn}^2}{m_N^2E_{Z^\prime}^2} \left[  e_k^3\mathcal{U}_{t1}^{(1,\mathrm{tot})}
    +e_k^4\mathcal{U}_{t2}^{(1,\mathrm{tot})}  \right] \nonumber \\&
    +\frac{f_{pn}^4}{m_N^2E_{Z^\prime}^2}\left[ e_l^3\mathcal{U}_{u1}^{(1,\mathrm{tot})}+
    e_l^4\mathcal{U}_{u2}^{(1,\mathrm{tot})}\right]\nonumber \\&
    + \frac{f_{pp}f_{nn}f_{pn}^2}{\left(\mathbf{k}^2+m_\pi^2\right)\left(\mathbf{l}^2+m_\pi^2\right)}\left[e_k\mathcal{U}_{tu1}^{(1,\mathrm{tot})}+f_k\mathcal{U}_{tu2}^{(1,\mathrm{tot})}\right.\nonumber \\
    & \left. +e_l\mathcal{U}_{tu1}^{(1,\mathrm{tot})}+e_l\mathcal{U}_{tu2}^{(1,\mathrm{tot})}+{\left(2f_k f_l-e_k e_l\right)}\mathcal{U}_{tu3}^{(1,\mathrm{tot})}\right]
    \biggl\} \,.
\end{align}
Then the terms ${\cal U}_{i}^{(1,\mathrm{tot})}$ (here $i = t1,\,t2, u1,\,u2,\,tu1,\,tu2,\,tu3$) can be further decomposed in terms of $E_{Z'}$ and $m_{Z'}$ as
\begin{align}
\mathcal{U}_{i}^{(1,\mathrm{tot})} \ = \ & E_{Z^\prime}^2\mathcal{D}_{i1}^{(1,\mathrm{tot})}+m_{Z^\prime}^2\left[\mathcal{D}_{i2}^{(1,\mathrm{tot})}+\frac{m_{Z^\prime}^2}{E_{Z^\prime}^2}\mathcal{D}_{i3}^{(1,\mathrm{tot})}\right] \,,
\end{align}
and the corresponding dimensionless coefficients are, respectively,
\begin{subequations}
\begin{align}
\mathcal{D}_{t11}^{(1,\mathrm{tot})} \ = \ & \mathcal{D}_{u11}^{(1,\mathrm{tot})} \ = \ \frac{16}{15}\left(1-c_{kl}\right) \, ,\\
\mathcal{D}_{t12}^{(1,\mathrm{tot})} \ = \ & \mathcal{D}_{u12}^{(1,\mathrm{tot})} \ = \ \frac{16}{15}\left(-12+7c_{kl}\right) \,, \\
\mathcal{D}_{t13}^{(1,\mathrm{tot})} \ = \ & \mathcal{D}_{u13}^{(1,\mathrm{tot})} \ = \ \frac{16}{15}\left(11-6c_{kl}\right) \,, \\
\mathcal{D}_{t21}^{(1,\mathrm{tot})} \ = \ & 
2\mathcal{D}_{t22}^{(1,\mathrm{tot})} \ = \ 
-\frac23 \mathcal{D}_{t23}^{(1,\mathrm{tot})} \ = \ 
\mathcal{D}_{u21}^{(1,\mathrm{tot})} \ = \ 
 2\mathcal{D}_{u22}^{(1,\mathrm{tot})} \ = \
-\frac23 \mathcal{D}_{u23}^{(1,\mathrm{tot})} \ = \
\frac{32}{15} \, ,\\
\mathcal{D}_{tu11}^{(1,\mathrm{tot})} \ = \ & \frac{8}{15}\big[9c_k c_l-10c_{kl}^2-c_{kl}\left(c_k c_l -2c_{kl}^2\right)\big] \,, \\
\mathcal{D}_{tu12}^{(1,\mathrm{tot})} \ = \ & \frac{8}{15}\big[-8c_k c_l+10 c_{kl}^2+7c_{kl}\left(c_k c_l -2c_{kl}^2\right)\big]\, ,\\
\mathcal{D}_{tu13}^{(1,\mathrm{tot})} \ = \ & -\frac{8}{15}\big[c_k c_l+6c_{kl}\left(c_k c_l c_{kl}-2c_{kl}^2\right)\big]\, ,\\
{\cal D}_{tu21}^{(1,\mathrm{tot})} \ = \ & 0 \,, \\
\mathcal{D}_{tu22}^{(1,\mathrm{tot})} \ = \ & -\mathcal{D}_{tu23}^{(1,\mathrm{tot})} \ = \ \frac{16}{3}\left(2c_{kl}^2-c_k c_l\right)\, ,\\
\mathcal{D}_{tu31}^{(1,\mathrm{tot})} \ = \ & \frac{32}{15}\left(-2c_{kl}^2+c_k c_l\right)\, ,\\
\mathcal{D}_{tu32}^{(1,\mathrm{tot})} \ = \ & \frac{16}{15}\left(3c_{kl}^2-4c_k c_l\right)\, ,\\
\mathcal{D}_{tu33}^{(1,\mathrm{tot})} \ = \ & \frac{16}{15}\left(c_{kl}^2+2c_k c_l\right)\, .
\end{align}
\end{subequations}

The decomposition of the $\mathcal{T}^{(1,{T})}$ term for the transverse mode is quite similar to that $\mathcal{T}^{(1,\mathrm{tot})}$, written as 
\begin{align}
\mathcal{T}^{(1,{T})}\ = \ &\left(\frac{2 m_N}{m_\pi }\right)^4\left[\frac{f_{nn}^2f_{pp}^2 }{ \left(\mathbf{k}^2+m_\pi^2\right)^2}\mathcal{T}^{(1,{T})}_t+\frac{f_{pn}^4}{ \left(\mathbf{l}^2+m_\pi^2\right)^2}\mathcal{T}^{(1,{T})}_u\right.\nonumber \\&\left.+\frac{f_{nn}f_{pn}^2f_{pp}}{ \left(\mathbf{k}^2+m_\pi^2\right) \left(\mathbf{l}^2+m_\pi^2\right)}\mathcal{T}^{(1,{T})}_{tu}\right] \,,
\end{align}
where the coefficients
\begin{align}
    \mathcal{T}_{i}^{(1,T)} \ = \   E_{Z^\prime}^2\mathcal{C}_{i1}^{(1,T)} +m_{Z^\prime}^2 \left[ \mathcal{C}^{(1,T)}_{i2} 
    +\frac{m_{Z^\prime}^2}{E_{Z^\prime}^2}\mathcal{C}^{(1,T)}_{i3}
    +\frac{m_{Z^\prime}^2}{\mathbf{k}_{Z^\prime}^2}\mathcal{C}_{i4}^{(1,T)} \right] \,,
\end{align}
with here $i = t,\,u,\,tu$, and the $m_{Z'}^2/{\bf k}_{Z'}^2$ terms coming from the $(k_{Z'})_i(k_{Z'})_j/\mathbf{k}_{Z'}^2$
in the transverse polarization sum given in Eq.~\eqref{eq.polarsum}. The corresponding dimensionless coefficients are, respectively,
\begin{subequations}
\begin{align}
    \mathcal{C}_{t1}^{(1,T)} \ = \ & \mathcal{C}_{u1}^{(1,T)} \ = \  \frac{4}{15} c_k^2 \Big[ 13 + 6 \left(8c_{kl} - c_k c_l\right) - 18 c_{kl}^2\Big] \,, \\
    \mathcal{C}_{t2}^{(1,T)} \ = \ & \frac{4}{15} c_k^2 \Big[ 53 - 102 c_{kl} + 2\left(7 c_k c_l + 11 c_{kl}^2\right) \Big] \,, \\
    \mathcal{C}_{t3}^{(1,T)} \ = \ & \frac{8}{15} c_k \Big[-13 c_k + 5 c_l - 15c_{kl} + 2c_k\left( 11 c_{kl} - 2 c_k c_l - c_{kl}^2\right)\Big] \,, \\
    \mathcal{C}_{t4}^{(1,T)} \ = \ & \mathcal{C}_{u4}^{(1,T)} \ = \ 0 \,, \\
    \mathcal{C}_{u2}^{(1,T)} \ = \ & \frac{4}{3} c_l^2 \left( 17 - 20 c_{kl} \right) \,, \\
    \mathcal{C}_{u3}^{(1,T)} \ = \ & \frac{8}{3} c_l^2 \left( 1 + 4 c_{kl} \right) \,, \\
    \mathcal{C}_{tu1}^{(1,T)} \ = \ & \frac{4}{15}\Big[-5\left( c_k c_l +6 c_{kl}^2\right) + 24 c_{kl}\left( c_k c_l - 3 c_{kl}^2 \right) - 6 \left(c_k^2 c_l^2 +c_k c_lc_{kl}^2  -6 c_{kl}^4\right) \Big] \,, \\
    \mathcal{C}_{tu2}^{(1,T)} \ = \ & \frac{8}{15}\Big[ -5\left( c_k + c_l -2 c_{kl}\right) -5 \left(3 c_k c_l - c_{kl}\left(4 c_k +4 c_l  -9 c_{kl}\right)\right)\Big. \nonumber \\
    &\Big. -c_{kl} \left(23 c_k c_l  - 74 c_{kl}^2 \right)+ 7 c_k^2 c_l^2 - 3 c_k c_lc_{kl}^2  - 22 c_{kl}^4 \Big] \,, \\
    \mathcal{C}_{tu3}^{(1,T)} \ = \ & \frac{2}{15}\Big[5\left( 2 c_k + 5 c_l - 4 c_{kl} \right)+ 5\left(35 c_k c_l - 4c_{kl}\left( c_k  + c_l  +11 c_{kl}\right)\right) \Big. \nonumber \\
    & \Big. + 8  c_{kl} \left(3 c_k c_l - 14 c_{kl}^2\right) - 8\left(2 c_k^2 c_l^2 - 3 c_k c_l c_{kl}^2 - 2 c_{kl}^4\right)\Big] \,, \\
    \mathcal{C}_{tu4}^{(1,T)} \ = \ & 2 \left( c_l+4 c_{kl} - 2 c_k c_l + 4 c_{kl}^2 \right) \,.
\end{align}
\end{subequations}
The decomposition of the term ${\cal U}^{(1,T)}$ is similar:
\begin{align}
    {\cal U}^{(1,{T})} \ = \ & \left(\frac{2m_N}{m_\pi}\right)^4 \biggl\{  \frac{f_{pp}^2f_{nn}^2}{m_N^2E_{Z^\prime}^2} \left[  e_k^3\mathcal{U}_{t1}^{(1,T)}
    +e_k^4\mathcal{U}_{t2}^{(1,T)}  \right] \nonumber \\&
    +\frac{f_{pn}^4}{m_N^2E_{Z^\prime}^2}\left[ e_l d_l^2\mathcal{U}_{u1}^{(1,\mathrm{tot})}+
    e_l^4\mathcal{U}_{u2}^{(1,\mathrm{tot})}\right]\nonumber \\&
    + \frac{f_{pp}f_{nn}f_{pn}^2}{\left(\mathbf{k}^2+m_\pi^2\right)\left(\mathbf{l}^2+m_\pi^2\right)}\left[e_k\mathcal{U}_{tu1}^{(1,\mathrm{tot})}+d_l\mathcal{U}_{tu2}^{(1,\mathrm{tot})} +{\left(2f_k f_l-e_k e_l\right)}\mathcal{U}_{tu3}^{(1,\mathrm{tot})}\right]
    \biggl\} \,,
\end{align}
where
\begin{align}
\mathcal{U}_{i}^{(1,T)}\ =& \ E_{Z^\prime}^2\mathcal{D}_{i1}^{(1,T)}+m_{Z^\prime}^2\left[\mathcal{D}_{i2}^{(1,T)}+\frac{m_{Z^\prime}^2}{E_{Z^\prime}^2}\mathcal{D}_{i3}^{(1,T)}\right] \,,
\end{align}
with the index $i = t1,\, t2,\, u1,\, u2,\, tu1,\, tu2,\, tu3$.
The corresponding dimensionless coefficients are, respectively, 
\begin{subequations}
\begin{align}
\mathcal{D}_{t11}^{(1,T)} \ = \ &\frac{16}{15}\left(1-c_{kl}\right) \, ,\\
\mathcal{D}_{t12}^{(1,T)} \ = \ &\frac{16}{15}\left(-11+c_{kl}\right) \, ,\\
\mathcal{D}_{t13}^{(1,T)} \ = \ & 16 \,, \\
\mathcal{D}_{t21}^{(1,T)} \ = \ & 
- \mathcal{D}_{t22}^{(1,T)} =
4 \mathcal{D}_{u21}^{(1,T)} = 
-4 \mathcal{D}_{u22}^{(1,T)} =
\frac{32}{15} \,, \\
{\cal D}_{t23}^{(1,T)} \ = \ & 
{\cal D}_{u23}^{(1,T)} \ = \ {\cal D}_{tu13}^{(1,T)} \ = \ 
{\cal D}_{tu33}^{(1,T)} \ = \ 0 \,, \\
\mathcal{D}_{u11}^{(1,T)}\ = \ &\frac{16}{15}\left(c_l^2-c_l^2 c_{kl}\right) \, ,\\
\mathcal{D}_{u12}^{(1,T)}\ = \ &\frac{16}{15}\left(15c_{kl}+10c_l-6c_l^2+c_l^2 c_{kl}\right) \, ,\\
\mathcal{D}_{u13}^{(1,T)}\ = \ &\frac{8}{3}\left(-3c_{kl}-2c_l+4c_l^2\right) \, ,\\
\mathcal{D}_{tu11}^{(1,T)}\ = \ &\mathcal{D}_{tu21}^{(1,T)}/{c_l}\ = \ \frac{8}{15}\big[-10c_{kl}^2+9c_k c_l+c_{kl}\left(2c_{kl}^2-c_k c_l \right)\big]\,, \\
\mathcal{D}_{tu12}^{(1,T)}\ = \ & \frac{8}{15}\big[-4c_k c_l-c_{kl}\left(2c_{kl}^2-c_k c_l \right)\big]\, ,\\
\mathcal{D}_{tu22}^{(1,T)}\ = \ &\frac{4}{15}\big[-5c_l\left(4c_{kl}+c_l\right)+2c_l\left(5c_{kl}^2+11c_k c_l\right)+2c_l^2c_{kl}\left(-2c_{kl}^2+c_k c_l\right)\big]\, ,\\
\mathcal{D}_{tu23}^{(1,T)}\ = \ &\frac{4}{3}\big[c_l\left( c_{kl}+c_l\right)-4c_k c_l^2\big] \, ,\\
\mathcal{D}_{tu31}^{(1,\mathrm{tot})} \ = \ & 
\mathcal{D}_{tu32}^{(1,\mathrm{tot})} \ = \
\frac{32}{15}\left(-2c_{kl}^2+c_k c_l\right) \,.
\end{align}
\end{subequations}

\section{Semi-Compton process for the $U(1)_{L_\mu-L_\tau}$ gauge boson}
\label{app:semi-Compton}

The production rate of $Z'$ in  the $U(1)_{L_\mu - L_\tau}$ model in the semi-Compton process $\gamma(k) + \mu^-(p_1) \rightarrow \mu^-(p_2) + Z'(k_{Z^\prime})$ can  be calculated by embedding the exact squared matrix element into a properly modified framework of the standard Compton scattering process. In the limit of $E_{Z'} \ll m_\mu$, the rate can be estimated by simply rescaling the standard Compton scattering $\gamma \mu \to \mu \gamma$~\cite{Lai:2024mse,Croon:2020lrf}. 
To provide a more rigorous treatment while maintaining analytical clarity, we operate within the same low-energy limit of $E_{Z^\prime} \ll m_\mu$, but explicitly recalculate the exact amplitude squared by preserving the finite mass $m_{Z'}$ in the final state. The matrix elements in the $s$- and $u$-channels are, respectively, 
\begin{subequations}
	\label{eqn:amplitudesa}
    \begin{align}
	{\cal M}_s \ = \ &
	\frac{e g_{\mu\tau}}{(p_2 + k_{Z^\prime})^2 - m_\mu^2}
	\Big[ \bar{u} (p_2) \gamma_\mu \left( \slashed{p}_2 + \slashed{k}_{Z'} + m_\mu \right) \gamma_\nu u(p_1) \Big] \varepsilon^{\mu\ast} (k_{Z^\prime})\varepsilon^\nu (k) \,,  \\
 \label{eqn:amplitudesb}
	{\cal M}_u \ = \ &
	\frac{e g_{\mu\tau}}{(p_1 - k_{Z^\prime})^2 - m_\mu^2}
	\Big[ \bar{u} (p_2) \gamma_\mu \left( \slashed{p}_1 - \slashed{k}_{Z'} + m _\mu\right) \gamma_\nu u(p_1) \Big]  \varepsilon^{\mu\ast} (k_{Z^\prime})\varepsilon^\nu (k) \,. 
    \end{align}
 \end{subequations} 
Using the polarization sum in medium in Eq.~(\ref{eq.polarsum}) and applying the trace technology, we can get the amplitude squared  for the transverse modes  
\begin{eqnarray}
    \sum_\mathrm{spins}|{\cal M} |^2_T & \ = \ & e^2 g^2_{\mu\tau} \left[ \frac{8m_\mu^2 (\mathbf{p_2}\cdot {\mathbf{k}_{Z^\prime}})^2}{{\mathbf{k}^2_{Z^\prime}}~( p_1\cdot k)^2}-\frac{8m_\mu^2 (\mathbf{p_2})^2}{({p_{1}}\cdot k)^2}+\frac{8(p_1\cdot k_{Z^\prime})}{p_1 \cdot k}+\frac{8(p_1 \cdot k)}{p_1\cdot k_{Z^\prime}} \right. \nonumber
    \\&& + m^2_{Z'} \left( \frac{4(\mathbf{p_2}\cdot {\mathbf{k}_{Z^\prime}})^2}{{\mathbf{k}^2_{Z^\prime}}(p_1\cdot k)(p_1 \cdot k_{Z^\prime})}+\frac{4m_\mu^2(p_1\cdot k-p_1\cdot k_{Z^\prime})^2}{(p_1\cdot k)^2(p_1\cdot k_{Z^\prime})^2} \right. \nonumber \\
    && + \left. \left. \frac{4( p_1 \cdot k)}{p_1 \cdot k_{Z^\prime}^2}-\frac{4 \mathbf{p_2}^2}{(p_1\cdot k) (p_1 \cdot k_{Z^\prime})}+\frac{4}{p_1\cdot k}+\frac{8}{p_1\cdot k_{Z^\prime}} \right) \right] \,,
\end{eqnarray}
The amplitude square in the longitudinal mode is
\begin{equation}
\sum_{\rm spins} |{\cal M}|_L^2 \ = \ \sum_{\rm spins} |{\cal M}|_{\rm tot}^2 - \sum_{\rm spins} |{\cal M}|_T^2 \,,
\end{equation}
with 
\begin{eqnarray}
    \sum_\mathrm{spins}|{\cal M} |^2_{\mathrm{tot}} & \ = \ & e^2 g^2_{\mu\tau} \left[ \frac{8m_\mu^4(p_1\cdot k-p_1\cdot k_{Z^\prime})^2}{(p_1 \cdot k)^2(p_1 \cdot k_{Z^\prime})^2}+\frac{16m_\mu^2(p_1 \cdot k_{Z^\prime}-p_1\cdot k)}{(p_1 \cdot k_{Z^\prime})(p_1\cdot k)} \right. \nonumber \\
    && +\frac{8((p_1 \cdot k)^2+(p_1 \cdot k_{Z^\prime})^2)}{(p_1 \cdot k_{Z^\prime})(p_1\cdot k)} 
    + m^2_{Z'} \left( \frac{8m_\mu^4(p_1\cdot k-p_1\cdot k_{Z^\prime})}{(p_1\cdot k)(p_1 k_{Z^\prime})^3} \right. \nonumber \\
    && \left. \left. -\frac{4m_\mu^2((p_1 \cdot k)^2-(p_1 \cdot k_{Z^\prime})^2)}{(p_1\cdot k)^2 (p_1\cdot k_{Z^\prime})^2}+\frac{4(p_1 \cdot k-p_1 \cdot k_{Z^\prime})^2}{(p_1 \cdot k)(p_1 \cdot k_{Z^\prime})} \right) \right] \,.
\end{eqnarray}

In the limit of low-energy photon with the incoming energy $E_\gamma \ll m_\mu$, the energy of $Z'$ is given by the modified Klein-Nishina formula 
\begin{equation}
    E_{Z'} \ \simeq \ \frac{E_\gamma + \frac{m_{Z^\prime}^2}{2m_\mu}}{1+\frac{E_\gamma}{m_\mu}(1-\cos\theta)} \,,
\end{equation}
with $\theta$ being the scattering angle. The differential scattering cross section turns out to be very simple: 
\begin{subequations}
\begin{align}
\frac{{\rm d}\sigma_{\rm sC|T}}{{\rm d}\cos\theta} \ = \ & \frac{\pi\alpha\alpha_{\mu\tau}}{m_\mu^2 E_\gamma^2} \Big( m_{Z'}^2 + 2E_\gamma^2\cos\theta \Big) \,, \\
\frac{{\rm d}\sigma_{\rm sC|L}}{{\rm d}\cos\theta} \ = \ & \frac{\pi\alpha\alpha_{\mu\tau}}{m_\mu^2 E_\gamma^2} \Big[ E_\gamma^2(1+\cos^2\theta) - \cos\theta(m_{Z'}^2+2E_\gamma^2) \Big] \,. 
\end{align}
\end{subequations}
Integrating over $\cos\theta$, we arrive at the cross sections in Eqs.~(\ref{eqn:xsec:sC}). 

\bibliographystyle{JHEP}
\bibliography{refs}

\end{document}